\documentclass[twocolumn]{aastex631}
\usepackage{hyperref}
\usepackage{subfigure}
\usepackage{natbib}
\usepackage{gensymb}
\usepackage{amsmath}
\usepackage{multirow}
\usepackage{afterpage}
\usepackage{capt-of}
\usepackage{chngcntr}

\shorttitle{Ivanov et al.}
\shortauthors{Ivanov et al.}

\begin{document}

\def\Msun{\hbox{M$_{\odot}$}}
\def\Lsun{\hbox{L$_{\odot}$}}
\def\kms{km~s$^{\rm -1}$}
\def\micron{$\mu$m}
\def\tCO{$^{13}$CO}
\def\nh3{NH$_{3}$}
\def\deg{$^{\circ}$}
\def\arcsec{$^{\prime\prime}$}
\def\arcmin{$^{\prime}$}
\def\Vlsr{\hbox{V$_{LSR}$}}

\title{Discovery of a Barred-Spiral Galaxy at $z_{\rm spec}=3.16$ I: Bar Identification and Properties}


\author[0009-0005-7495-3298]{Daniel Ivanov}
\affiliation{University of Pittsburgh, 3941 O'Hara Street, Pittsburgh, PA 15260, USA}
\affiliation{University of Massachusetts Amherst, 710 North Pleasant Street, Amherst, MA 01003-9305, USA}

\author[0000-0002-7831-8751]{Mauro Giavalisco}
\affiliation{University of Massachusetts Amherst, 710 North Pleasant Street, Amherst, MA 01003-9305, USA}

\author[0000-0001-8551-071X]{Yingjie Cheng}
\affiliation{Department of Astronomy, University of Washington, Seattle, WA 98195, USA}
\affiliation{University of Massachusetts Amherst, 710 North Pleasant Street, Amherst, MA 01003-9305, USA}

\author[0000-0002-4162-6523]{Yuchen Guo}
\affiliation{Department of Astronomy, The University of Texas at Austin, Austin, TX, USA}
\affiliation{Cosmic Frontier Center, The University of Texas at Austin, Austin, TX 78712}

\author[0000-0001-6820-0015]{Luca Costantin}
\affiliation{Centro de Astrobiolog\'ia (CAB), CSIC-INTA, Ctra de Ajalvir km 4, Torrej\'on de Ardoz, 28850, Madrid, Spain}

\author[0000-0003-2676-8344]{Elena D'Onghia}
\affiliation{Department of Astronomy, University of Wisconsin, Madison, WI 53706, USA}

\author[0000-0003-1614-196X]{John R. Weaver}\thanks{Brinson Prize Fellow}
\affiliation{MIT Kavli Institute for Astrophysics and Space Research, 70 Vassar Street, Cambridge, MA 02139, USA}
\affiliation{University of Massachusetts Amherst, 710 North Pleasant Street, Amherst, MA 01003-9305, USA}
\email{john.weaver.astro@gmail.com}

\author[0000-0002-1590-0568]{Shardha Jogee}
\affiliation{Department of Astronomy, The University of Texas at Austin, Austin, TX, USA}

\author[0000-0001-7160-3632]{Katherine E. Whitaker}
\affiliation{University of Massachusetts Amherst, 710 North Pleasant Street, Amherst, MA 01003-9305, USA}

\begin{abstract}
The formation of stellar bars is an important milestone in the secular evolution of spiral galaxies, which typically indicates the presence of a massive rotationally supported disk. Determining when these structures first appeared in the early universe is crucial to constraining the timeline of galactic disk assembly. Here, we report the discovery of COSMOS-74706, a barred spiral galaxy at $z_{\rm spec} = 3.159$. Imaging of COSMOS-74706 with JWST/NIRCam indicates a disk-like morphology and spiral structure with an elongated central feature aligned between the spiral arms, most conspicuously visible in the F200W, F277W, and F356W filters. Three independent methods all support the presence of a bar: visual inspection of residuals from S\'ersic-profile fitting shows a linear structure, isophotal ellipse-fitting displays characteristic profiles of ellipticity and position angle consistent with a bar signature, and Fourier decomposition of the galaxy produces a central bisymmetric mode above a threshold strength calibrated to $z=1-3$ barred spirals. Leveraging archival Keck/MOSDEF spectroscopy overlapping with a blue clump on the edge of the galaxy, a robust redshift is inferred, with photometric constraints indicating that this structure lies at the same redshift as the main spiral. This spectroscopic evidence placing an unlensed barred spiral at $z>3$ supports the idea that galaxies with rotationally supported disks and disk-halo properties that are conducive to bar formation were already in place within 2 Gyr after the Big Bang.


\vspace{1cm}

\end{abstract}

\section{Introduction}
\label{sec:intro}

\begin{figure*}[t]
    \centering

    \newlength{\threesubfigwidth}
    \setlength{\threesubfigwidth}{\dimexpr \textwidth/3 - 0.34pt \relax}

    \subfigure
    {
        \includegraphics[width=\threesubfigwidth]{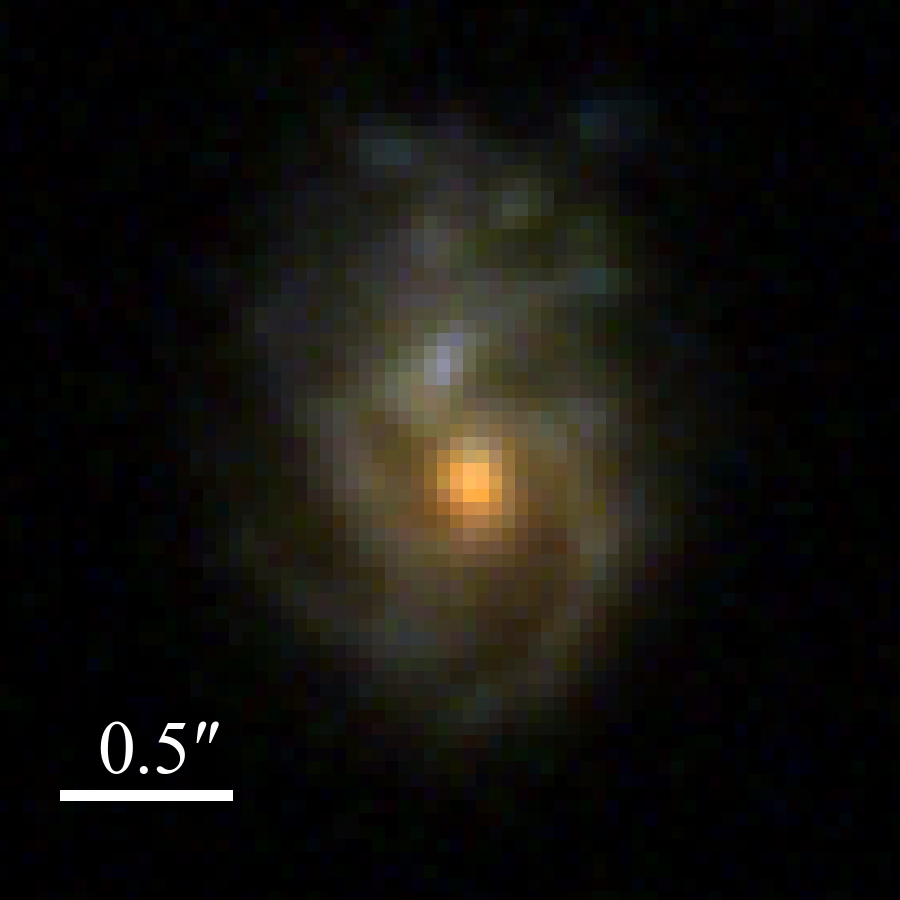}
        \label{fig:panel_left}
    }%
    \subfigure
    {
        \includegraphics[width=\threesubfigwidth]{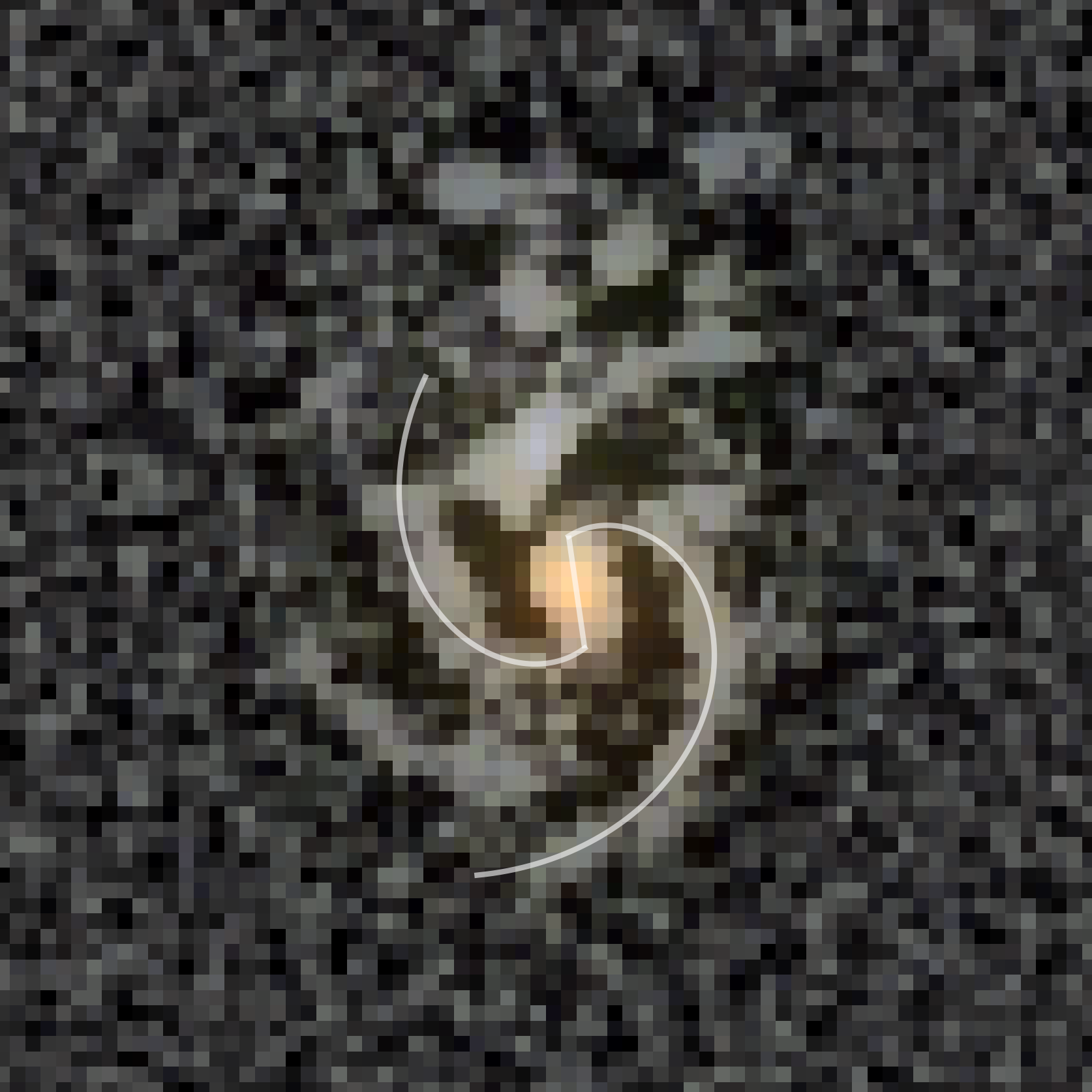}
        \label{fig:panel_middle}
    }%
    \subfigure
    {
        \includegraphics[width=\threesubfigwidth]{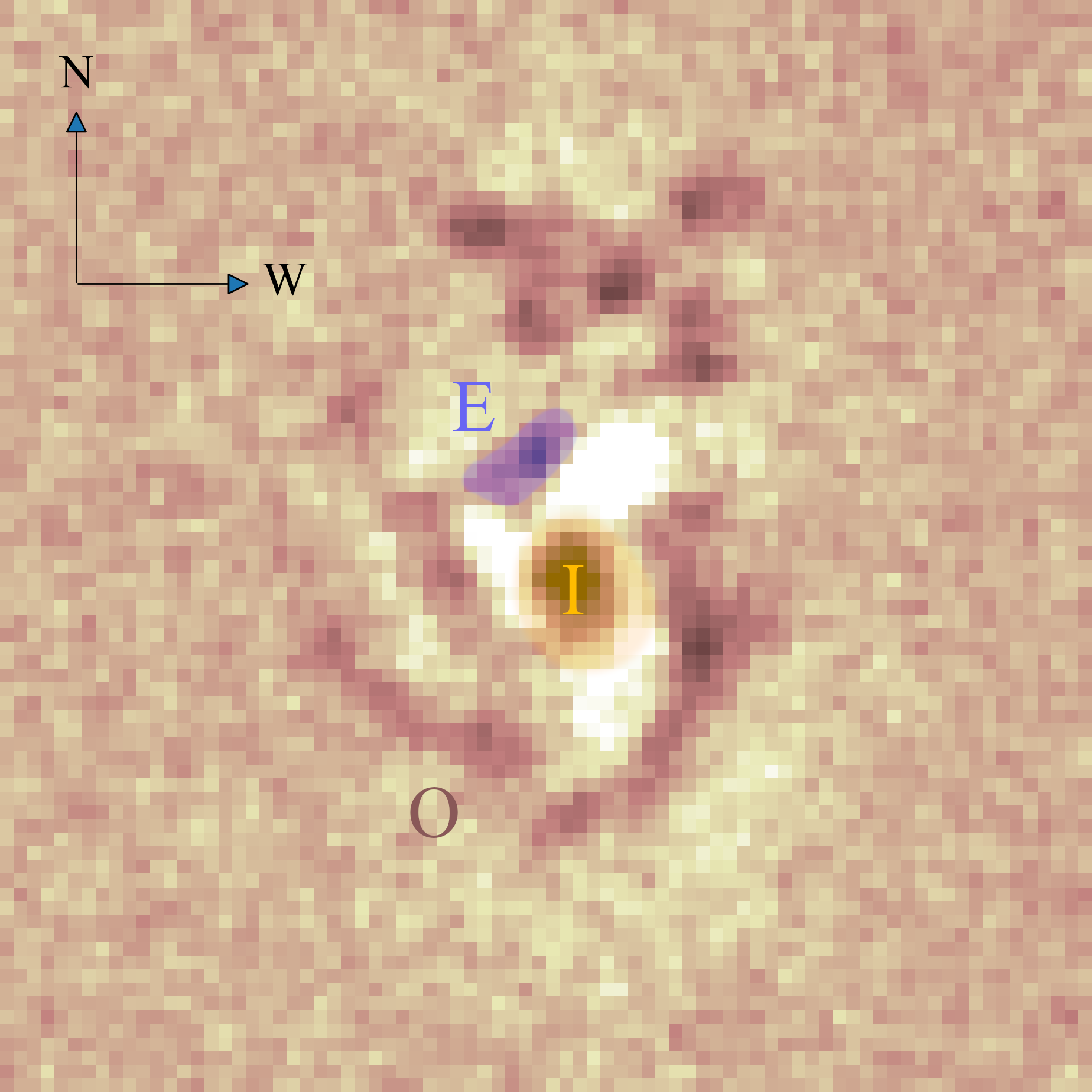}
        \label{fig:panel_right}
    }

    \caption{The panels are images of COSMOS-74706 which visually highlight the prominent spiral arms and prospective central bar structure. Each picture covers 2.6 arcseconds, corresponding to a physical scale of approximately 20 kpc along each side.
    \textit{Left panel}: A composition of COSMOS-74706 in the F200W, F277W, and F356W filters shows the spiral arms and a pronounced bulge as well as the blue northern element directly above it. \textit{Middle panel}: An unsharp mask overlaid onto the F200W, F277W, and F356W filter composition. The white lines are logarithmic spirals fitted to points along the arm structures and a line segment fitted to the approximately North to South aligned bar structure.
    \textit{Right panel}: The GALFIT residual of a one-component S\'ersic fit in the F277W filter with I, O, and E denoting the inner light and bulge region, the outer disk, and the northern element respectively.}
    \label{fig:morphology_panels}
\end{figure*}

Disk galaxies represent a substantial fraction of the overall galaxy population in the local universe, likely constituting more than half of all massive galaxies in the present epoch \citep{Willett_2013, Bamford_2009}. Among them the most prominent non-axisymmetric features are stellar bars. In the local Universe, potentially more than $60\%$ of large disk galaxies host strong bars \citep{Barazza_2008, Menendez_Delmestre_2007, Marinova_2007} with their prominence being so significant that early morphological classification schemes dedicated an entire branch of the Hubble tuning fork to this specific population of disks. 

Stellar bars are large-scale non-axisymmetric structures formed by orbital resonances resulting from perturbations to the disk structure. They can arise repeatedly during a galaxy's lifetime \citep{Bournaud_2002, Bournaud_2005}, potentially remaining stable for billions of years by transferring angular momentum from the disk to the dark matter halos \citep{Debattista_1998, Athanassoula_2002, Athanassoula_2003, Weinberg_2007}. In the presence of such dark matter halos with significant angular momentum, even dynamically hot stable disks can produce bar structures \citep{Saha_2013}. Bars often form once a rotationally supported disk becomes sufficiently massive and dynamically cold to undergo a global instability \citep{Ostriker_1973, Athanassoula_2013}, however, bar formation is a complex process and depends on the halo and disk properties, such as the disk mass fraction compared to the halo; the dynamical conditions of the disk, the relative importance of rotational energy and the gas fraction \citep{Bland_Hawthorn_2023, Bi_2022, Rosas_Guevara_2022, Beane_2023, Ghosh_2023, Romeo_2022}. The triggering mechanism for such instabilities may differ but is likely to be primarily tidal in nature \citep{Noguchi_1987, Bi_2022, Rosas_Guevara_2022, Guo_2025}. Once a disk becomes unstable, gravitational resonances naturally lead stellar orbits in the galaxy's central region to align and elongate, forming a coherent rotating density wave that constitutes the large-scale feature observed as a stellar bar \citep{Sellwood_1993}. 

Bars play a key role in their host galaxy's secular evolution promoting star formation and redistributing gas and angular momentum outwards \citep{Kormendy_2013, Athanassoula_1992, Kormendy_2004, Jogee_2005} while substantively altering the neutral hydrogen gas fraction \citep{Mukundan_2023}, possibly indirectly feeding the supermassive black hole (\citealt{Carles_2016, Jogee_2006} and references therein), and possibly contributing to the assembly of bulges or pseudo-bulges \citep{McClure_2025, Zhou_2020, Kormendy_2004}. Understanding when these crucial structures began to emerge is vital for constraining the timeline of early galaxy assembly and disk settling. 

Observations of bar formation at earlier cosmic epochs have historically been complicated by limitations in spatial resolution and the availability of morphological classification \citep{Simmons_2014}. As a result, prior to the advent of the James Webb Space Telescope (JWST), observations suggested that the Hubble tuning-fork morphology sequence, including well-defined spirals, was largely unestablished prior to $z\sim 2$. This lack of consensus on whether bars were at all common in disks at cosmic noon was foremost driven by an absence of instrumentation capable of finding such features \citep{Conselice_2014}. Before JWST's launch in 2021, the highest redshift barred spiral was reported at $z=1.97$ by \cite{Simmons_2014}. JWST however has due to its unprecedented resolution and sensitivity to the rest-frame optical to near-infrared opened a wider window into the high-redshift universe, significantly enhancing our ability to characterize the morphology of galaxies out to $z \geq 2$ \citep{Guo_2025, Guo_2023, Costantin_2023, Huang_2023, Amvrosiadis_2025, LeConte_2024, McKinney_2024, Salcedo_2025, Geron_2025, Huang_2025, Huertas_2025, LeConte_2025} and revealing the existence of well-defined spiral galaxies at significantly earlier epochs \citep{Tsukui_2021, Jain_2024}.

The increasing availability of higher resolution instrumentation has led to rapid advancements in the ability to identify bars at earlier times and has allowed numerous bars at $z>2$ to be uncovered \citep{Guo_2023}, with the highest redshift spectroscopic confirmation at $z=2.467$ \citep{Huang_2023}. \citealt{Costantin_2023} pushed the earliest known definite barred spiral further to a redshift of $z_{\rm phot}\simeq 3.03$ and \citealt{Guo_2025} has presented a strong candidate at $z_{\rm spec}=3.10$. The unambiguous identification of such features, however, remains a challenge. Most recently, observations in strong gravitationally lensed galaxies have revealed features at even earlier times that are highly suggestive of early onset bar formation \citep[$z=3.76$ and $z=4.26$;][]{Smail_2023, Amvrosiadis_2025}. Evidence now indicates that spiral galaxies likely constitute more than $20\%$ of the massive galaxy population at $z \sim 3$ \citep{Kuhn_2024} and that the bar fraction may exceed $5\%$ in this epoch, with suggestions that barred galaxies might already exist at redshift $z \geq 3.5$ \citep{Guo_2025, Huertas_2025, Geron_2025}.

These discoveries suggest that disk galaxy assembly and the emergence of stable bar structures can occur on relatively short cosmic timescales. Stellar kinematics simulations indicate that bar formation should be largely absent at $z>6$ due to the irregularity and turbulent nature of galaxies in this epoch \citep{Zana_2022}. High gas fractions in early galaxy disks are also expected to suppress bar formation \citep{Bland-Hawthorn_2024} or limit the slowing down and growth of bars \citep{Beane_2023}. Early JWST/NIRCam bar observations following this era are thus key to testing feedback models and constraining the characteristic length of time required for bars to begin to emerge.

The ascertainment of barred features in NIRCam imaging requires robust and unambiguous analysis. The most common technique is that of visual inspection, usually with the aid of ``force multipliers'' such as the use of multi-spectral imaging or of multiple participants. More quantitative techniques include multicomponent profile fitting in which S\'ersic profiles are fit to a galaxy's projected image to disentangle the bar component \citep{Gadotti_2009, Weinzirl_2009}, isophotal ellipse-fitting, in which a bar is identified by the distinctness of its shape in comparison to the host \citep{Jogee_2004, Marinova_2007}, and Fourier decomposition, in which the bisymmetry of the galaxy is directly measured \citep{Considere_1988, Ohta_1990}. The consistent identification of a bar across multiple independent methods provides the most secure evidence for its existence.

In this paper, we present the discovery of COSMOS-74706 (\autoref{fig:morphology_panels}), a barred spiral galaxy at $z>3$ identified by visual inspection of multi-component \texttt{GALFIT} \citep{Peng_2010} profile-fitted residuals, isophotal ellipse-fitting, and Fourier multiplicity decomposition. The sky-coordinates of this galaxy are given in \autoref{tab:coords}. Spectroscopic observations obtained with the Multi-Object Spectrometer For Infra-Red Exploration (MOSFIRE; \citealt{McLean_2012}) instrument on Keck I as part of the MOSFIRE Deep Evolution Field (MOSDEF; \citealt{Kriek_2015}) survey provide unambiguous line detections that anchor a bright spatially associated Northern component at redshift $z=3.159$. In a companion study (\citealt{Cheng_2026}; hereinafter Paper II), we present a detailed analysis of COSMOS-74706's SED that robustly confirms the main barred galaxy lies at the same redshift as this component. We describe the data we utilized throughout the course of this study in \S\ref{sec:data}. In \S\ref{sec:method}, we outline the procedures adopted for target selection, data reduction, morphological classification, and bar verification. Then, \S\ref{sec:results} details the results of our analysis. We discuss the broader implications of this discovery for our understanding of disk galaxy assembly and early bar formation in \S\ref{sec:discussion}. Finally, we summarize our conclusions in \S\ref{sec:conclusion}.

We adopt a flat $\Lambda$CDM cosmology with $H_0 = 69.6\textrm{ km/s/Mpc}$ with $\Omega_m = 0.286$ and $\Omega_\Lambda = 0.714$. At $z=3.159$ this cosmology gives a proper distance scale of $7.733 \textrm{ kpc/}$\arcsec and a universe age of 2.049 Gyr. All magnitudes are given in the AB system.

\begin{table}[h]
  \centering
  \setlength{\tabcolsep}{8pt}
  \begin{tabular}{ l c c }
    \hline
    \textbf{Source} & \textbf{RA} & \textbf{DEC}  \\
    \hline
    COSMOS-74706   & 150.1007649  & 2.341443311 \\
    3DHST-15810    & 150.100800   & 2.3414753   \\
    \hline
    \multicolumn{3}{c}{\textbf{Derived properties}} \\
    \hline
    \multicolumn{2}{l}{$z_{\rm spec}$}                 & $3.1591$ \\
    \multicolumn{2}{l}{$\log(M_\star/M_\odot)$}        & $10.63 \pm 0.13$ \\
    \multicolumn{2}{l}{SFR ($M_\odot\,\mathrm{yr}^{-1}$)} & $\sim 40$--$50$ \\
    \multicolumn{2}{l}{Mass-weighted age (Gyr)}        & $0.91 \pm 0.16$ \\
    \multicolumn{2}{l}{$A_V$ (mag)}                    & $1.13^{+0.08}_{-0.11}$ \\
    \multicolumn{2}{l}{$\log(Z/Z_\odot)$}              & $-0.54^{+0.16}_{-0.14}$ \\
    \hline
  \end{tabular}
  \caption{J2000 sky coordinates of the sources corresponding to the target galaxy, and SED-derived physical properties of COSMOS-74706 from Paper II.}
  \label{tab:coords}
\end{table}

\section{Data}
\label{sec:data}

This study is based primarily on JWST/NIRCam imaging from the PRIMER survey. We used HST/ACS imaging to construct a mask for a blue contaminant feature that partially intruded into the bar region and may be associated with either clumpy star formation or a foreground interloper galaxy. Ground-based spectroscopy from Keck I was used to constrain the redshift. 

\subsection{JWST PRIMER Imaging}
\label{subsec:primer_data}

The core imaging data used in this work are from the JWST Public Release IMaging for Extragalactic Research (PRIMER) Treasury Programme (PID 1837; \citealt{Dunlop_2021}). PRIMER is a large area (400 arcmin$^2$) multi-wavelength GO ``galaxies'' imaging survey undertaken by JWST in Cycle~1 that is designed to provide infrared MIRI and NIRCam imaging in two fields that overlaps with existing archival optical/UV imaging taken with the ACS instrument on HST through the Cosmic Assembly NIR Deep Extragalactic Legacy Survey (CANDELS) program \citep{Grogin_2011, Koekemoer_2011}. Our target COSMOS-74706 lies in the equatorial 200 arcmin$^2$ COSMOS field \citep{Scoville_2007}, and is also covered by COSMOS-Web \citep{Casey_2023}. The NIRCam observations were obtained in seven wide-band filters F090W, F115W, F150W, F200W, F277W, F356W, and F444W, as well as one medium-band filter F410M, spanning a wavelength range 0.9\micron - 4.44\micron $\textrm{ }$ over integration times of 14-84 minutes. Additionally, we obtain MIRI photometry in the F770W band. At $z=3.159$, these filters probe from the rest-UV to rest-IR wavelengths (0.21\micron - 1.85\micron).

The NIRCam data products were processed with the \texttt{Grizli} pipeline \citep{Brammer_2019} which is a modified version of the standard JWST science-calibration pipeline that includes steps for detector-level corrections, striping removal, flat-fielding, photometric calibration, masking of snowball and wisp artifacts, and removal of cosmic rays. To ensure co-location of features across overlapping frames, the calibrated dithered exposures were aligned to a mutual astrometric frame using the $\textit{Gaia}$ Data Release 3 \citep{Gaia_Collaboration_2023} catalog as a basis, then drizzled and stacked together to correct for geometric distortions and improve sampling of the PSF \citep{Fruchter_2002}. The coadded images reproduce the same fine pixel scale of 0.04\arcsec/pixel across all filters. We used the fully reduced mosaic science images and corresponding weight maps modeling the observational noise from the v2.0.0 PRIMER data release.

For the morphological fitting with \texttt{GALFIT} described in \S\ref{subsec:fitting}, it is necessary to model the response of the instrument to a point source. We utilized the empirical PSFs provided by the PRIMER collaboration, which were constructed individually for each NIRCam filter by directly stacking isolated and unsaturated stars from the science mosaics \citep{Weaver_2024, Cutler_2024}.

\subsection{HST CANDELS Imaging}
\label{subsec:candels_data}

Our target is spatially coincident with the source 3DHST-15810 from the 3D-HST survey \citep{Brammer_2012, Momcheva_2016, Skelton_2014} which corresponds in location to the blue northern element. This survey provides a low-resolution WFC3/G141 grism redshift for the source 3DHST-15810 at $z_{\rm grism} = 3.24$, but no slit spectroscopic redshift. We discuss the grism redshift further in \autoref{subsec:redshift}.

In order to mask the blue light from the blue northern element, we used data from the Advanced Camera for Surveys (ACS) Wide Field Channel (WFC) with F606W (V-band, 0.606\micron) and F814W (I-band, 0.814\micron) filters providing a spatial resolution of 0.06\arcsec/pixel. Due to its Balmer-break nature at $z\sim 3.1$, COSMOS-74706 is virtually undetected at these wavelengths, save for the much bluer northern element. The HST data were reduced using the standard HST pipeline and drizzled with \verb|AstroDrizzle| \citep{Fruchter_2010}.

\subsection{MOSFIRE Spectroscopy}
\label{subsec:keck_data}

In order to constrain the redshift of COSMOS-74706 we utilized ancillary spectroscopic data from the 10-meter Keck I telescope in Hawaii. MOSDEF targeted the galaxy corresponding to the HST source 3DHST-15810 and obtained near-infrared spectra in the four bands Y (0.97\micron-1.12\micron), J (1.15\micron-1.35\micron), H (1.46\micron-1.81\micron), and K (1.93\micron-2.45\micron) with the MOSFIRE instrument. Because the HST coordinates of the galaxy are coincident with the blue northern element, the spectra obtained correspond to this feature. The MOSDEF survey data were processed with a custom IDL pipeline including steps for telluric correction and sky subtraction \citep{Kriek_2015}.

\section{Methodology}
\label{sec:method}

In this section we outline the approach we used to identify COSMOS-74706 as a target of interest and explain the analysis of its morphology. We first describe the criteria adopted to select the galaxy out of a parent catalog in \S\ref{subsec:target_selection}. Then we detail the process of creating and applying the masks we used in \S\ref{subsec:masking}, followed by describing the suite of three modes of analysis which were used to confirm the presence of a bar in \S\ref{subsec:fitting}, \S\ref{subsec:isophotal}, and \S\ref{subsec:fourier} respectively.

\subsection{Target Selection}
\label{subsec:target_selection}

We use the v2.0.0 PRIMER-COSMOS internally consistent photometric catalog \citep{Cutler_2024} which is generated from a long-wavelength detection image created from a noise-equalized stack of the F277W, F356W, and F444W science mosaics. Photometric redshifts, stellar masses, and star formation rates (SFRs) are derived via the EAZY code \citep{Brammer_2008} including its extensions for SPS-parameter inference \citep{Gould_2023}. The EAZY-derived parameters are cross-matched with spectroscopic redshifts when available. We note that while systematic offsets compared to other physically motivated SED frameworks are known to exist for EAZY-derived parameters, our main goal is to ascertain galactic morphology and thus is not sensitive to these uncertainties. Galaxies in the redshift range $4.0 \geq z\geq 1.0$ of mass $\log M_\odot \geq 10$, and showing potential disk-like morphologies are flagged for follow-up with automated isophotal and Fourier-based bar-identification analysis, as detailed in \S\ref{subsec:fourier}, as well as fit with a single S\'ersic profile using GALFIT. A full description of this parent sample and its properties will be presented in a forthcoming paper (Ivanov et al., in prep). We single out COSMOS-74706 for further inspection due to (1) an initial favorable assessment of barred character in the autonomous assessment, (2) its well-defined disk component in multi-filter imaging, (3) a spectroscopic redshift identification of $z=3.159$, and (4) visual evidence of bar structure in the F277W filter.

By adjusting the contrast and brightness levels in DS9 \citep{Joye_2003} to optimize the visibility of low-surface-brightness features, we manually inspect COSMOS-74706 in the NIRCam filters F150W, F200W, F277W, F356W, F410M, and F444W. We further ascertain the presence and coherence of spiral arms, as well as identify any possible contamination from foreground sources or diffraction spikes.

\begin{figure*}[t]
    \centering

    \subfigure
    {
        \includegraphics[width=1.0\textwidth]{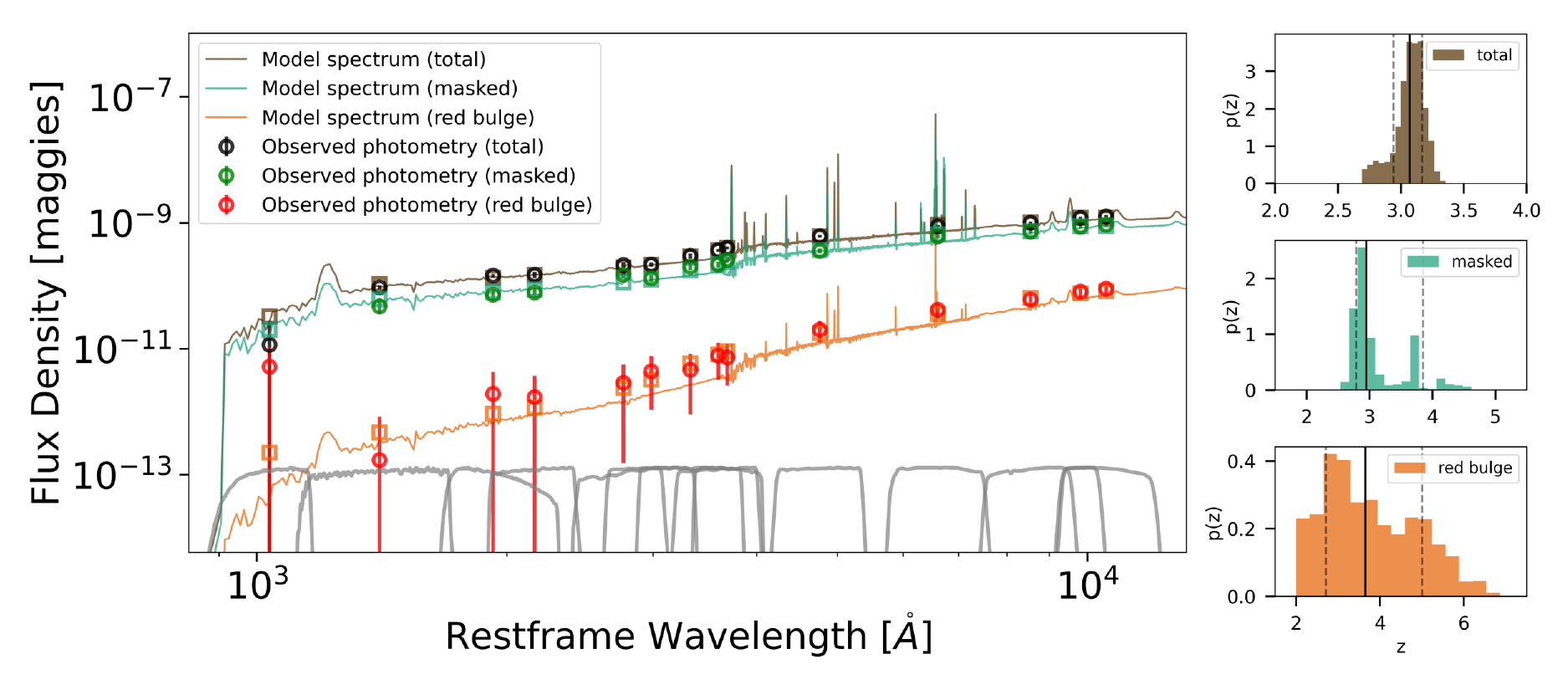}
    }
    \hspace{\fill}

    \caption{Best-fit SEDs of COSMOS-74706 are shown at left, including fits to the entire galaxy (black), with the blue northern element masked (green), and on the bulge region only (red). The gray curves at the bottom depict the transmission of the available filters.
    Histograms on the right side depict the corresponding redshift probability distributions $p(z)$ with the solid black vertical lines depicting the best fit redshift, and the dashed vertical gray lines depicting the $1\sigma$ error. The photometric redshifts produced by the fitting of each component are respectively $z_{\rm phot}(\text{total})=3.07^{+0.10}_{-0.13}$, $z_{\rm phot}(\text{masked})=2.95^{+0.90}_{-0.16}$, and $z_{\rm phot}(\text{bulge})=3.65^{+1.36}_{-0.94}$, all of which are consistent with $z_{\rm spec}$.}
    \label{fig:sed_pz}
\end{figure*}

\subsection{Redshift Determination}
\label{subsec:redshift}

The MOSFIRE spectra of the blue northern element show strong emission lines [OIII]$\lambda$5007 and H$\beta$4861 in the K-band as well as the unresolved [OII]$\lambda$3729 doublet in the H-band. As detailed in Paper II, these lines unambiguously place the redshift of this feature at $z_{spec} = 3.1591$. The 3D-HST survey independently provides a WFC3/G141 grism redshift of $z_{\rm grism} = 3.24^{+0.01}_{-0.03}$ for the source which is broadly consistent with, though offset from, the MOSFIRE determination, likely reflecting the limited spectral resolution ($R\sim 100$) and susceptibility to contamination inherent to slitless grism extractions \citep{Momcheva_2016}. We thus adopt the higher resolution ($R \sim 3500$) MOSFIRE determination for the spectroscopic redshift.

The combination of strong lines in the MOSFIRE spectrum and the blue color implies that the feature contains young stellar populations and may be gas-rich. This combination of qualities leaves open the question of whether the blue element is a bright compact star-forming complex within COSMOS-74706 or a foreground galaxy. To robustly resolve this, we perform detailed SED modeling which is depicted in \autoref{fig:sed_pz} and is presented in greater detail in Paper II. The figure shows the best-fit templates and the redshift PDFs associated with the SED fits to the red central bulge region, and to the whole galaxy both with and without masking the northern element.

Two key findings strongly support the conclusion that the blue northern element lies at the same redshift as the spiral
\begin{enumerate}
    \item SED fitting of the full system including the northern element yields a photometric redshift of $z_{\rm phot} = 3.07^{+0.10}_{-0.13}$, which is fully consistent with the spectroscopic determination from MOSFIRE
    \item Re-fitting upon masking the northern element yields $z_{\rm phot} = 2.95^{+0.90}_{-0.16}$, which strongly disfavors the possibility that the spiral is at a much higher redshift than the blue element.
\end{enumerate}

The scenario in which COSMOS-74706 is a foreground galaxy is also disfavored, as the high surface brightness of the blue northern element shows minimal dust obscuration or reddening, which would be expected if its light had to pass through a dusty disk. Furthermore, no two strongly detected independent sets of lines are found as would be expected if two spatially coincident galaxies existed at markedly different redshifts, implying that if the northern element is a foreground interloper, it exists at a very similar redshift to COSMOS-74706. Finally the redshift solution is qualitatively supported by COSMOS-74706's lack of detection in the F606W and F814W Hubble filters corresponding well to the Balmer break expected for a $z\sim 3$ galaxy.

This places COSMOS-74706 at $z_{spec} = 3.159$. While the exact nature of the northern element remains unresolved, all indicators we interrogate point to both components co-existing at the same epoch.

\begin{figure*}[t]
    \centering

    \subfigure
    {
        \includegraphics[width=1.0\textwidth]{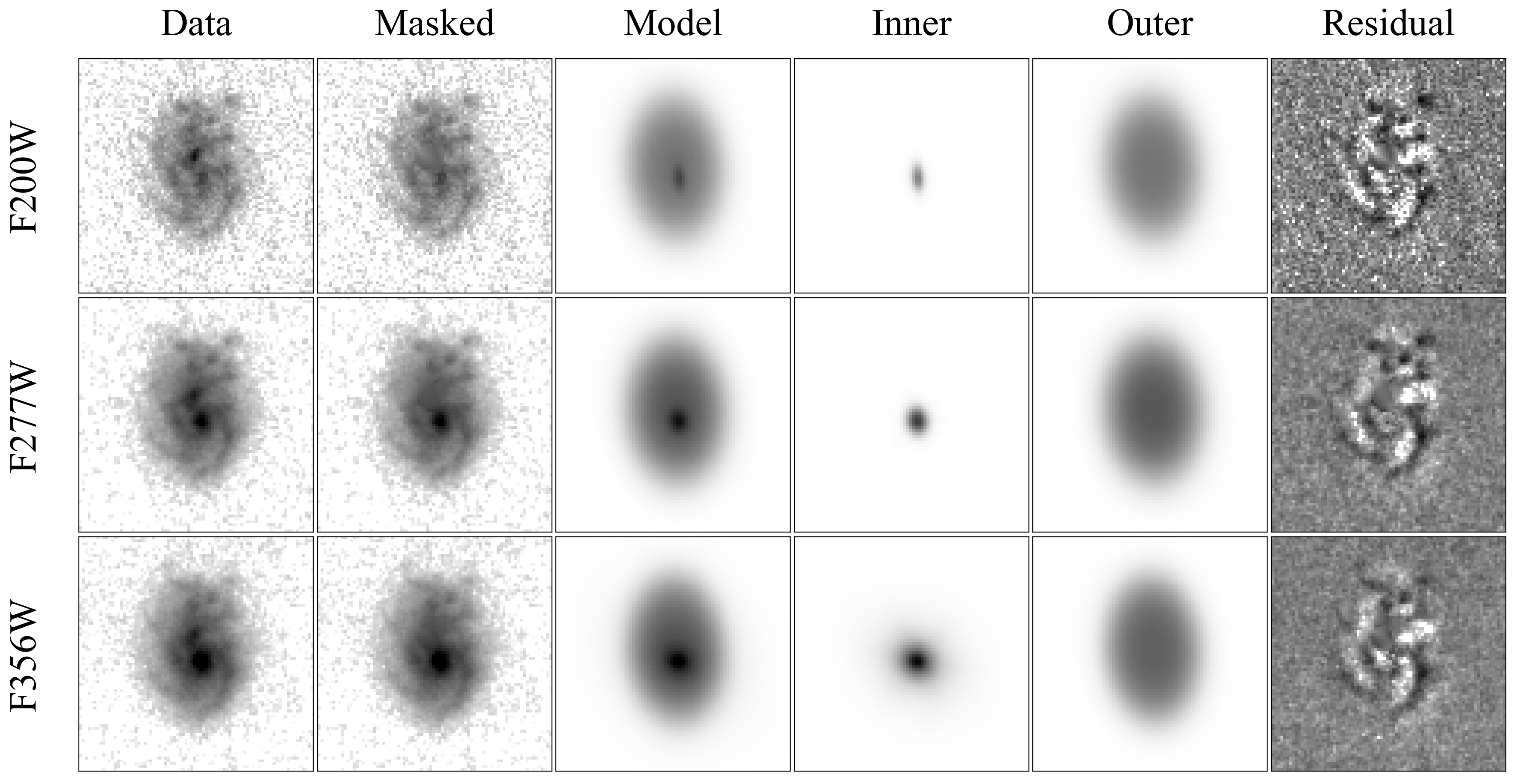}
    }
    \hspace{\fill}

    \caption{GALFIT decompositions of COSMOS-74706 found by MCMC exploration in three different JWST/NIRCam filters as indicated on the left. The columns display from left to right (1) the science mosaic image, (2) the same image but with the northern element masked, (3) the full two-component GALFIT model of the light profile, (4) the inner component of the model, (5) the outer component, and (6) the model residual with the disk and inner light fit-away.}
    \label{fig:decomp}
\end{figure*}

\subsection{Masking}
\label{subsec:masking}

In order to prevent the contribution of the bright, blue northern element from biasing the results of the morphological assessment, we use 3DHST ACS imaging to create a set of masks. The almost non-detection of COSMOS-74706 in the HST/ACS imagery contrasts with the visibility of the bright and compact northern element making these data ideal for reliably isolating the feature. In comparison, in the NIRCam imaging the light from the Northern feature is blended with light from the inner disk. We utilize a stack of the F606W and F814W HST/ACS images from which we use a source-detection algorithm to produce segmentation maps of the contaminant selecting pixels above a certain threshold SNR. We use noise level cutoffs of $2.5\sigma$, $3\sigma$, $3.5\sigma$, and $4\sigma$ to create a total of four masks of the northern element. Upon overlaying each of these onto the NIRCam images, the masked pixels are excised and their values replaced with a 2D interpolation using the \verb|griddata| function from \verb|scipy.interpolate| with the method set to linear. We then conduct three independent searches for signatures of a stellar bar in COSMOS-74706 using these masked images, consisting of a visual identification, an isophotal analysis, and a Fourier analysis.

\subsection{Light Profile Modeling}
\label{subsec:fitting}

In order to isolate non-axisymmetric features for further visual inspection we model the symmetric light of the galaxy alongside an inner nuclear component using the two-dimensional fitting software GALFIT \citep{Peng_2002, Peng_2010}. GALFIT is designed to take guessed input parameters for a number of S\'ersic profiles and run a minimization to find the best fit to a galaxy's light profile. Once GALFIT converges the model can be subtracted from the data to generate a residual image that effectively highlights any features not well captured by the fit. In particular for diffuse light profiles like that of a disk or bulge component, this includes any spiral arms or central bar. To best interpret the galaxy's morphological characteristics with this method, we perform all fitting in the F200W, F277W, and F356W filters so as to capture emission where the bar is highly distinct as well as where a large quantity of red light is emitted. We use the $3\sigma$ noise-level cutoff mask where the bright Northern source is effectively suppressed while the masked area does not significantly overlap the bar region.

In the local universe spiral galaxies generally have several components to their light profiles including contributions from a classical bulge ($n \sim 4$) or a pseudobulge ($n \lesssim 2$), an outer disk ($n \sim 1$), and often a potential bar ($n \sim 0.75$). A single S\'ersic parameterized profile thus generally will not suffice to describe the light profile. We therefore used GALFIT to model the galaxy both with two components (central component + disk) and three components (bar + bulge + disk), allowing the parameters to vary within wide physically sensible bounds. To search the large parameter space of possible input parameters for those which produce the optimal model fit, we ran a Markov-Chain Monte-Carlo algorithm for $N=10000$ steps in each filter which was sufficient to fully explore the posterior and obtain errors on the fitted quantities. The MCMC code was allowed to explore guess inputs within the same wide physical bounds constraining the GALFIT optimizer, as displayed in \autoref{tab:galfit}.

While the three-component model consistently achieved a marginally better reduced chi-squared of $\chi_\nu = (1.096, 1.024, 1.004)$ for the F200W, F277W, and F356W filter than the two-component model with $\chi_\nu = (1.101, 1.124, 1.167)$, visual inspection of the three-component model and its residuals revealed that this statistical improvement was due to overfitting. The bulge component did not converge on a bulge-like structure and instead contorted to absorb flux from the inner region of the spiral arms. Adjusting the bounds to be more constraining resulted in the fit repeatedly hitting the boundaries of the prior. We therefore adopt the more robust and physically interpretable two-component model for the analysis in \autoref{sec:results}. 

The lowest $\chi_\nu$ two-component model is displayed in \autoref{fig:decomp} which shows the light profiles resulting from the fitting and the corresponding residuals. The fits are effective at isolating both the disk and inner light with the spiral arm structure particularly apparent in the residuals. 

\subsection{Isophotal Analysis}
\label{subsec:isophotal}

Following the procedure in \citet{Jogee_2004} and adapted by \citet{Guo_2023} for JWST data, we perform an isophotal ellipse-fitting analysis on every interpolated mask and filter combination image of COSMOS-74706 using \verb|photutils| \citep{Bradley_2024}. This method characterizes the shape of the galaxy by fitting a series of concentric ellipses to contours of constant surface brightness.

In order for the isophotal analysis to be an effective indicator of bar presence, it is crucial to accurately determine the point at which to fix the isophotes. This is typically selected by first running \verb|Ellipse.fit_image| with a free center and then taking the average centroid of all of the ellipses to be the center of the galaxy. In our case, since we are examining the same galaxy in multiple filters and need to be able to compare the resulting isophotes, we cannot determine the center this way as it changes between filters. In the shorter wavelength filters the average center of free isophotes is strongly pulled north of the bulge due to a brighter patch and the influence of the outer regions where the masked light from the northern element still strongly affects the apparent morphology. To avoid this bias, we adopt as the fixed center for the F277W, F356W, F410M, and F444W fits the brightest pixel in the F410M, F356W, and F277W masked images which is consistent across those filters. For the F150W and F200W filters this pixel falls well south of the high surface brightness area and the brightest pixel in each of the filters is located at the edge of the bright region rather than at its center. Thus for those two filters we fix the center to the visually identified midpoint of the bright area.

Thereafter we run \verb|Ellipse.fit_image| with the center fixed at this spot, geometrically increasing the semi-major axis in increments of a factor of 1.1. At each step, we employ a $3\sigma$ clipping to reject outlier pixels along the isophotes path so as to improve the robustness of the fit. The fitting is terminated when the signal-to-noise ratio along an isophote drops below 3 upon reaching the galaxy’s faint outer regions. This produces a set of $\sim 40$ elliptical isophotes of outwardly increasing radii. Using these, we obtain radial profiles of surface brightness ($\mu$), ellipticity (e), and position angle (PA). 

We evaluate the presence of a bar using the four established criteria described in \citet{Jogee_2004, Marinova_2007} requiring:
\begin{enumerate} 
\item A smooth rise in ellipticity within the putative bar region, reaching $e>0.25$.
\item A roughly constant position angle (within $30^\circ $) about the same region.
\item A subsequent drop in ellipticity of $\Delta e > 0.1$ in the outer disk region. 
\item A shift in the position angle of at least $10^\circ$ between the outer disk isophotes and the bar region. 
\end{enumerate} 
To best encompass the characteristics of a bar signature across the full ensemble of filters, we define our bar region by a conservative union. The inner boundary is set at the outermost radius isophote in any filter at which the ellipticity begins to rise monotonically while the outer boundary is defined by the innermost radius isophote where the ellipticity has peaked. The fulfillment of all four criteria is required for a given filter-mask combination in order to constitute a positive indicator of the presence of a bar.

While the F444W filter is more commonly used in isophotal analysis so as to trace the smoother light of the older stellar populations which are most readily present in the bar region, we also opt to analyze the shorter-wavelength filters in this manner. These are more affected by the clumpy star formation probed at shorter optical wavelengths, but this detriment is offset by the gained resolution and the increased prominence the COSMOS-74706 visual bar exhibits in the F200W-F356W filters. This multi-spectral approach further allows us to confirm if the same feature is present across multiple bands.

\subsection{Fourier Decomposition}
\label{subsec:fourier}

For the third independent test we conduct a Fourier analysis to help further ascertain the presence of a nuclear bar in accordance with the methodology first developed by \cite{Garcia_Gomez_1991} and later refined in \cite{Garcia_Gomez_2017}. This technique leverages the fact that logarithmic spirals well-approximate typical galactic spiral arms by decomposing the galaxy image in the basis of logarithmic spirals. The Fourier profile in the $m=2$ multiplicity mode which is sensitive to bisymmetric structures is then used to reveal the presence of a bar. A crucial initial step in this method requires the deprojection of the source angle. In our case the presence of the northern element as a possible foreground interloper makes a reliable deprojection highly uncertain and risks introducing spurious signals and significant artifacts that can distort or otherwise affect an existing bar signature. Furthermore, the edge of the disk is not very well defined and we know little about the scale height of disks at high redshift which introduces additional complications.

The GALFIT model seen in \autoref{fig:decomp} shows an elliptical outer disk with axis ratio $q \sim 0.64-0.67$ depending on the filter. This ellipticity presents two possibilities: (1) an approximately circular disk viewed at moderate inclination or (2) an intrinsically triaxial ellipsoid disk viewed at low inclination. Statistical studies of large galaxy samples have shown that spiral galaxy disks are often not perfectly circular, with a mean intrinsic axis ratio of around $\langle q \rangle \approx 0.8$ and a distribution exhibiting significant scatter with $\sigma_q \approx 0.2$ \citep{Rodriguez_2013} with such high intrinsic ellipticities increasingly common for redder galaxies \citep{Ryden_2004}. Since the outer disk shape is consistent with an intrinsic ellipticity and given that any deprojection attempt would be highly uncertain due to the contaminant feature, as well as an observed absence of significant foreshortening of the galaxy's spiral structure, we proceed by assuming an approximately face-on orientation. We discuss the implications of this assumption in \autoref{sec:discussion} where we also note that the choice not to deproject does not affect our results.

Next we re-map the galaxy image from cartesian coordinates $I(x, y)$ to a log-polar grid $I(u, \theta)$ with $u=\ln r$ using 1000 discrete angular steps and 512 logarithmic radial steps extending to a maximum radius of 39 pixels or 1.56\arcsec. Then in order to avoid aliasing and ensure the bulge does not dominate the decomposition, we taper the galaxy's previously-determined center by multiplying the pixel values within $k$ pixels of the center by a profile of $\sin^2(\pi r/2k)$ \citep{Garcia_Gomez_2017}. We adopt a value of $k=5$ so as to minimally obscure the bar region and use the same center in all images as determined from the longer wavelength filters.

Following \cite{Garcia_Gomez_2017} the two-dimensional Fourier transform is then taken as
\begin{equation}
    A(p, m) = \iint I(u, \theta)e^{i(pu+m\theta)} \, d\theta \, du
\end{equation}
Where $A(p,m)$ gives the transformed map as a function of $p$ which is called the logarithmic pitch angle parameter and corresponds to the tightness of the spiral wrapping, as well as $m$ which is a multiplicity giving the number of arms. The Fourier amplitudes are computed for integer multiplicities $m=1, \, 2, \, 3, \, 4$ each at 2048 discrete points over the range of pitch angle parameters $p=\pm 128$.

To isolate the bar's contribution, the resulting one-dimensional $m=2$ power spectrum $|A(p,m)|$ is modeled as a sum of Gaussian functions. We employ an iterative fitting procedure beginning by inputting to \verb|scipy.optimize.curve_fit| a quantity of guess Gaussians equal to the number of local maxima in the spectrum with peaks at least 10\% of the global maximum. If the number of such peaks exceeds 15, the fitting is stopped and the image is judged to lack evidence of a bar -- in our testing, such noisy decompositions are usually an indicator of very low bisymmetry and are often a symptom of poor SNR. Assuming this does not occur, a residual spectrum is then calculated by subtracting the best-fit model from the original data and a new Gaussian is added at the location of the maximum residual. Thereafter the fitting process is rerun now including the new component, allowing the parameters of all existing Gaussians to adjust in response. This iterative process of calculating residuals, adding a new component, and re-fitting the model is continued until either the model achieves an excellent goodness of fit with an R-squared value $R^2 > 0.995$ or a maximum of 15 components has been reached at which point the iteration procedure is halted in order to avoid over-fitting.

Once the best fit is finalized the bar component is identified as the highest amplitude Gaussian centered between $p=-1.0$ and $p=1.0$. If no such component exists then the image is judged to be not indicative of the presence of a bar. Assuming such an element is present, we then integrate the area of the $m=2$ Fourier profile in p-space and derive the \cite{Garcia_Gomez_2017} bar strength indicator $B_{M}$ termed the bar modulus which is given by
\begin{equation}
    B_{M} = C_B \int_{-p_{\text{max}}}^{p_{\text{max}}} \exp \left( - \frac{(p - p_B)^2}{ 2 \sigma_B^2} \right) dp
\end{equation}
With $p_B$ the bar pitch angle given by the center of the Gaussian it corresponds to, $\sigma_B$ the standard deviation of this component, and $C_B$ its amplitude.

\begin{figure}
    \centering
    \includegraphics[width=1.0\linewidth]{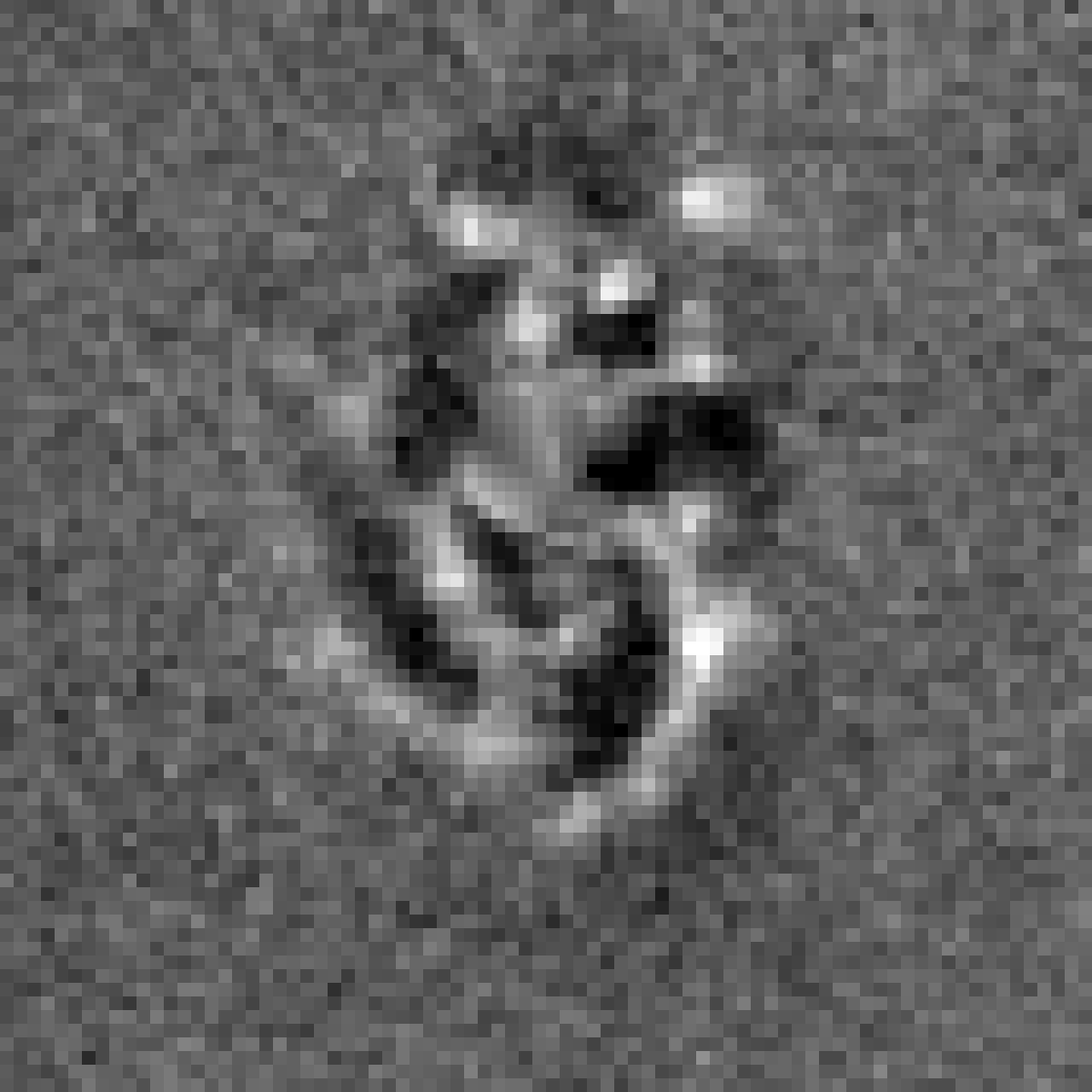}
    \medskip
    \includegraphics[width=1.0\linewidth]{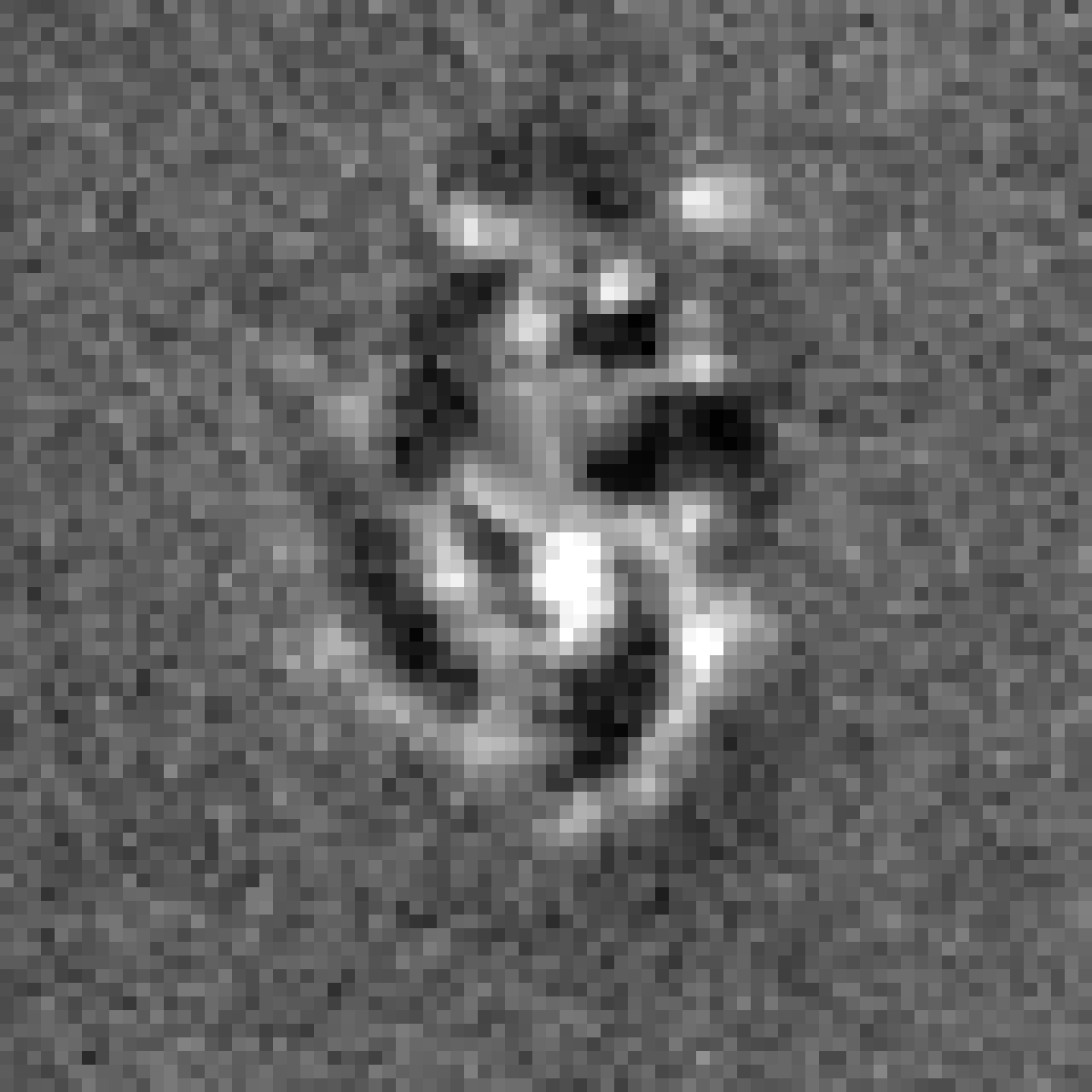}
    \caption{\textit{Upper panel:} The sum of the best-fit 2-component S\'ersic light profile models for all three filters subtracted from the sum of the three images at the $3\sigma$ mask cutoff level. A bar-like structure can be seen connecting the residuals of the spiral arms. \textit{Lower panel:} 
    As upper panel but with the inner light profiles added back in with a different stretch to highlight the elongation without saturating the image.}
    \label{fig:barred_unbarred}
\end{figure}

To establish a quantitative threshold for bar identification, we perform a calibration using an ensemble of 396 visually identified disk galaxies taken from the parent catalog detailed in \S\ref{subsec:target_selection}. This calibration sample is constructed to contain 99 galaxies identified as barred by both isophotal analysis and visual inspection, 99 each that are identified by one method and not the other, and finally 99 galaxies identified as unbarred by both methods. All galaxies are deprojected using the axis ratio of their one-component GALFIT S\'ersic profile fit prior to computing the bar strength.

We compute $B_M$ alongside the suite of related bar parameters $B_P$, $B_{M0}$, $B_{P0}$, $B_{M2}$, and $B_{P2}$ from \cite{Garcia_Gomez_2017}. We use slightly different definitions of these related parameters compared to those presented in the original paper; the exact equations used for our implementation can be found in \autoref{app:bar_metrics}. Thereafter we determine the optimal cutoff threshold for each parameter by simultaneously maximizing agreement with both the isophotal and visual classifications. 

All parameters perform similarly well achieving $55\pm1\%$ agreement with each of the isophotal and visual samples individually, and each agrees with 70-75\% of the combined ensemble of 792 classifications (396 visual and 396 isophotal). $B_M$ is the only metric whose optimal threshold produces a significant bar-fraction similar to the 50\% found by both the isophotal and visual classifications, identifying 35\% of the sample as barred. This is significantly closer to the baseline than the other metrics which all classify as barred around 10-25\% of the sample. We therefore adopt the calibrated best threshold of $B_M \geq 4.9$ to indicate the presence of a bar. Finally we compute the $B_{M}$ for COSMOS-74706 in every filter and mask combination determining those combinations which meet the prerequisites for identification and where the $B_{M}$ value is greater than the cutoff threshold as indicative of the presence of a bar.

\section{Results}
\label{sec:results}

\begin{table*}
\begin{center}
\caption{Means of the GALFIT model fitting parameters found by the MCMC exploration across all filters. Uncertainty is 1$\sigma$ of the fitted values over the posterior. Magnitude is the integrated light over the entire profile component. $R_e$ is the effective radius of the S\'ersic profile in pixels and $n$ is the dimensionless S\'ersic index while $q$ gives the functions ellipticity. PA indicates position angle in degrees from north with counter-clockwise positive.}
\label{tab:params_all}
\begin{tabular}{llccc}
\hline
\textbf{Component} & \textbf{Parameter} & \textbf{F200W} & \textbf{F277W} & \textbf{F356W} \\
\hline
\multirow{5}{*}{\textbf{\shortstack{Inner \\ Light}}} & mag & $27.442 \pm 0.007$ & $26.180 \pm 0.0003$ & $24.348 \pm 0.006$ \\
& $R_e$ (px) & $3.043 \pm 0.019$ & $2.389 \pm 0.001$ & $7.99 \pm 0.02^*$ \\
& $n$ & $0.54 \pm 0.01$ & $0.636 \pm 0.001$ & $3.845 \pm 0.008$ \\
& $q$ (b/a) & $0.336 \pm 0.003$ & $0.6386 \pm 0.0002$ & $0.813 \pm 0.001$ \\
& PA ($\degree$) & $3.59 \pm 0.27$ & $7.918 \pm 0.001$ & $57.73 \pm 0.12$ \\
\hline
\multirow{5}{*}{\textbf{\shortstack{Outer \\ Disk}}} & mag & $23.415 \pm 0.001$ & $23.0608 \pm 0.00002$ & $23.163 \pm 0.001$ \\
& $R_e$ (px) & $15.948 \pm 0.002$ & $15.4478 \pm 0.0001$ & $15.442 \pm 0.003$ \\
& $n$ & $0.323 \pm 0.001$ & $0.3323 \pm 0.00002$ & $0.2881 \pm 0.0004$ \\
& $q$ (b/a) & $0.6556 \pm 0.0001$ & $0.6722^\dagger$ & $0.6442 \pm 0.0001$ \\
& PA ($\degree$) & $3.330 \pm 0.010$ & $2.5321 \pm 0.0001$ & $3.470 \pm 0.006$ \\
\hline
\end{tabular}
\end{center}
\footnotesize
$^*$Parameter value converged to a prior boundary for the best model.
\newline
$^\dagger$The posterior distribution for this parameter has zero variance as all samples converged to the same value.
\end{table*}

The best two-component GALFIT model is highly effective at isolating the inner light. Our most direct evidence for the presence of a bar from the light profile decomposition comes from its effective subtraction of the centrally concentrated light oriented between the spiral arms in the F200W and F277W filters, which strongly indicates that it has the character of a S\'ersic-profile. 

We create an average image of COSMOS-74706 among the F200W, F277W, and F356W filters and subtract from it the average of all best-fit models. The resulting average residual image is depicted in the upper panel of \autoref{fig:barred_unbarred} and shows a lighter band oriented along the North-South direction between the spiral arms indicating that there is some amount of inefficiently isolated central light with further substructure not well described by the model fits. The lower panel of \autoref{fig:barred_unbarred} shows the same average of residual images across filters but now with the addition of the average inner-light S\'ersic profile, effectively displaying the galaxy with only the disk light subtracted off. This visualization reveals a much clearer elongated, bisymmetric structure at the galaxy's center spatially coincident with the inner anchor points of its two main spiral arms and with the light vertical band of inner substructure. Across the three filters the inner light is collectively elongated along the North-South axis.

The mean quantitative parameters derived from our GALFIT model fitting are summarized in \autoref{tab:params_all}. These provide an effective means of characterizing the morphology of COSMOS-74706 across filters. The MCMC exploration accepted 2398 values in the F200W filter's posterior, 3272 values for F277W, and 2895 values for F356W. For the axis ratio parameter $q$ of the outer disk in F277W, all steps converged to the same value which prevents us from reporting an error. The F356W effective radius $R_e$ of the inner light hit the boundary in $\sim 78\%$ of accepted steps and hence its error should be interpreted with caution.

\begin{figure*}[t]
    \centering

    \subfigure
    {
        \includegraphics[width=1.0\textwidth]{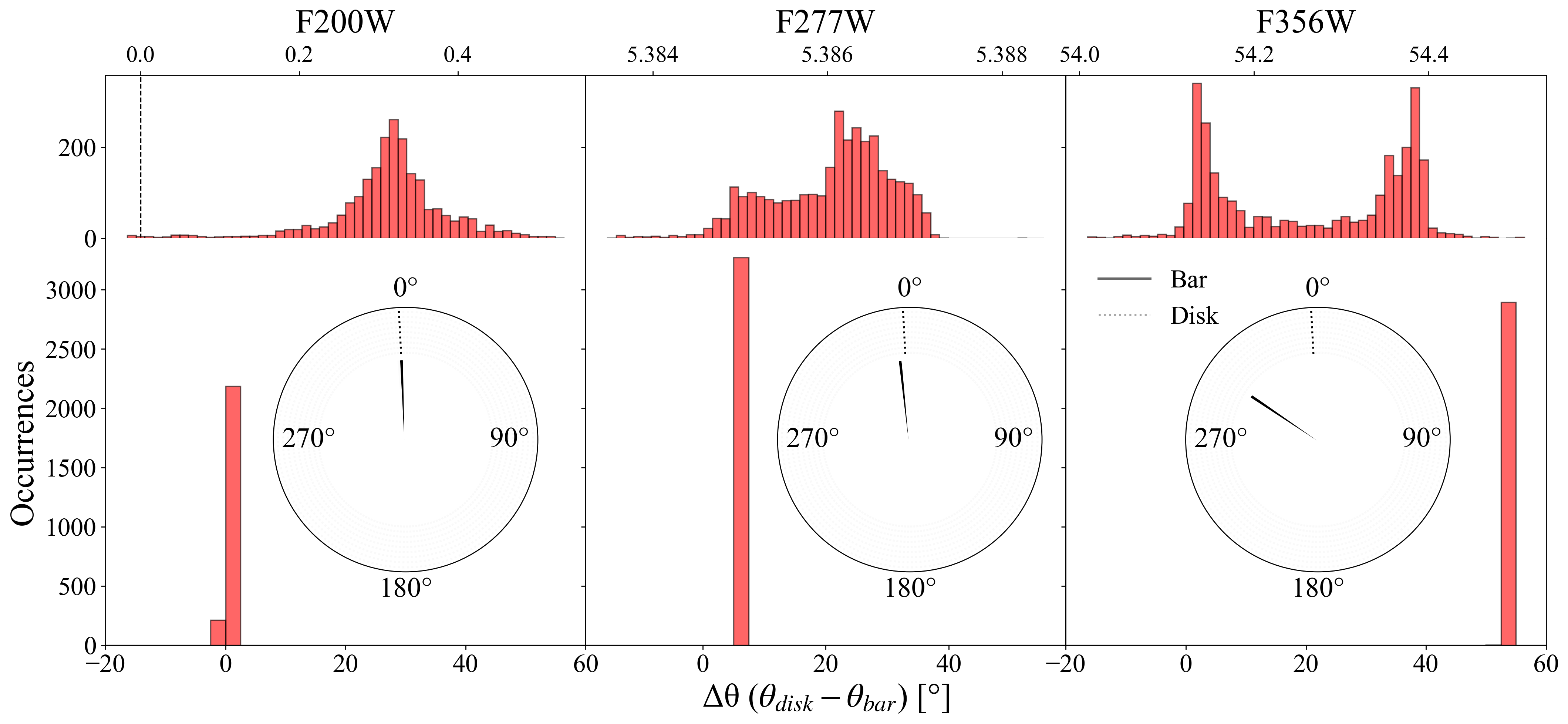}
    }
    \hspace{\fill}

    \caption{The three panels depict histograms of the distribution of differences between the position angle of the disk and bar components ($\theta_{disk}-\theta_{bar}$) in the posterior of the two-component S\'ersic models of COSMOS-74706. The top part of each panel shows a zoomed-in view of the distribution. The circular insets depict the position angles themselves with the inner solid lines representing the orientation of the profiles fitting the inner light and the outer dashed lines representing the orientation of the outer disk components. All three filters are host to a systematic difference between the orientations of the disk and inner component indicating that any detection of a bar is not the product of a projection artifact.}
    \label{fig:circs}
\end{figure*}

The disk properties are consistent across filters with a $q\approx 0.65$ and an orientation PA$\approx 3\degree$. It has the brightest integrated magnitude of the two components in the fitting owing to the significantly more spatially concentrated nature of the inner light. The S\'ersic index parameter $n$ is very low compared to local $n\sim 1$ disks indicating an unusually diffuse halo with a slow fall-off in light that is imperfectly fit by a classic exponential profile.

The inner light displays a significant bifurcation in its behavior between filters. For F200W and F277W the inner light is best matched by a low S\'ersic index $n\sim 0.6$ which is consistent with local bars \citep{Gadotti_2009}. In addition, these two filters show a low $R_e \sim 3$ and a significant elongation on the order of $q\sim 0.3$ for the F200W filter and $q\sim 0.6$ in F277W. Both components are elongated roughly along the North-South axis with PA$\sim 3\degree-8\degree$ consistent with the band of substructure apparent in the residuals and with the orientation of the bar anticipated from visual inspection.

The F356W filter shows a wholly different behavior of the inner light with a much larger $R_e$ that strains against the upper boundary of the prior and appears broadly inconsistent with the much more compact profile of the F200W and F277W filters. In this filter the inner light shows a profile with $n\sim 3.8$ which is slightly less compact than that of a classical de Vaucouleurs bulge but otherwise essentially consistent with local bulges. This component is relatively spherical with $q\approx 0.8$ and is not co-aligned or perpendicular to either of the inner components in the other two filters. Due to this orientation it is improbable for it to be the progenitor of any induced bar-shaped artifact caused by the over-fitting away of a portion of the inner light. It is therefore noteworthy that the residual for the F356W filter clearly shows a North-South aligned inner feature left over from the fitting as can be seen in the lower right panel of \autoref{fig:decomp}. This is suggestive of a distinct bar and bulge component co-existing in the inner region which cannot be easily disentangled by the MCMC exploration of the three-component model due to the greater propensity of the optimizer to fit the residual light from the spiral arms rather than that from the bar.

The orientation of the disk and inner light components is not consistent with each other within their errors from the MCMC exploration in any filter. \autoref{fig:circs} depicts histograms of the difference in orientation between these components for each fit and within each filter. The circular insets depict the distribution of actual orientations of the central and outer components with the inner solid line indicating the orientation of the inner profile and the outer dashed line giving that of the disk. These clearly show a lack of divergently oriented inner profiles in the sample.

In the F277W filter every single model in the posterior has a difference between the two orientations of $5\degree$-$7.5\degree$ while the F200W filter shows a more modest difference of $0\degree$-$2.5\degree$. The raw F200W posterior formally admits $\Delta\theta = 0$ within its $1\sigma$ scatter ($0.26\degree \pm 0.27\degree$); however, this uncertainty is inflated by a tail of $\sim$ 350-400 outlier samples out of the 2398 sample posterior. Applying either iterative $3\sigma$ clipping or Tukey's fences ($1.5\times \textrm{IQR}$) removes this tail and yields $\Delta \theta = 0.32\degree \pm 0.06\degree$, where the $1\sigma$ interval now securely excludes zero. The inner-light position angle itself tightens from $3.59\degree \pm 0.27 \degree$ to $3.65\degree \pm 0.05\degree$, while the disk orientation is essentially unchanged, indicating that the inflated scatter originates from the inner component. In the top panels of \autoref{fig:circs} the zoomed-in view of each histogram can be seen, showing the scatter of the profile orientations. In particular, a bimodality can be seen in the F356W filter which is indicative of the multiple well-defined minima for the orientation of the inner light profile which can also be seen on the corner plot for this filter in \autoref{fig:corner_356}. A Welch t-test gives a very low value of $p \ll 0.01$ for all three filters, strongly indicating that any potential bar is not co-aligned with the disk.

The differing orientations of the inner light and disk components in the two filters, where the inner profile seems to correspond to a bar morphology, strongly suggest that such a potential bar is not an artifact caused by a projected circular disk and bulge. In such a case, the two would have orientations consistent with one another. The insets also display a change in the orientation of the inner light across different filters with a counterclockwise shift towards the longest wavelengths. This may also be related to a transition between the dominance of the light from the bulge in longer filters to the light from the bar in shorter filters which implies a difference in the dynamic environment of the two components.

\begin{figure}
    \centering
    \includegraphics[width=1.0\linewidth]{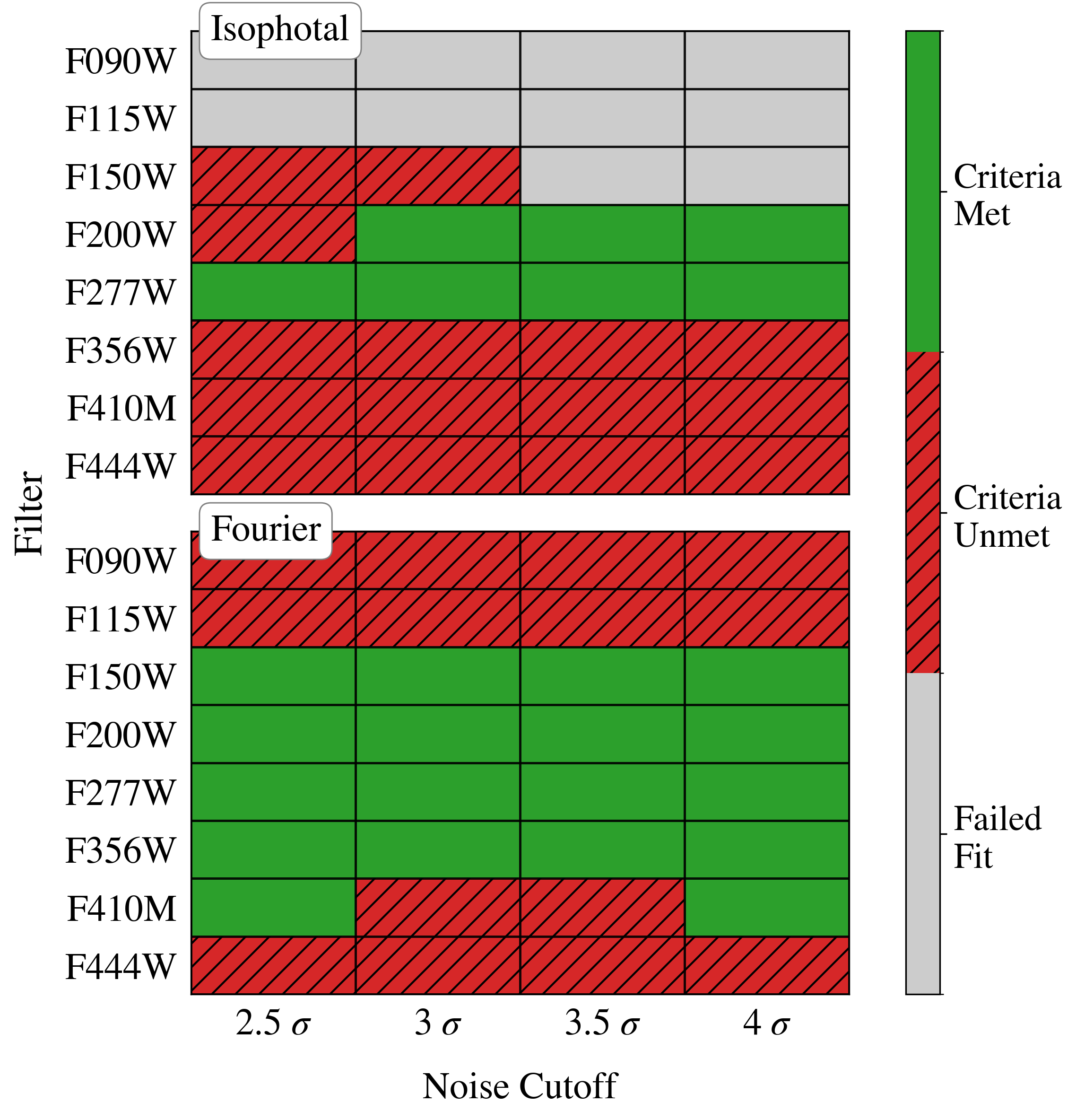}
    \caption{The outcomes of the Isophotal and Fourier analysis modes are displayed on the top and bottom panel, sorted by JWST filter and mask noise-level-cutoff. Green indicates for the isophotal analysis that all four bar criteria were met while hatched red indicates that at least one of the criteria was not met. For the Fourier analysis mode green indicates that the $B_{M}$ value (see \S\ref{sec:method}) for that mask-noise-cutoff and filter combination was above the cutoff value while hatched red indicates that it was not. Gray indicates that the code failed to finish.}
    \label{fig:analysis_panels}
\end{figure}

\autoref{fig:analysis_panels} shows the results of the Isophotal and Fourier analysis. The two modes collectively display the most positive bar identifications in the F200W and F277W filters where the bar is most easily visually ascertainable. In fact the Fourier analysis shows positive bar identifications across all mask noise-cutoffs in the F150W, F200W, F277W, F356W, and F410M filters as well as one positive identification at the $4\sigma$ noise-cutoff in F444W. The F090W filter analysis gives a negative result due to an excessively noisy Fourier spectrum though it also produces no central Gaussians that result in a value above the cutoff. The F115W filter, on the other hand, registers as unbarred as it never produces a strong central component in its spectrum. For the isophotal analysis there are positive identifications for all noise cutoffs in the F277W filter and in all but the $2.5\sigma$ cutoff for F200W. Most of the isophote fits in shorter wavelengths fail due to the low SNR of the data and the choppy irregular light profile.

Our results show strong agreement between the two modes of analysis. In particular of the 7 filter-mask combinations which return a positive verdict on the presence of a bar in the isophotal analysis, none show a $B_{M}$ value below the cutoff. In the F200W and F277W filters where the bar is most easily discernible visually, all mask combinations for both filters produce a $B_{M}$ value above the cutoff indicating the presence of a bar. For these filters, out of the 8 combinations for which the isophote fits converged, 7 of them show evidence of barred morphology.

\begin{figure*}[t]
    \centering

    \subfigure
    {
        \includegraphics[width=0.31\textwidth]{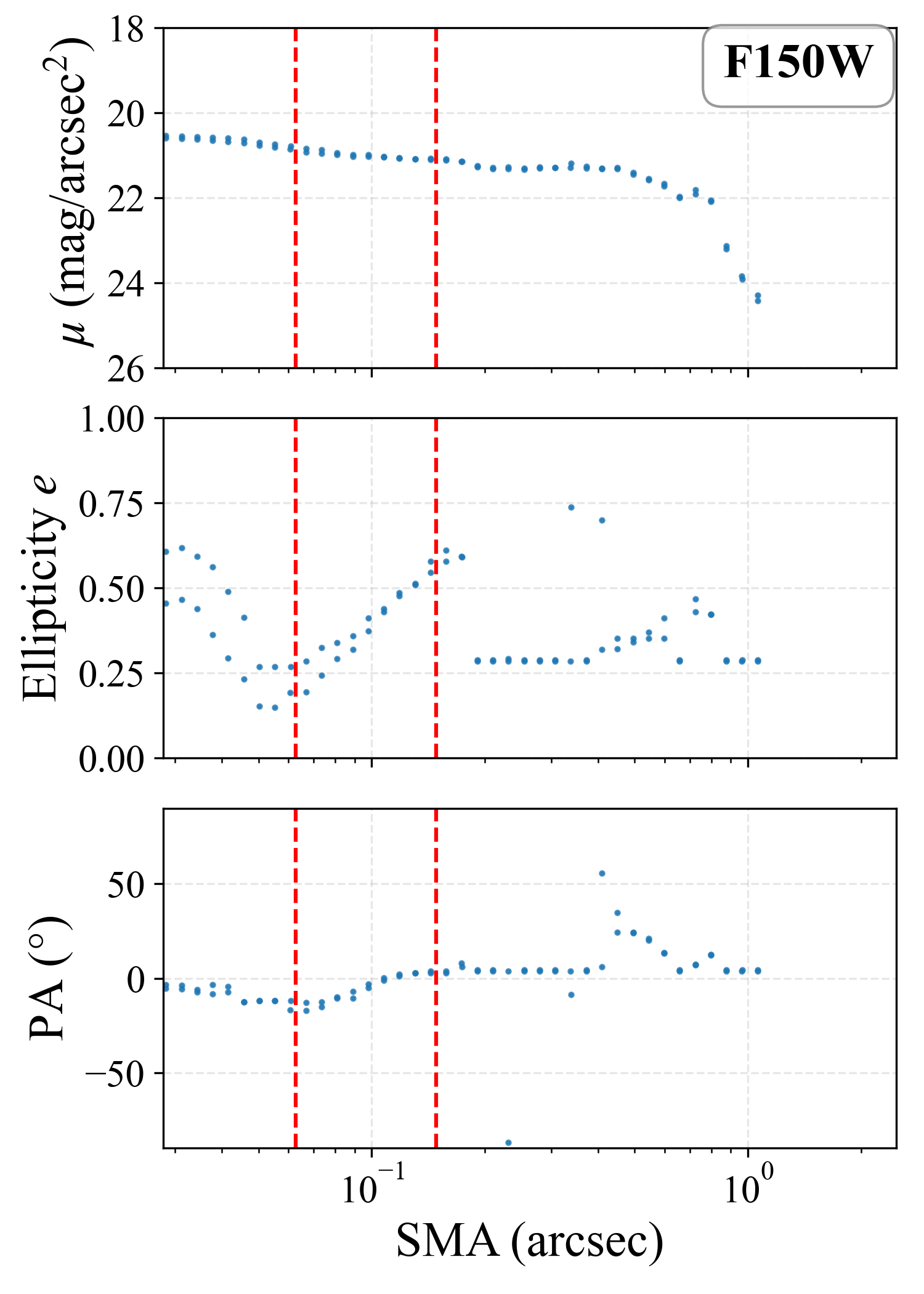}
        \label{fig:panel_composite}
    }
    \hspace{\fill}
    \subfigure
    {
        \includegraphics[width=0.31\textwidth]{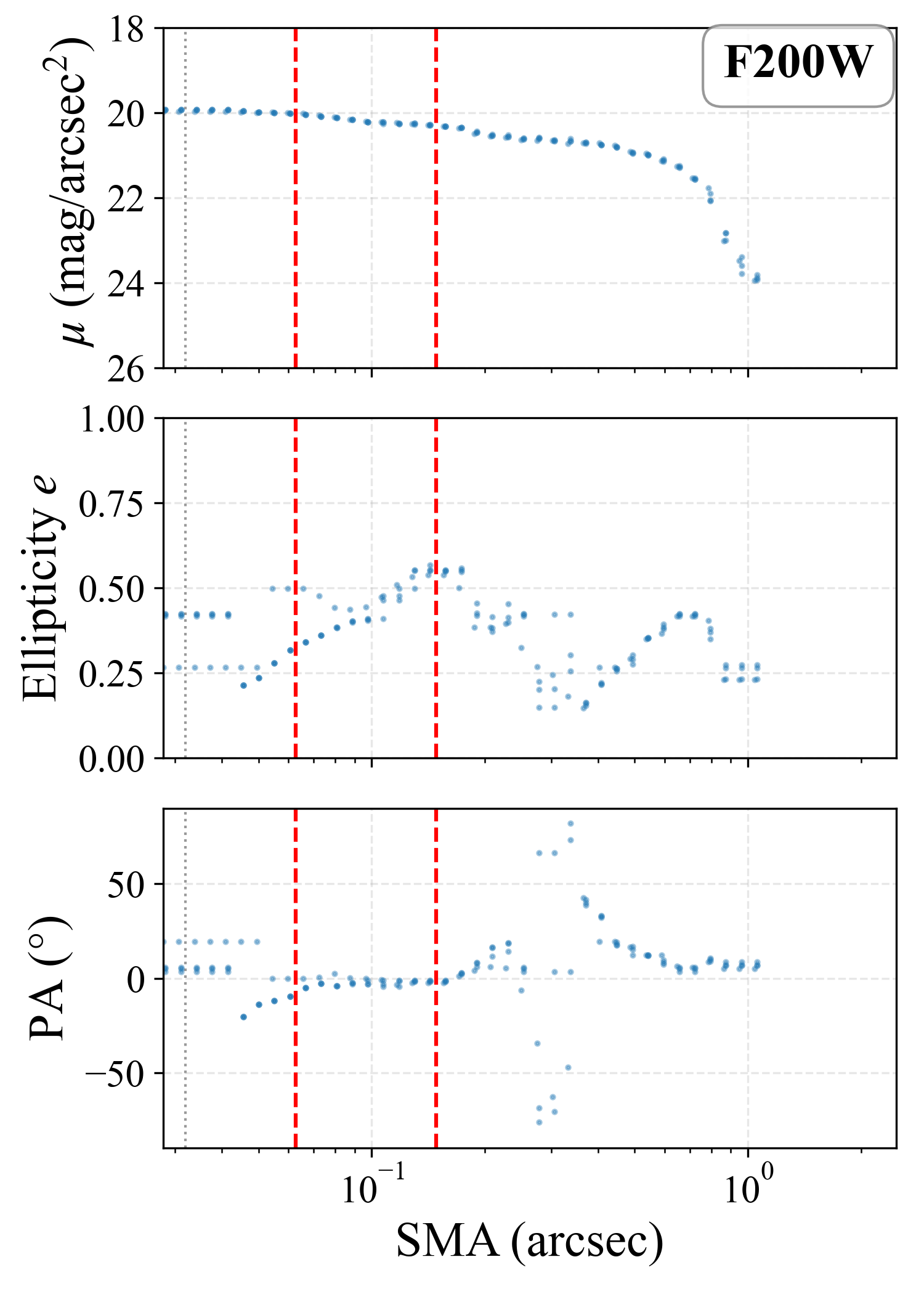}
        \label{fig:panel_unsharp}
    }
    \hspace{\fill}
    \subfigure
    {
        \includegraphics[width=0.31\textwidth]{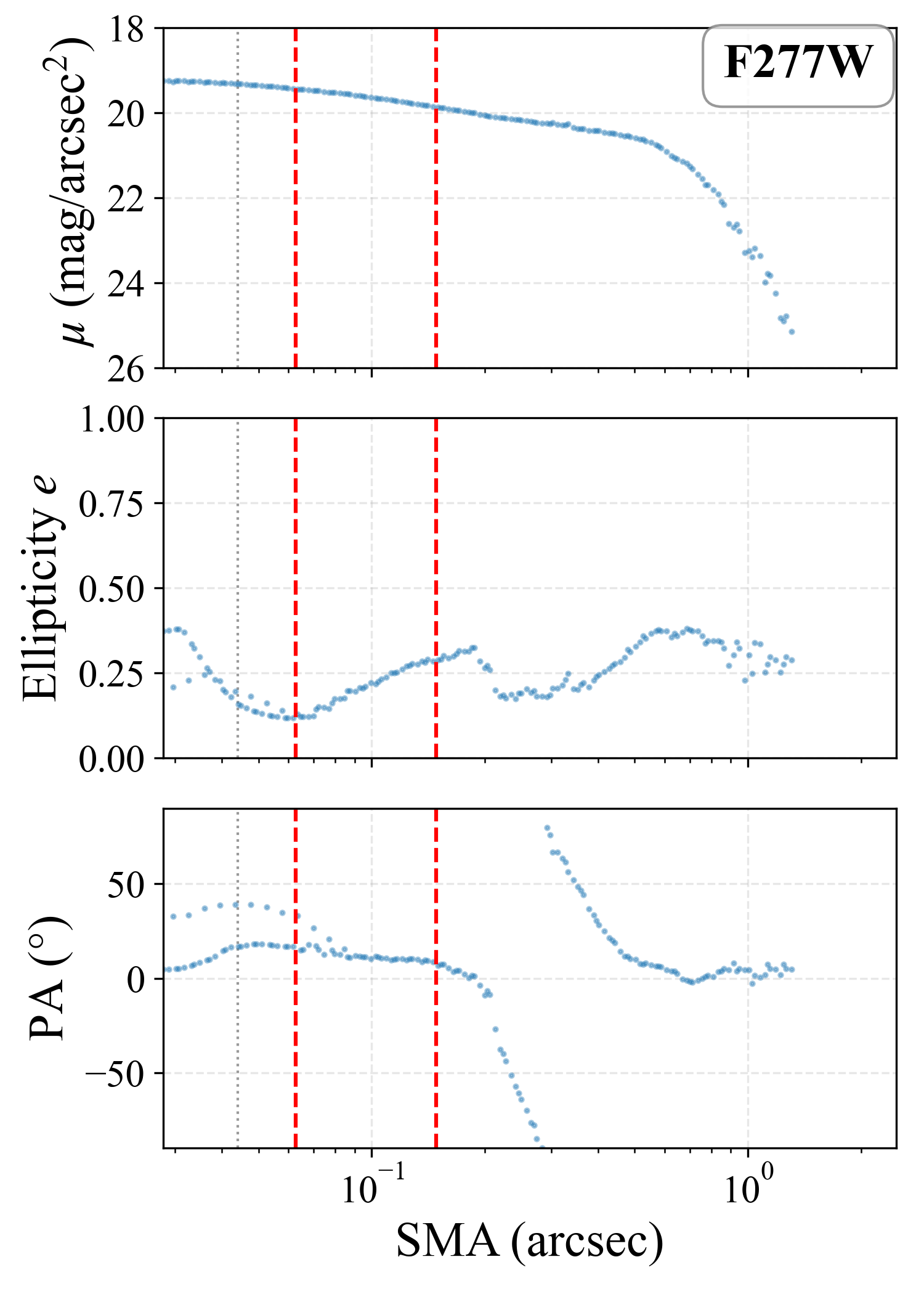}
        \label{fig:panel_residual}
    }
    \hspace{\fill}
    \subfigure
    {
        \includegraphics[width=0.31\textwidth]{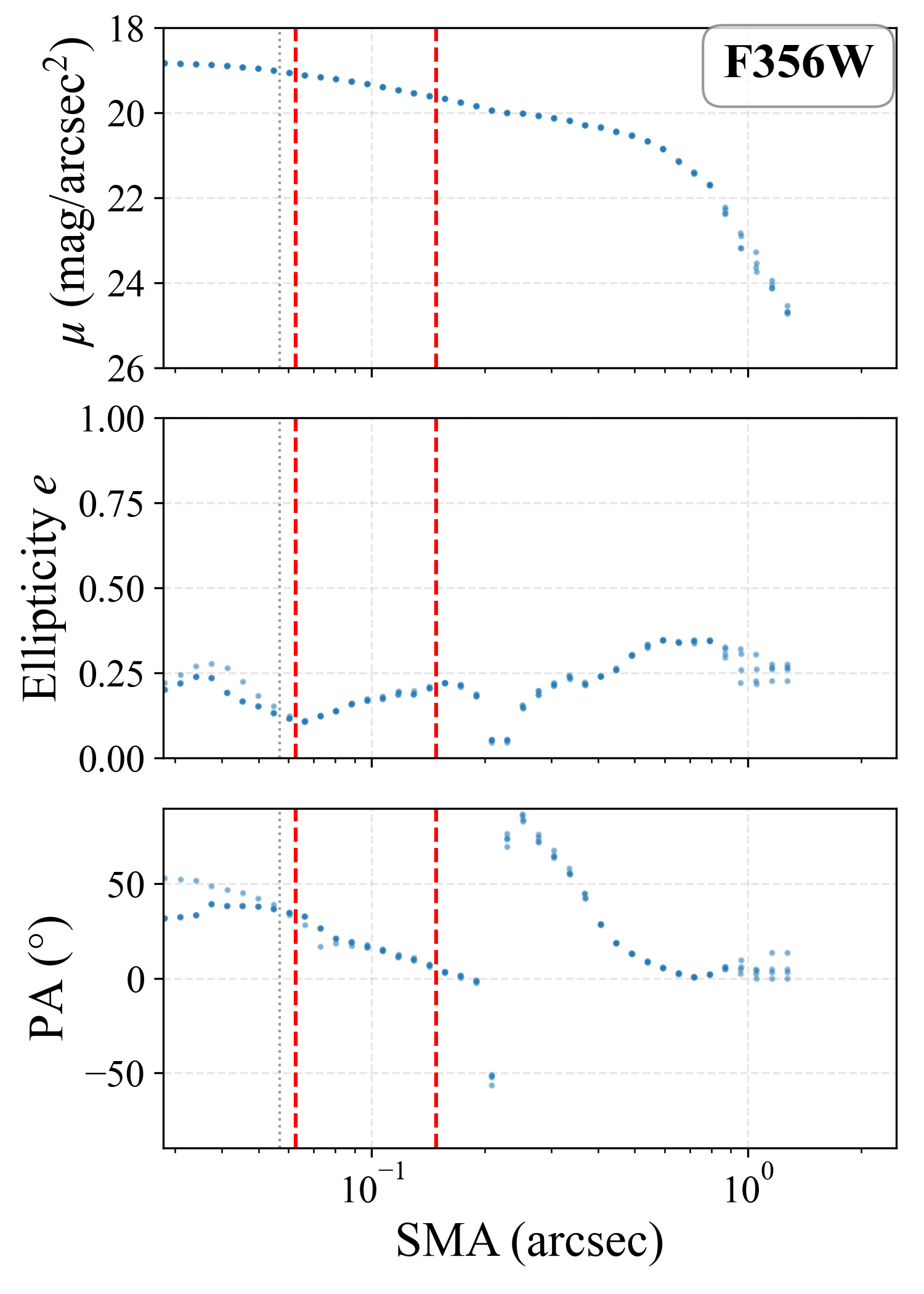}
        \label{fig:panel_composite}
    }
    \hspace{\fill}
    \subfigure
    {
        \includegraphics[width=0.31\textwidth]{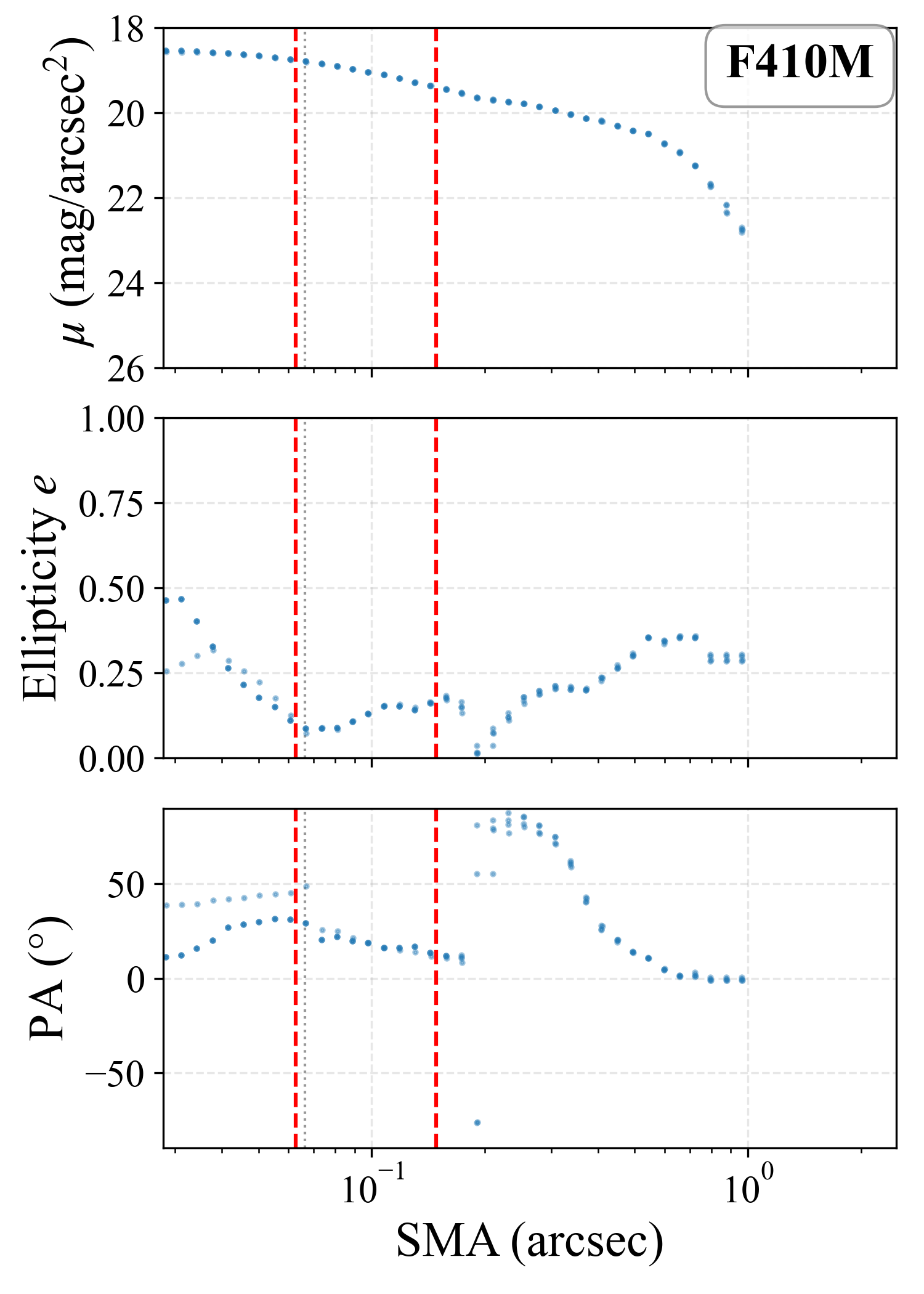}
        \label{fig:panel_unsharp}
    }
    \hspace{\fill}
    \subfigure
    {
        \includegraphics[width=0.31\textwidth]{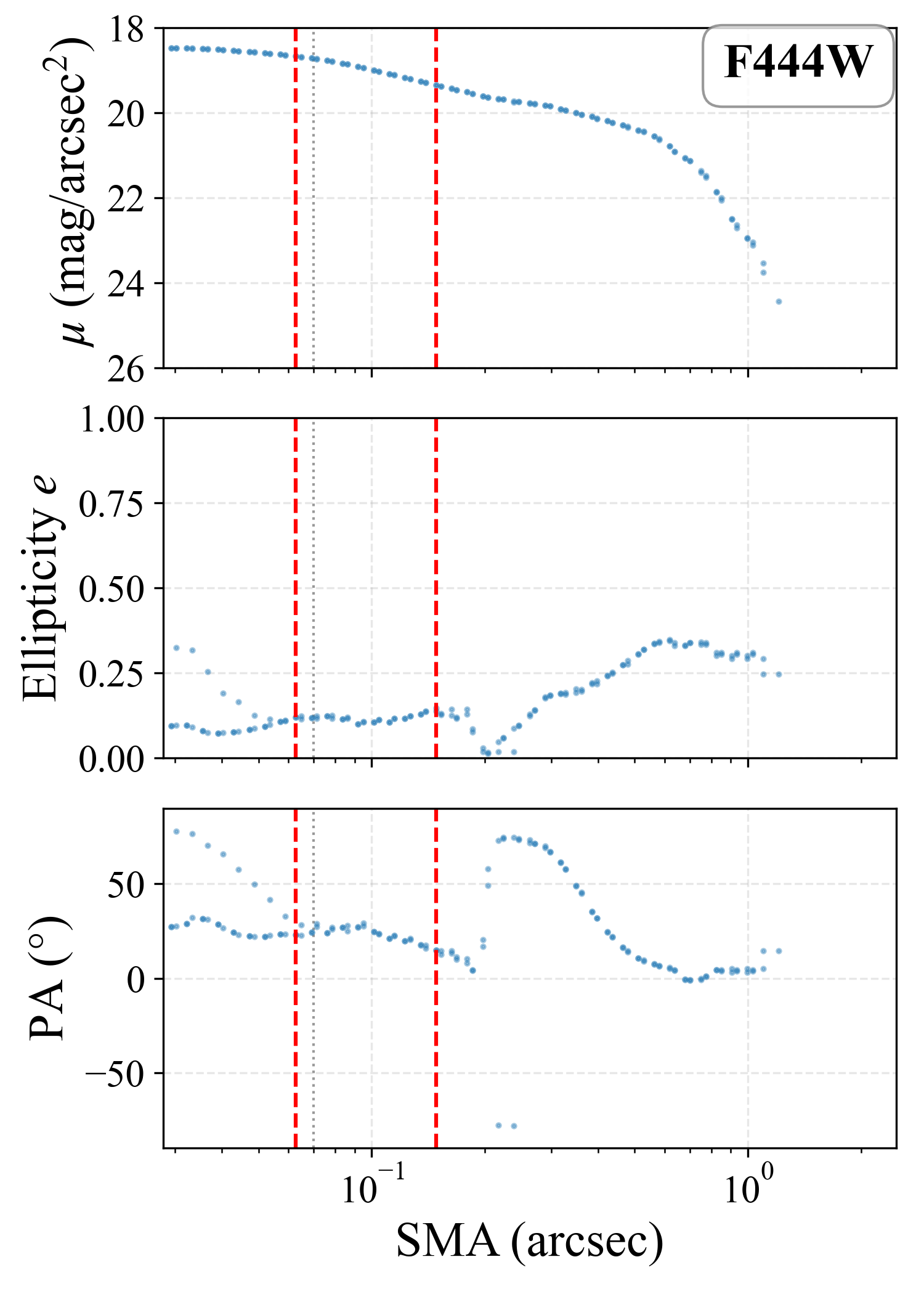}
        \label{fig:panel_residual}
    }

    \caption{The panels display the results of the isophotal analysis of COSMOS-74706 in six JWST/NIRCam filters as labeled. The various mask noise-level cutoffs are overlaid on top of one another in each panel, showing from top to bottom the surface brightness, ellipticity, and position angle of the isophotes moving radially outward. A PA of 0$\degree$ corresponds to an ellipse whose semi-major axis lies along the East-West axis with increasing PA corresponding to counterclockwise rotation. The red dashed vertical lines indicate the radial extent of the bar region and the dotted gray lines indicate the HWHM of the PSF. The signature of a bar can be seen most strongly in the F200W and F277W panels as a rise in ellipticity to $e>0.25$ with a constant position angle over the bar region followed by a drop in $e$ and a change in PA.}
    \label{fig:isophotes}
\end{figure*}

Detailed results of our isophotal ellipse fitting are shown in \autoref{fig:isophotes}. The ellipticity profiles for the F150W, F200W, F277W, F356W, and F410M filters all show a smooth rise over the bar region, but only in the F150W, F200W and F277W does this elongation extend above $e=0.25$ as necessary for a bar identification. There are also varying levels of change of the position angle of the isophotes over the regions between filters, with the bar isophotes holding a constant PA in the F277W filter but rising smoothly in the F200W filter and falling in F356W and F410M. The changing PA in F200W is the one criterion that is not met in the $2.5\sigma$ noise level. The bar extent is several times the HWHM of the PSF in the shorter wavelengths and is well above the resolution limit in all filters.

In the isophotal analysis, the filters as a whole show minimal variation in the fitted isophotes between masks with the largest variations generally seen interior to the bar region. The differences are relatively minor and only change the resulting determination in the case of the $2.5\sigma$ noise cutoff in the F200W filter where the ellipticity does not increase monotonically over the bar region in this mask. The ensemble of higher noise-cutoff masks shows a unified behavior over the bar region which does meet the smooth increase criterion.

The F200W filter sees the largest rise in ellipticity over the bar region of $\Delta e \sim 0.2$ with a smaller change seen towards longer wavelengths dropping to $\Delta e \sim 0$ by F444W. The maximum ellipticity seen over the bar region likewise drops dramatically from $e \sim 0.55$ at F150W to $e \sim 0.15$ by F444W, decreasing further between each filter. The surface brightness profile in contrast retains its shape throughout all filters with the exception of a shallower decline in the inner region at shorter wavelengths, particularly visible in the F150W and F200W filters.

In the Fourier analysis the $B_{M}$ values for all masks in the F150W, F200W, F277W, F356W, and F410M filters are above the threshold value and the two filters F150W and F200W consistently show high values in excess of $B_{M} = 10.0$ more than double the threshold, which is significant as in our testing very high values of $B_M$ correlate strongly with particularly visually barred galaxies. No noise-cutoff for the F277W filter has a $B_{M}$ value which falls below 6.8 while for the F356W filter the lowest $B_{M}$ is found for the $3.5\sigma$ mask where we measure $B_{M} \simeq 6.1$. The $B_{M}$ values for the filters F150W, F200W, F277W, F356W, F410M, and F444W are tabulated in \autoref{tab:bm_values}. It can be seen that there is a consistent drop-off in the values of $B_M$ towards higher noise-level cutoffs for the mask as less of the northern element is obscured, leading to a reduction in symmetry and hence decreased strength in the $m=2$ Fourier signal. 

Aside from the positive Fourier identifications in the three filters in which the bar is easily visible by eye (F200W, F277W, F356W), all of the noise cutoffs for our mask also result in positive bar identifications in the F150W and F410M filters. This may be due to the high sensitivity of the Fourier decomposition to large-scale structure which allows for inference of bar presence even in cases where the light is very clumpy as in F150W, or where the non-axisymmetry is difficult to pick out by eye as in F410M.

The reported uncertainties on $B_M$ in \autoref{tab:bm_values} were taken from the scatter achieved by introducing random variation in pixel values across the image in accordance with the noise level implied by the weight maps. We performed 10 determinations with different randomized variations for each image. The uncertainties are consistent within filters and generally small compared to the $B_M$ values except for the noise-prone F150W filter.

\begin{table}[h]
  \centering
  \setlength{\tabcolsep}{8pt}
  \begin{tabular}{ c c c }
    \hline
    \textbf{Filter} & \textbf{Noise Cutoff} & \textbf{$\boldsymbol{B_{M}}$}  \\
    \hline
    Threshold   & --    & \textbf{4.9} \\
    \hline
    F150W   & 2.5 $\sigma$  & 13.46 $\pm 3.70$ \\
    F150W   & 3.0 $\sigma$  & 13.17 $\pm 4.22$ \\
    F150W   & 3.5 $\sigma$  & 14.20 $\pm 5.08$ \\
    F150W   & 4.0 $\sigma$  & 11.15 $\pm 4.28$\\
    \hline
    F200W   & 2.5 $\sigma$  & 12.24 $\pm 1.09$ \\
    F200W   & 3.0 $\sigma$  & 10.44 $\pm 1.35$ \\
    F200W   & 3.5 $\sigma$  & 10.83 $\pm 1.14$ \\
    F200W   & 4.0 $\sigma$  & 11.66 $\pm 1.20$ \\
    \hline
    F277W   & 2.5 $\sigma$  & 9.40 $\pm 1.04$ \\
    F277W   & 3.0 $\sigma$  & 9.05 $\pm 1.01$ \\
    F277W   & 3.5 $\sigma$  & 9.18 $\pm 0.67$ \\
    F277W   & 4.0 $\sigma$  & 6.88 $\pm 1.15$\\
    \hline
    F356W   & 2.5 $\sigma$  & 6.48 $\pm 0.68$ \\
    F356W   & 3.0 $\sigma$  & 6.34 $\pm 0.68$  \\
    F356W   & 3.5 $\sigma$  & 6.16 $\pm 0.66$ \\
    F356W   & 4.0 $\sigma$  & 6.19 $\pm 0.74$ \\
    \hline
    F410M   & 2.5 $\sigma$  & 5.96 $\pm 1.95$ \\
    F410M   & 3.0 $\sigma$  & 6.93 $\pm 0.49$ \\
    F410M   & 3.5 $\sigma$  & 5.77 $\pm 1.78$ \\
    F410M   & 4.0 $\sigma$  & 5.75 $\pm 0.57$ \\
    \hline
    F444W   & 2.5 $\sigma$  & 4.31 $\pm 1.32$\\
    F444W   & 3.0 $\sigma$  & 4.25 $\pm 0.89$ \\
    F444W   & 3.5 $\sigma$  & 4.73 $\pm 1.13$ \\
    F444W   & 4.0 $\sigma$  & 5.05 $\pm 1.05$ \\
    \hline
  \end{tabular}
  \caption{Values of $B_{M}$ for different filters and mask noise-level cutoffs. Uncertainties represent the $1\sigma$ values derived from bootstrap variation of the image noise.}
  \label{tab:bm_values}
\end{table}

\begin{figure}
    \centering
    \includegraphics[width=1.0\linewidth]{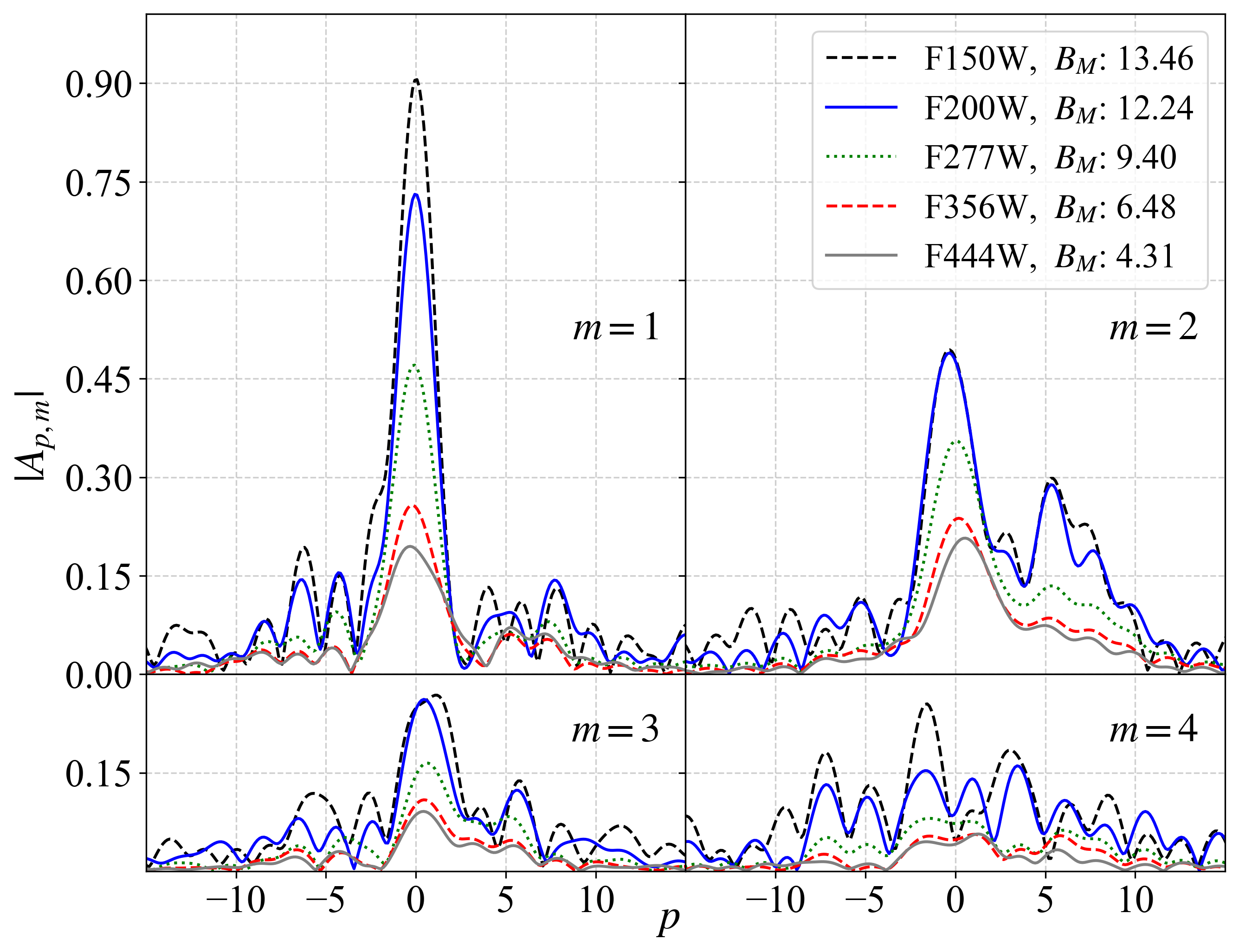}
    \caption{Fourier amplitudes $|A(p, m)|$ from the decomposition of COSMOS-74706 into logarithmic spirals. Each panel shows a different multiplicity mode $m$ as a function of the pitch angle parameter $p$. The different line styles represent various JWST/NIRCam filters with the 2.5$\sigma$ mask. Corresponding bar strength parameters $B_{M}$ are listed in the legend for each filter and are commensurate with each filter's strength in the $m=2$ mode which traces bisymmetric features. Notably, there is significant strength in the $m=1$ mode due to the presence of considerable non-antisymmetric features.}
    \label{fig:avp_plot}
\end{figure}

The resulting Fourier amplitude for different winding angles across filters is plotted in \autoref{fig:avp_plot} where it can be seen that the $m=2$ multiplicity mode is of significant magnitude compared to the $m=1$ mode, showing an increase in relative strength towards longer wavelengths. The $m=1$ multiplicity is stronger in all filters except F444W, likely due to the ``smoothing'' effect of the Rayleigh criterion flattening asymmetries and making the galaxy as a whole appear more homogeneous. The $m=2$ multiplicity also weakens towards longer wavelengths due to the loss in resolution decreasing the prominence of the bar. Nonetheless all filters shortward of F444W show a sufficient central $m=2$ feature to produce a $B_{M}$ above the threshold value. In the shorter filters, the $m=2$ Fourier spectrum also shows a secondary bump at $p\approx 5$ which corresponds directly to the two main spiral arms. The $m=3$ spectrum is weak but contains a central feature while the $m=4$ spectrum is effectively devoid of any consistently identifiable elements.

\vspace{1cm}

\section{Discussion}
\label{sec:discussion}

The discovery of a well-defined barred spiral before the onset of cosmic noon has significant implications for our understanding of disk settling and bar formation theories. Here we place COSMOS-74706 within the context of those theories and characterize its physical qualities. Then in \S\ref{subsec:cons} we critically assess the robustness of our methodology and address potential sources of uncertainty. Finally we conclude with an overview of possible alternate interpretations of the results.

\subsection{Observed Properties}
\label{subsec:prop}

Using the F200W filter where the bar is most exposed and thus appears strongest, we visually identify the ends of the bar and measure a bar semi-major axis of 0.18\arcsec that at $z=3.159$ corresponds to 1.4 kpc, that is, roughly half as long as the bar of the Milky Way \citep{Lucey_2023}. The existence of such a strong bar at an epoch when the universe was less than 2 Gyr old provides compelling evidence that the physical conditions needed for mature disk settling had already been achieved by massive galaxies in this epoch. Furthermore, our two-component model light profile decompositions in the F356W filters yield a bulge component with a classical characteristic S\'ersic index of $n \sim 4.0$, likely indicative of a mature classical merger-built bulge \citep{Kormendy_2004}.

\begin{figure}
    \centering
    \includegraphics[width=1.0\linewidth]{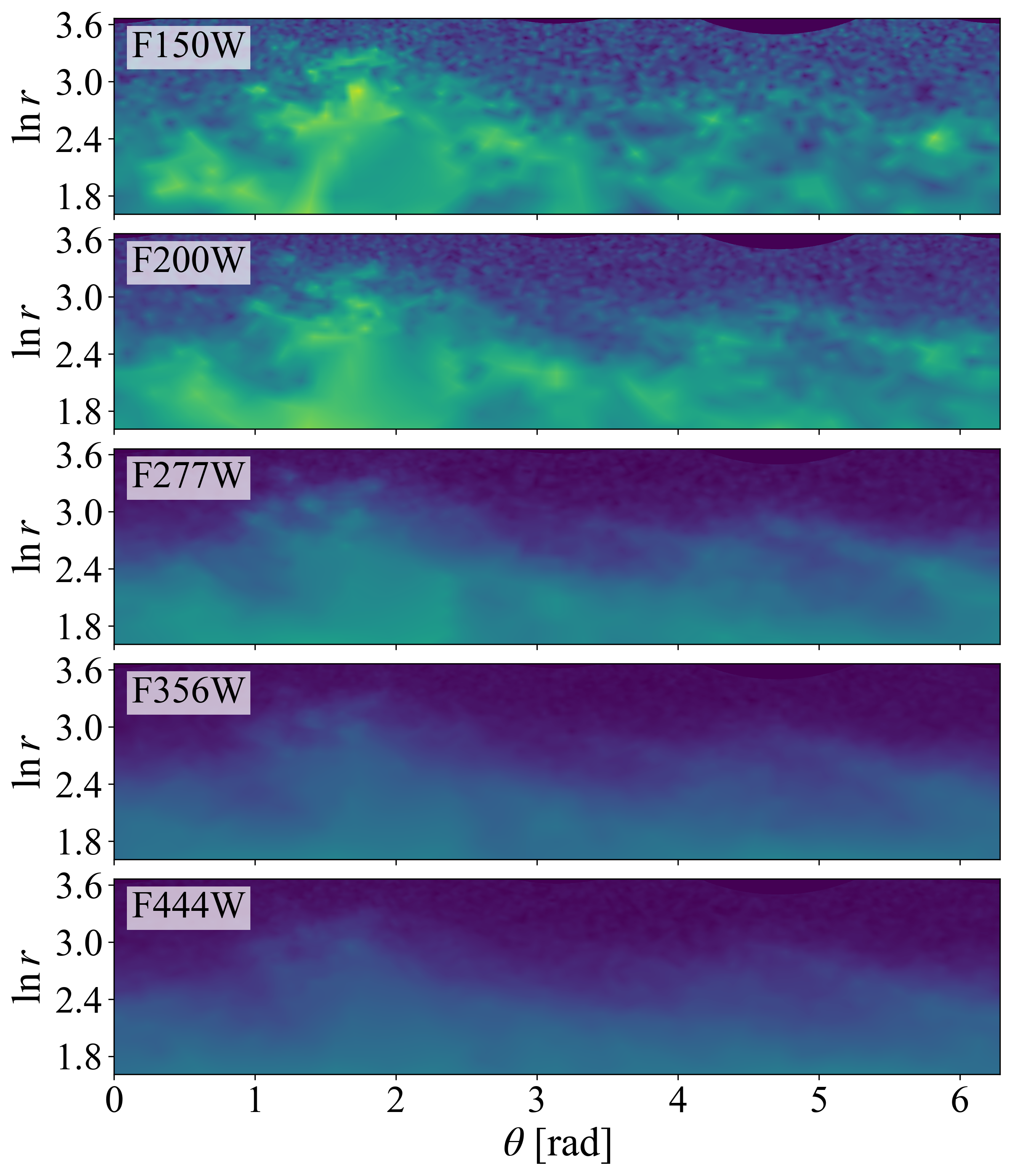}
    \caption{A log-polar unwrapping of COSMOS-74706 across filters shows the surface brightness plotted by azimuthal angle $\theta$ and radius $\ln r$. The tapered inner region is not depicted. The apparent decreasing surface brightness towards longer filters is a consequence of increased contrast between the surface brightness at the center and outer regions of the galaxy.}
    \label{fig:lnr_plot}
\end{figure}

The spiral arms of COSMOS-74706 are narrow and tightly wrapped but exhibit kinked and clumpy segments. \autoref{fig:lnr_plot} shows the surface brightness of the galaxy across different filters plotted by $\ln r$ and $\theta$. Linear features in this space correspond to logarithmic spirals in Cartesian space with the slope of the line corresponding directly to the pitch angle. Two main linear structures are apparent, corresponding to the spiral arms at $\theta \sim 2-3$ radians and $\theta \sim 5-6$ radians. Several deviations from linearity are notable, suggesting breaks and small changes in wrapping angle within the arm structure. The most prominent kink in the arms is visible in the eastern limb where a segment of the arm straightens before breaking out of visibility. The northwestern clumps are another significant deviation from the morphology of local grand-design spirals possibly reflective of the unsettled environment typical in high redshift disks \citep{Genzel_2008}. These clumps may be evidence of fragmented star formation or minor mergers which are common in this epoch \citep{Dekel_2009}. Nonetheless the arms are broad and highly collimated manifesting as a significant signal in the $m=2$ Fourier spectrum around the wrapping angle characterized by $p=5$, and are thus very unlikely to be misidentified tidally disrupted streams.

The mildly asymmetric and slightly irregular nature of COSMOS-74706's spiral arms is suggestive of some degree of tidal perturbation. We do not see an obvious companion in the JWST imaging, so if the blue northern element is a nearby galaxy at the same redshift, it would be the most likely perturber, though it cannot explain the disruption if it is a region of the spiral. The Southeastern arm-like feature manifests similarly in appearance to a spiral arm but does not seem to be connected to the bar, raising the possibility that it is a distorted minor arm, a disturbance caused by supernova feedback, or a remnant structure from an encounter event with another galaxy. Alternatively, this limb may be an $m=4$ minor arm as is commonly seen in mature spirals in the local universe. The symmetric counterpart of this arm may be obscured by the northern element leading to the disjointed appearance in the north-western quadrant of the image and to a lack of detection as a signal in the $m=4$ multiplicity Fourier mode.

\subsection{Theoretical Interpretation}
\label{subsec:theory}

Bar formation in a dynamically primed disk can be efficiently triggered by a gravitational perturbation from a fly-by, as is evidenced by the presence of nearby companions in many barred spirals at cosmic noon (\citealt{Guo_2025}, Ivanov et al. in prep). Given the suggestion of tidal perturbation and the apparent presence of a classical bulge, it is plausible that the bar of COSMOS-74706 was triggered by an encounter, possibly even with the northern element if it is a distinct object. Such a tidal interaction would also compress gas and induce a starburst which COSMOS-74706 experienced $\sim 1$ Gyr before observation (paper II) lending support, alongside the bar's significant strength, to the possibility that the observed bar may have formed as early as $z\sim 5$.

For such a tidal trigger to occur, the disk must already be able to support a bar, thereby requiring it to be baryon-dominated \citep{Ostriker_1973}, though the disk mass fraction alone does not discriminate between barred and unbarred systems indicating that while baryon domination is necessary, it is not on its own a sufficient condition \citep{Romeo_2022}. Recent hydrodynamic simulations show that the bar formation timescale within such disks obeys the Fujii relation \citep{Fujii_2018}, with the length of time needed to generate a bar falling exponentially as the ratio of disk to total mass fraction increases \citep{Bland_Hawthorn_2023}. In particular, \citet{Frosst_2026} have shown that even disks with elevated radial dispersions representative of high-redshift conditions can form bars within a few Gyr provided they are sufficiently baryon-dominated. Observations suggest that baryons dominate the centers of galaxies at $z=1-3$ \citep{Genzel_2017, Genzel_2020, Price_2021}, implying that bars form promptly in this epoch. In addition, the formation timescale may be further shortened in the highly turbulent gas-rich disks expected at higher redshift \citep{Bland-Hawthorn_2024}. The existence of a well-developed 1.4 kpc bar in COSMOS-74706 lends observational support for this fast-formation model and is suggestive of significant ongoing bar formation already being in place before $z\sim 3$.

The rapid-formation picture in baryon-dominated disks is consistent with the redshift-evolution of the observed bar fraction. Massive disks at $z\sim 0.8$ already host bars with fractions approaching those of the local universe, with bar-bulge coexistence preferred at higher redshifts, indicating downsizing in the maturation of disk dynamics \citep{Sheth_2008}. Mass-limited JWST samples with $\log M_\star/M_\odot \gtrsim 10$ give a visually-classified bar fraction of $\sim 15\%$ at $2 \leq z \leq 3$ and $\sim 5\%$ at $3 \leq z \leq 4$, with both values held to be lower limits on the intrinsic fraction owing to systematics causing shorter and weaker bars to go largely undetected \citep{LeConte_2024, Guo_2025}. COSMOS-74706 is a $\log M_\star/M_\odot \sim 10.6$ disk galaxy appearing to exhibit both a stellar bar and a coexisting $n\sim 4$ classical bulge component at $z>3$, more than 4 Gyr earlier than the $z\sim 0.8$ horizon where the dynamical maturity downsizing regime was previously established, but serving as a pristine example of the exact sort of system which the paradigm predicts should reach a bar-supporting state earliest. Finally, we observe that an early onset of bar formation is consistent with the recent age dating of thin disk stars in the Milky Way, which itself has a strong bar now fully identified in the stellar density field \citep{Queiroz_2021}, placing their formation to $z>4$ \citep{Nepal_2024}, in line with the identification of disks galaxies at redshift as early as $z\approx 5$ by JWST \citep{Robertson_2023}.

The dynamical mass budget of COSMOS-74706 can be constrained using swing amplification theory, wherein the dominant mode of the spiral arms provides a clue to the disk-to-halo mass ratio $f_{\rm disk}$. The relation $f_{\rm disk}(R) \approx 1/ (2mX)$, with $1 \leq X \leq 3$ \citep{Toomre_1981, DOnghia_2015}, allows us to predict the disk fraction required for specific arm multiplicities. For a dominant mode of $m=4$, one expects $f_{\rm disk}\approx 0.4-0.5$. Thus, the observed bar and multi-armed spiral structure in COSMOS-74706 can also be explained by secular evolution in a regime with a high disk-mass fraction. This requires the disk to be sufficiently cold; however, turbulence and high gas fractions may still play a complex role.

Such an early secular evolution would contrast with predictions from recent simulations which demonstrate that high gas fractions in early galaxy disks stifle bar growth \citep{Beane_2023} and suppress formation mechanisms \citep{Bland-Hawthorn_2024, Athanassoula_2013b}. The existence of a robust bar in COSMOS-74706 then suggests that the system may have depleted its gas unusually early, that the bar is tidally triggered, or that our understanding of secular bar formation in gas-enriched environments requires refinement.

\subsection{Methodological Considerations}
\label{subsec:cons}

A critical aspect of our two-component GALFIT models is the luminosity contrast between the extended disk and nuclear component seen in \autoref{tab:params_all}, with the integrated light from the disk being 1-4 magnitudes brighter than the corresponding inner component across filters with the largest contrast occurring at shorter wavelengths. Given that the morphology of the inner light corresponds to that of a bar in the F200W and F277W filter, we can use this to resolve upper limits to the contribution of a bar to the total light. In the F200W filter, the inner component contributes 2.4\% of the light from the model while in the F277W filter this rises to 5.3\%. This is somewhat lower than but still consistent with the $9\pm 5\%$ contribution of the bar light typical of galaxies in the local universe within the same wavelength range \citep{Gadotti_2011} and aligns well with the broader $3-30\%$ range observed in near-infrared surveys \cite{Weinzirl_2009}. The very low S\'ersic index $n\sim 0.3$ of the outer disk component allows it to fit a substantial portion of the inner regions' surface brightness. The low-luminosity fraction of the bar may then be an artifact of how the light is partitioned between the model components and not necessarily a reflection of the bar's physical significance. Although these modeling degeneracies mean the relative magnitudes of the individual components should be interpreted with caution, the GALFIT decomposition provides an accurate tracer of the galactic morphology as evidenced by the low $\chi_\nu \sim 1$ achieved in all three filters.

\begin{figure}
    \centering
    \includegraphics[width=1.0\linewidth]{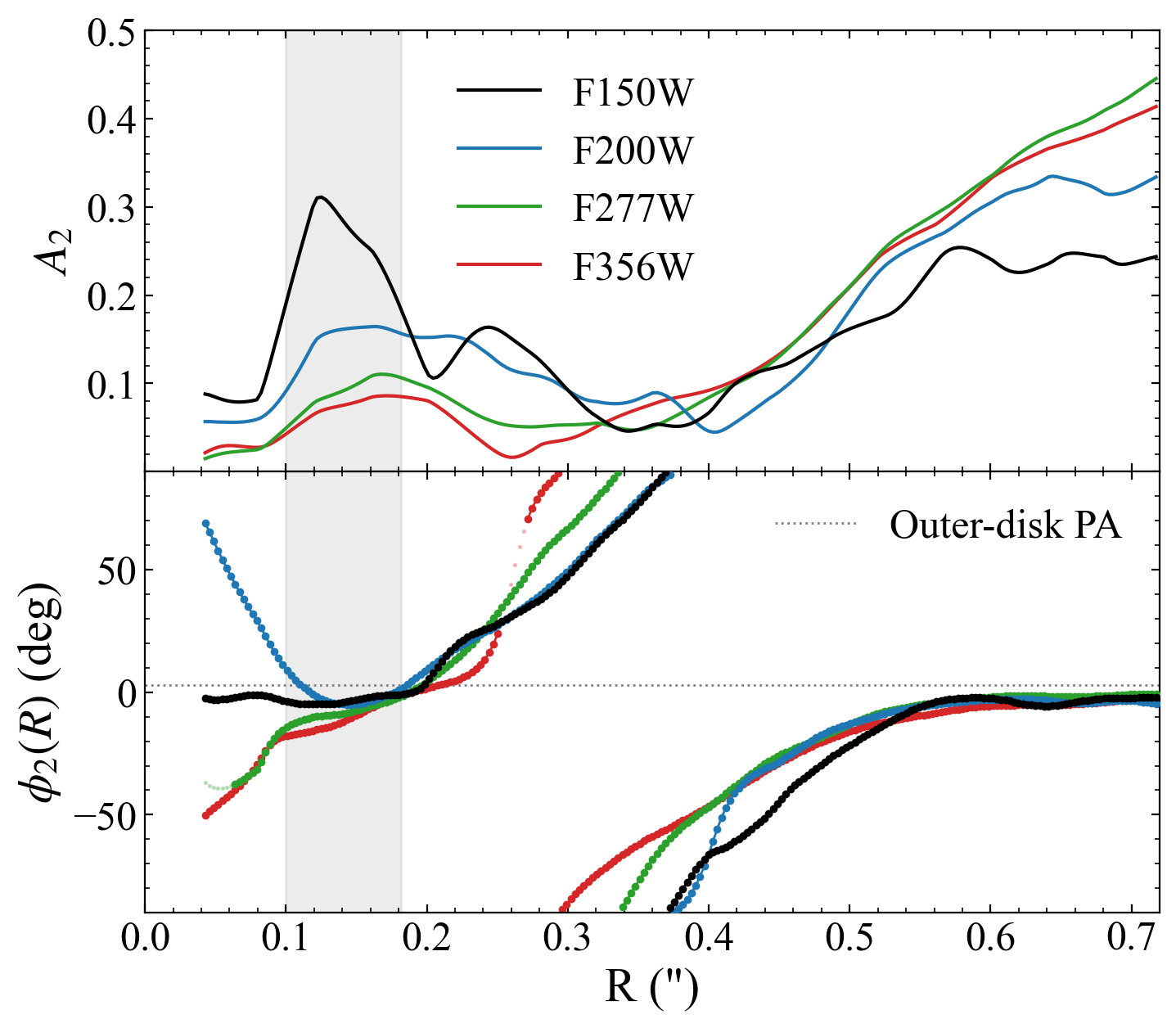}
    \caption{Radial Fourier decomposition of the $m=2$ mode for COSMOS-74706 in F150W, F200W, F277W, and F356W. The gray shaded band demarcates the bar region and the dotted line on the bottom panel indicates the disk PA. \textit{Top:} normalized $m=2$ amplitude $A_2$ as a function of galactocentric radius. \textit{Bottom:} $m=2$ phase $\phi_2(R)$ expressed as a position angle measured east of north. Faded markers indicate radii where $A_2$ falls below the noise threshold and the phase is unreliable.}
    \label{fig:phase}
\end{figure}

To further evaluate the nature of the bisymmetric signal, we examine the
radial dependence of the $m=2$ mode. We compute the angular Fourier
coefficient at each radius as
\begin{equation}
    a_m(R) = \frac{1}{2\pi}\int_0^{2\pi} I(\theta, R)\, e^{-im\theta}\, d\theta
    \label{eq:am_radial}
\end{equation}
Which preserves the radial information collapsed by $A(p, m)$. The
normalized amplitude $A_2 \equiv 2|a_2(R)|/a_0(R)$ then measures the fractional
$m=2$ contrast at each radius, where the factor of 2 in the numerator is introduced so as to match the convention of \citet{Aguerri_2009} where the $m=0$ mode differs in normalization from higher multiplicity terms by a factor of a half. The corresponding position angle is then given by
\begin{equation}
    \phi_2(R) = \frac{1}{2}\arg a_2(R)
\end{equation}
These qualities are plotted for the $2.5\sigma$ masks of a range of filters in \autoref{fig:phase}. Within the prospective bar region the $m=2$ pattern angle $\phi_2(R)$ varies smoothly with radius in all four filters ruling out a clumpy, unstructured origin for the inner light $m=2$ signal. In particular, the inner-light phase in the F150W filter tracks that of the longer-wavelength filters, indicating that the same coherent bisymmetric structure is present and dominates the $m=2$ signal within the inner region at rest-UV wavelengths. The total circular phase swing across the bar region, $\Delta\phi_2$, ranges from $\sim 4\degree$ in F150W, $\sim 14\degree$ in F200W, to $\sim 12\degree$ in F277W, and up to $\sim 16\degree$ in F356W.

While the \citet{Aguerri_2009} threshold of $\Delta \phi_2 \lesssim 20\degree$ is met for all filters, the F200W and longer filters phase changes over the bar region modestly exceed the $\Delta \phi_2 \lesssim 10\degree$ working tolerance adopted by \citet{Durbala_2009} as a definition of bar extent in their sample of well-resolved, deprojected, isolated local Sb-Sc galaxies, following the qualitative phase-coherence criterion of \citet{Laurikainen_2005} and references therein. Across all wavelengths the bisymmetry amplitude shows a pronounced peak in $A_2$ within the bar region which is characteristic of a linear non-axisymmetric structure, with peak values $A_2 \sim 0.1-0.3$ with the F200W and F277W filters in particular somewhat weaker than the $A_2 \gtrsim 0.2$ threshold that is typically used for identification of local strong bars \citep{Aguerri_2009}. Notably, the F150W filter exceeds the threshold of $A_2$ for a bar identification while remaining phase-consistent over the requisite region, whereas the F200W filter has a peak just below $A_2 \sim 0.2$, though its phase changes on the order of $\Delta \phi_2 \sim 15\degree$ over the same range.

Several effects important at $z\sim 3$ act to dilute the radial $m=2$ phase coherence relative to the local-universe baseline against which the $\Delta \phi_2 \lesssim 10\degree$ tolerance was calibrated. Foremost, the bar and spiral $m=2$ components are not spatially separated resulting in signal-mixing. \citet{Salo_2010} found that the local amplitude of the spiral arm instability modes depends on the degree of tangential forcing which the bar exercises. Baryon domination which as described in \autoref{subsec:prop} is probably the mass condition of COSMOS-74706 suppresses the azimuthally averaged radial force and thus enhances the tangential forcing $Q_b$ expected at this redshift such that high amplitude at inner radii and hence significant contamination by spiral arms is likely. Spiral contamination is further evidenced by the subtle change in the characteristic slope of $\phi_2(R)$ at approximately the end of the bar region most evident in the F277W filter but also expressed in F200W and F150W filters. This behavior corresponds to the region where the flux from the arms is no longer in superposition with the flux from the bar and the phase is now dominated by the orientation of the arms. Such contamination is observed in local barred spirals such as NGC 1808 \citep{Block_2004} and NGC 6907 \citep{Buta_2009} and prevents the $\Delta \phi_2 \lesssim 10\degree$ criterion from being met in such cases. Moreover, because of the differentiated stellar populations in the bar and arm components, the degree of arm contamination is expected to be wavelength-dependent which is consistent with the variable coherence of the prospective bar across filters. This effect may be further enhanced by PSF smearing of the arm light into the bar region at its outer edges which thus simultaneously affects the $A_2$ amplitude. Potentially more importantly, the intrinsically bright outer disk contributes $m=0$ power at all radii raising $a_0$ and thus substantially depressing the normalized contrast and hence depleting the peak $A_2$ in the bar region below the canonical threshold.

\begin{figure}
    \centering
    \includegraphics[width=1.0\linewidth]{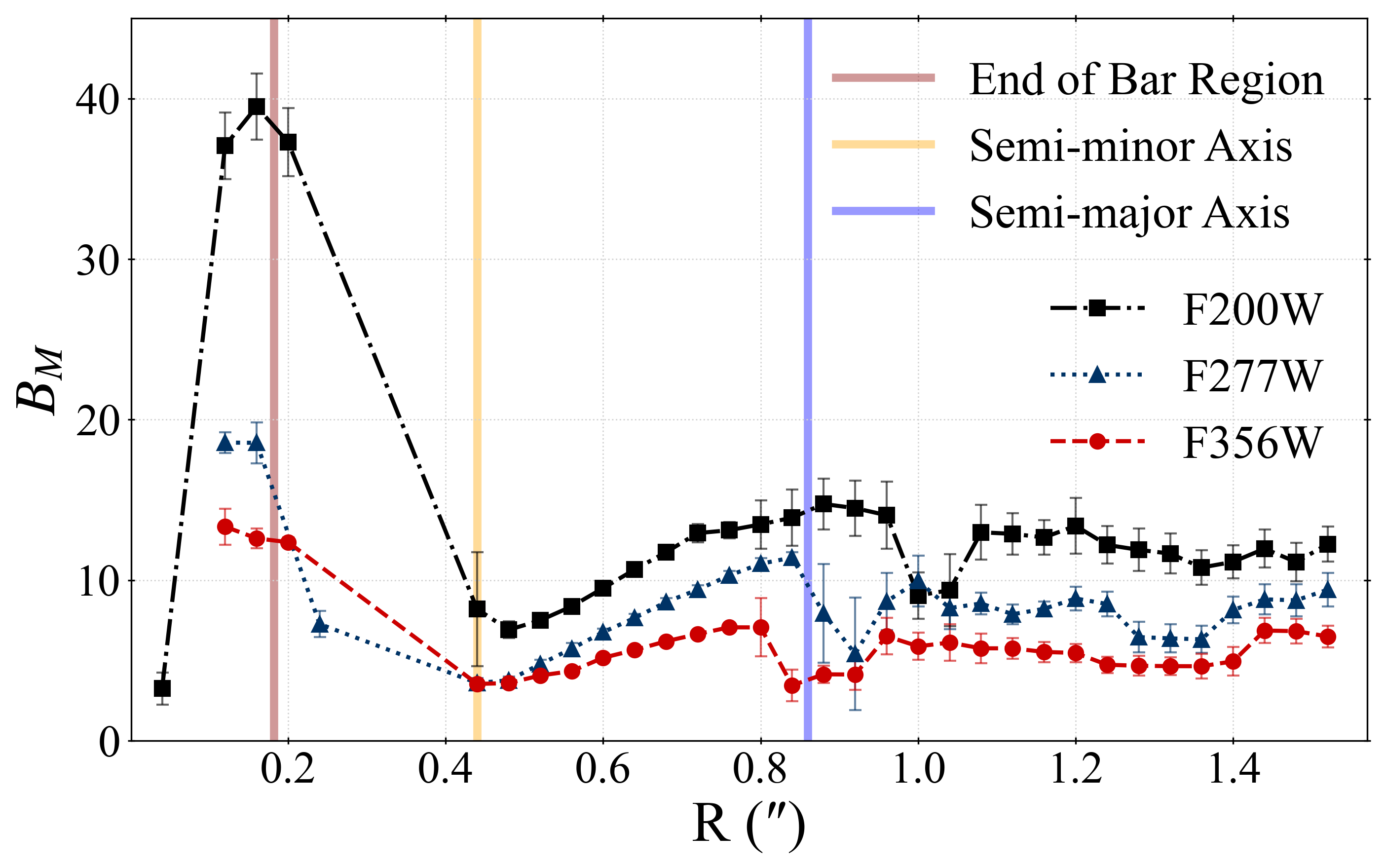}
    \caption{The $B_M$ computed for different values of the outermost radius included in the Fourier decomposition of COSMOS-74706 using the $2.5\sigma$ masked images in each of three filters. Radii for which $B_M$ was not computed due to the absence of a central Gaussian are excluded. The identified end of the bar region is the same as in \autoref{fig:isophotes}.}
    \label{fig:Bm_by_r}
\end{figure}

The $A_2$ signal in \autoref{fig:phase} clearly rises with radius starting at $R \gtrsim 0.4$\arcsec which corresponds to the extent of the semi-minor axis. In this regime the $A_2$ signal is dominated by the massive contrast in flux over the annulus between the disk and surrounding medium leading to the large inflation in signal which is purely a consequence of the intrinsic ellipticity of the disk. Its value in excess of the maximum observed in the bar region is not an indicator of its relative contribution to the $B_M$ signal as the latter metric is flux weighted, while the $A_2$ bisymmetry metric renormalizes the flux at each annular radius. The bar region's higher surface brightness thus means it contributes disproportionately to $B_M$ relative to its $A_2$ amplitude as we further demonstrate quantitatively in \autoref{fig:Bm_by_r}.


In using the methodology of \citealt{Garcia_Gomez_2017}, we have skipped deprojecting COSMOS-74706 and have hence implicitly assumed that it is in an approximately face-on orientation. To ensure this choice does not affect our results, we performed trial deprojections of the 2.5$\sigma$ mask F200W image for inclination angles of 10-70 degrees. Among these only the 50 degree inclination resulted in a value of $B_M$ below the threshold for bar identification. We note that this angle corresponds to the inclination implied by the observed axis ratio $q\approx 0.65$ under the assumption of a thin circular disk geometry. This suggests that the global bisymmetry is maximized by the galaxy's projected shape, though the bar remains robustly detected at all other angles. The full results of these deprojections can be found in \autoref{tab:deproj}.

While the outer disk's observed axis ratio of $q\approx 0.6$ is consistent with a face-on intrinsically elliptical ellipsoid, a small inclination nonetheless cannot be ruled out entirely. However, if the central elongation was the result of projecting a tilted circular disk, the feature would necessarily be aligned with the disk's major axis. We have demonstrated that the bar is oriented at a different angle than the disk and so it cannot be a projection artifact. Nonetheless, our calibration of the $B_M$ threshold was performed on a sample of deprojected galaxies whose disks were assumed to be circular. As the disk is intrinsically elliptical, it is itself, as seen in \autoref{fig:phase}, a source of bisymmetry and is expected to contribute significantly to the $m=2$ Fourier power. A high $B_M$ value for the galaxy thus cannot completely disentangle the contributions of the elliptical disk and stellar bar. 

In order to ensure that the $B_M$ value exceeds the threshold because of the presence of a bar rather than due to the intrinsic ellipticity of the disk, we repeated the Fourier decomposition for each of the filters now varying the outermost included radius for each filter. \autoref{fig:Bm_by_r} shows how $B_M$ varies by radius from the center in the $2.5\sigma$ mask. Notably $B_M$ rises to its highest value in the F200W filter where the bar is most exposed, and reaches a local maximum of $B_M>10.0$ in all three filters just before the end of the bar region where the shape of the disk has no impact on the degree of bisymmetry. As the radius increases the increased prominence of a closely spaced second bisymmetric signal with $p \approx 5$ shifts the center of the Gaussian that is fitting the bar signal to $p>1$ leading to no further $B_M$ identifications in any of the F200W, F277W, or F356W filters until $r>0.44$\arcsec when the outer radius exceeds the projected semi-minor axis causing the contribution from the shape of the outer disk to become relevant. Thereafter, as more of the outer disk is included in the decomposition, the value of $B_M$ gradually increases. Between 0.8-1.0\arcsec from the galactic center, the outer disk dwindles off. Our selected center is slightly closer to the Southern edge of the disk than to the Northern edge which accounts for a drop and subsequent increase in the bisymmetry as the outer edges of the disk on the North side begin to be included. The bar modulus does not significantly change beyond this radius.

The maximum value of $B_M$ for all three filters lies inside the bar region, which supports the significance of the bar to the overall signal. While the final $B_M$ achieved in the outer regions is a non-trivial sum of the contribution from the bar and from the shape of the disk, the minimum contribution of the bar can be estimated using the lowest value taken by the modulus immediately after the semi-minor axis is reached. This is the radius at which the bar no longer totally dominates the image, but the shape of the disk is not yet an important factor. For F200W, F277W, and F356W this value is respectively $B_M = 8.2, \, 3.6, \, 3.5$, thus significantly exceeding the bar threshold in F200W with minimal contribution from the shape of the disk. In F277W and F356W the minimum contribution is still a substantial fraction of the threshold though some additional bisymmetry from the shape of the disk substantially contributes to the threshold being exceeded in the outer regions.  Finally, since Fourier decomposition is inherently a flux-weighted procedure, we can expect that the highest surface brightness regions contribute most strongly to the resultant signal and hence it is physically motivated to conclude that a substantial fraction of the Fourier strength originates from the structure of the inner light and hence from the bar region.

\subsection{Alternate Possibilities}
\label{subsec:possibilities}

Our bar identification rests on three independent tests. However, because all three methods are ultimately an assessment of the geometry of the galaxy, each is susceptible to degeneracies with physical structures that superficially resemble a bar while having altogether different physical and dynamic properties. The most substantive alternate explanation to the observed central linear bisymmetry in COSMOS-74706 is that the inner light component represents not a bar but a small rotationally supported inner disk embedded concentrically within the larger outer disk. Such structures are common in the local universe, forming naturally as a consequence of gas inflow and nuclear star formation in disk galaxies \citep{Erwin_2002}. They manifest as a smaller flattened component sharing its rotational axis and therefore the position angle with the host disk. The near-coincidence of the inner-component and outer-disk position angles in the GALFIT decomposition up to $\Delta \theta \lesssim 1\degree$ in our highest resolution F200W fit is consistent with this interpretation in particular in light of the low likelihood of such a precise alignment between a bar and intrinsically elliptical outer disk where the PA of the two components should in principle be independent. Several lines of evidence taken together nonetheless favor the bar interpretation.

First, the F277W posterior shows a clean $\Delta\theta \gtrsim 5\degree$ separation between the inner and outer PA in every accepted MCMC sample, which is unlikely to arise for a coaxial nested disk. Second, the inner-light S\'ersic index bifurcates across filters showing a much lower $n\sim 0.6$ in the F200W and F277W decompositions as compared to the more centrally concentrated $n\sim 3.8$ seen in F356W. This wavelength dependence is naturally explained by the presence of two distinct co-existing structures whose relative flux contributions shift with rest wavelength physically overlapping in the central region, but is difficult to reconcile with a single inner-disk structure that should yield a consistent S\'ersic profile across bands \citep{Gadotti_2009}. Third, the spiral arms anchor in a $180\degree$ symmetric fashion about the ends of the inner elongation which is a geometry that is naturally produced by interior orbital trappings and is uncommon in unbarred grand design spirals, though such systems do exist in the local universe \citep{Elmegreen_1985}. Though less conclusive, in the $m=2$ mode at all filters are phase-coherent up to the \citet{Aguerri_2009} $\Delta\phi_2 \lesssim 20\degree$ tolerance whereas we expect that significant non-coherence would indicate the presence of an inner disk. Only the F150W filter however stays within the more strict $\Delta\phi_2 \lesssim 10\degree$ limit and while the corresponding image does show an inner bar-type feature visually, bars are intrinsically difficult to identify at rest-UV wavelengths where compact star-forming complexes dominate the galactic anatomy \citep{Menendez_Delmestre_2024}, warranting caution in its interpretation. Notably the F200W filter falls in the rest frame optical and is not subject to the same concern. Regardless none of these arguments is individually decisive, the variations in PA offset might for instance reflect dust or population gradients, but the collective consistency with the bar scenario, alongside the Fourier and isophotal evidence favors this interpretation as the most parsimonious explanation. Nonetheless, resolution limits at this redshift restrict how definitively the various diagnostics we employ can serve as structural differentiators, and so imaging alone does not allow a conclusive discrimination between a bar and a circumnuclear disk in this system.

The wavelength-dependent character of the bar signature bears scrutiny in the context of bandshifting. Due to the influence of kinematic fractionation driving younger stellar populations to more elongated orbits along the bar's leading edge \citep{Debattista_2017} supplemented by the diminishing surface brightness contribution of the bulge at shorter wavelengths, bars tend to appear longer and more elliptical in bluer bands \citep{Menendez_Delmestre_2024}. The observed monotonic reduction in the peak isophotal ellipticity of COSMOS-74706 in the inner light region as seen in \autoref{fig:isophotes} is qualitatively consistent with this trend, though we do not observe a significant evolution in the radius at which a turnover occurs in either the PA or ellipticity which would indicate a lengthening effect. The lack of such a lengthening may be an artifact of the need to shift the isophotes center for the F200W and F150W filters to avoid biasing the isophotes by the change in the brightest parcel of the central region, or a physical consequence of the redistribution of stellar orbits that resulted from the central starburst experienced by the system $\sim 1$ Gyr prior to observation (paper II). The recency of such a starburst concentrated in the inner regions simultaneously provides a physically meaningful explanation for the relative strength of the bar in bluer bands. The bifurcation of the inner-light S\'ersic index between the F200W/F277W and F356W is plausibly then a direct manifestation of the bulge becoming substantially more prominent at rest-NIR wavelengths, making the diminished bar signature at longer wavelengths a genuine bandshifting effect, compounded by the larger FWHM of the PSF at these wavelengths reducing the structural contrast of the bar.

A further concern is the possibility of contamination in the inner light by an unresolved source such as a dim foreground star with an SED peaking in the observer frame near-IR blueward of F150W. Both the bar region and the larger northern element are well resolved in excess of the PSF in all filters and so neither could directly correspond to such an object. The inner light is additionally uniform, elongated in blue filters, and significantly dimmer in those same bands all of which is inconsistent with a foreground single-star contribution. A star-forming region is also disfavored to account for the inner light due to its symmetry and aforementioned uniformity. The northern element on the other hand is composed of several non-uniform asymmetric clumps with mutually consistent color characteristics, disfavoring a local stellar contribution to its flux. A nature as a star-forming complex within COSMOS-74706 however remains a viable possibility.

\section{Conclusion}
\label{sec:conclusion}

We present a morphological analysis of a spiral galaxy at z=3.159 that shows evidence of a stellar bar. SED fitting and a spatially coincident detection of redshifted [OIII] in the MOSDEF spectrum of a spatially coincident object both point to this redshift. The barred character of COSMOS-74706 is inferred through three independent lines of evidence.

\begin{itemize}
\item After the subtraction of axisymmetric bulge and disk elements, GALFIT residuals consistently reveal an elongated central structure aligned approximately along the North-South axis. This result does not depend strongly on the initial guess parameters for the subtraction and visual inspection confirms that the bar-structure is aligned between a pair of spiral-arm features. 

\item Isophotal ellipse fitting also supports this finding, showing consistent fulfillment of the bar criteria in the F200W and F277W filters where the bar structure is most easily ascertainable by eye. A consistent rise in ellipticity across the bar region is observable in other filters as well and is associated with a changing PA.

\item A Fourier decomposition analysis of the galaxy reveals a central bisymmetric signal where the flux-weighted bar modulus parameter $B_{M}$ \citep{Garcia_Gomez_2017} exceeds the threshold value we calibrated for identifying bars in the F150W, F200W, F277W, F356W, and F410M filters in all noise-level cutoffs of the mask. Radial decomposition of the $m=2$ mode further reveals relative phase-coherence in the F150W filter alongside wavelength-dependent arm-contamination at longer wavelengths. A confounding factor is the intrinsic shape of the outer disk which contributes to the bisymmetry and dilutes the peak $A_2$ while magnifying $B_M$.

\end{itemize}

Future spectroscopic follow-up of COSMOS-74706 would help confirm its redshift and the mass-assembly history of the inner disk. In particular, spatially resolved kinematics of ionized or cold gas would provide a validating diagnostic of bar-driven non-circular motions in the disk. A comparison to a larger sample of barred spiral galaxies at cosmic noon is also important to constrain the frequency at which bars appear in the early universe and to further ascertain their formation time-scale as well as how their presence correlates to the intergalactic environment, dark matter halo mass, and interior gas fraction.

The question remains to what extent COSMOS-74706 is a representative of a broader population of barred disk galaxies lying at $z>3$. This galaxy was discovered in the course of work to construct a catalog of barred spiral galaxies near cosmic noon. In that work we will also look to examine the broader statistics of barred spiral galaxies in this epoch to help answer this question. 

Even as a single case study, however, and independent of its rarity COSMOS-74706 demonstrates that massive, rotationally supported disks capable of forming both bars and multi-armed spirals already existed at $z>3$. Any comprehensive theory of disk evolution must account for such systems.

\section{Acknowledgments}
This work is based in part on observations made with the NASA/ESA/CSA James Webb Space Telescope. The data were obtained from the Mikulski Archive for Space Telescopes, which is operated by the Association of Universities for Research in Astronomy, Inc., under NASA contract NAS 5-03127 for JWST. These observations are associated with program \#1837. Support for program \#1837 was provided by NASA through a grant from the Space Telescope Science Institute.

This research made use of data that were obtained at the W. M. Keck Observatory, operated jointly by the California Institute of Technology, the University of California, and the National Aeronautics and Space Administration. The Observatory's work is made possible by generous financial support from the W. M. Keck Foundation.

We also made use of Astropy, a community-developed core Python package for Astronomy \citep{astropy:2013, astropy:2018, astropy:2022}, and of Photutils, an Astropy package for detection and photometry of astronomical sources \citep{Bradley_2024}. In addition, we would like to thank the EAZY \citep{Brammer_2008} and GALFIT \citep{Peng_2010} development teams for making their software publicly accessible.

J.R.W. acknowledges that support for this work was provided by The Brinson Foundation through a Brinson Prize Fellowship grant.

Lastly we would like to thank the anonymous referee for their time and helpful comments.

\bibliography{refs.bib}{} 

\appendix

\vspace{-0.5cm}

\counterwithin{figure}{section}
\counterwithin{table}{section}

\section{Bar Strength Metrics}
\label{app:bar_metrics}

We used the same metrics for the bar strength as the ones introduced in \cite{Garcia_Gomez_2017}. The formulae reported here are slightly different from theirs due to an algebraic error 
\begin{equation}
    B_{P} = C_B^2 \int_{-p_{\text{max}}}^{p_{\text{max}}} \exp \left( - \frac{(p - p_B)^2}{\sigma_B^2} \right) dp
\end{equation}

\begin{equation}
    B_{M0}
= \frac{ C_B \displaystyle \int_{-p_{\max}}^{p_{\max}} 
\exp\!\left(-\frac{(p - p_B)^2}{2\sigma_B^2}\right)\, dp }
{ \displaystyle \int_{-p_{\max}}^{p_{\max}} \lvert A(p,0) \rvert \, dp }
\end{equation}

\begin{equation}
    B_{M2}
= \frac{ C_B \displaystyle \int_{-p_{\max}}^{p_{\max}} 
\exp\!\left(-\frac{(p - p_B)^2}{2\sigma_B^2}\right)\, dp }
{ \displaystyle \int_{-p_{\max}}^{p_{\max}} \lvert A(p,2) \rvert \, dp }
\end{equation}

\begin{equation}
    B_{P0}
= \frac{ C_B^2 \displaystyle \int_{-p_{\max}}^{p_{\max}} 
\exp\!\left(-\frac{(p - p_B)^2}{\sigma_B^2}\right)\, dp }
{ \displaystyle \int_{-p_{\max}}^{p_{\max}} \lvert A(p,0) \rvert^2 \, dp }
\end{equation}

\begin{equation}
    B_{P2}
= \frac{ C_B^2 \displaystyle \int_{-p_{\max}}^{p_{\max}} 
\exp\!\left(-\frac{(p - p_B)^2}{\sigma_B^2}\right)\, dp }
{ \displaystyle \int_{-p_{\max}}^{p_{\max}} \lvert A(p,2) \rvert^2 \, dp }
\end{equation}

\section{Alternate Stack}
\label{app:alt_stack}

Throughout the course of this paper all masks we reference have used a single stack composed using the F606W and F814W HST/ACS images. We refer to this as stack 2. An alternate stack was also created in the course of testing using the same two ACS filters alongside the F090W JWST/NIRCam image. After noise-equalizing and stacking to the same pixel scale we produced segmentation maps of the blue northern element with stack 3 and created four additional masks using $2.5\sigma$, $3\sigma$, $3.5\sigma$, and $4\sigma$ noise cutoffs in the segmentation map respectively. The addition of the NIRCam filter did not alter our results and we include it here for completeness.

\begin{figure*}
        \centering
    
        \subfigure
        {
            \includegraphics[width=0.31\textwidth]{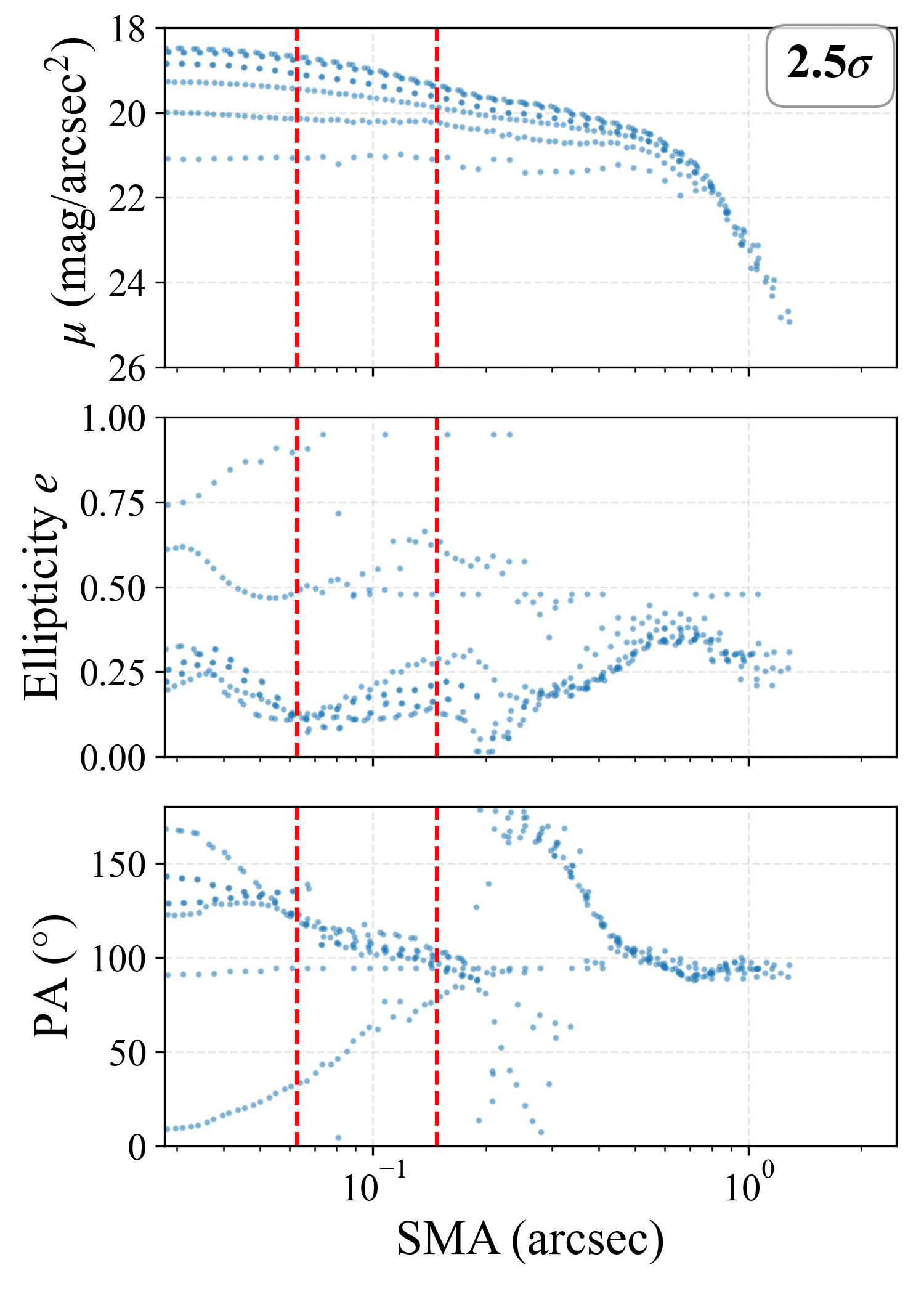}
        }
        \hspace{\fill}
        \subfigure
        {
            \includegraphics[width=0.31\textwidth]{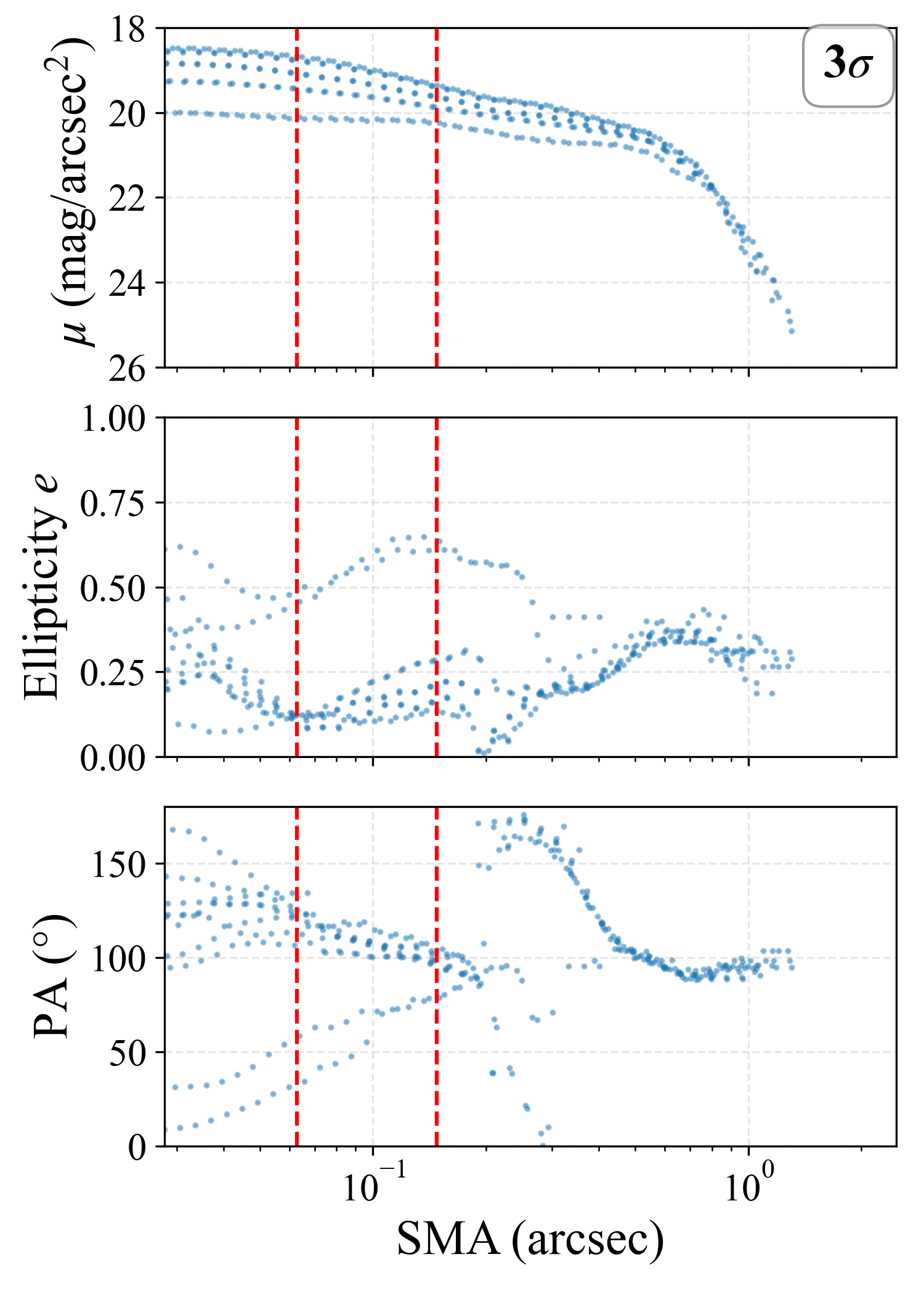}
        }
        \hspace{\fill}
        \subfigure
        {
            \includegraphics[width=0.31\textwidth]{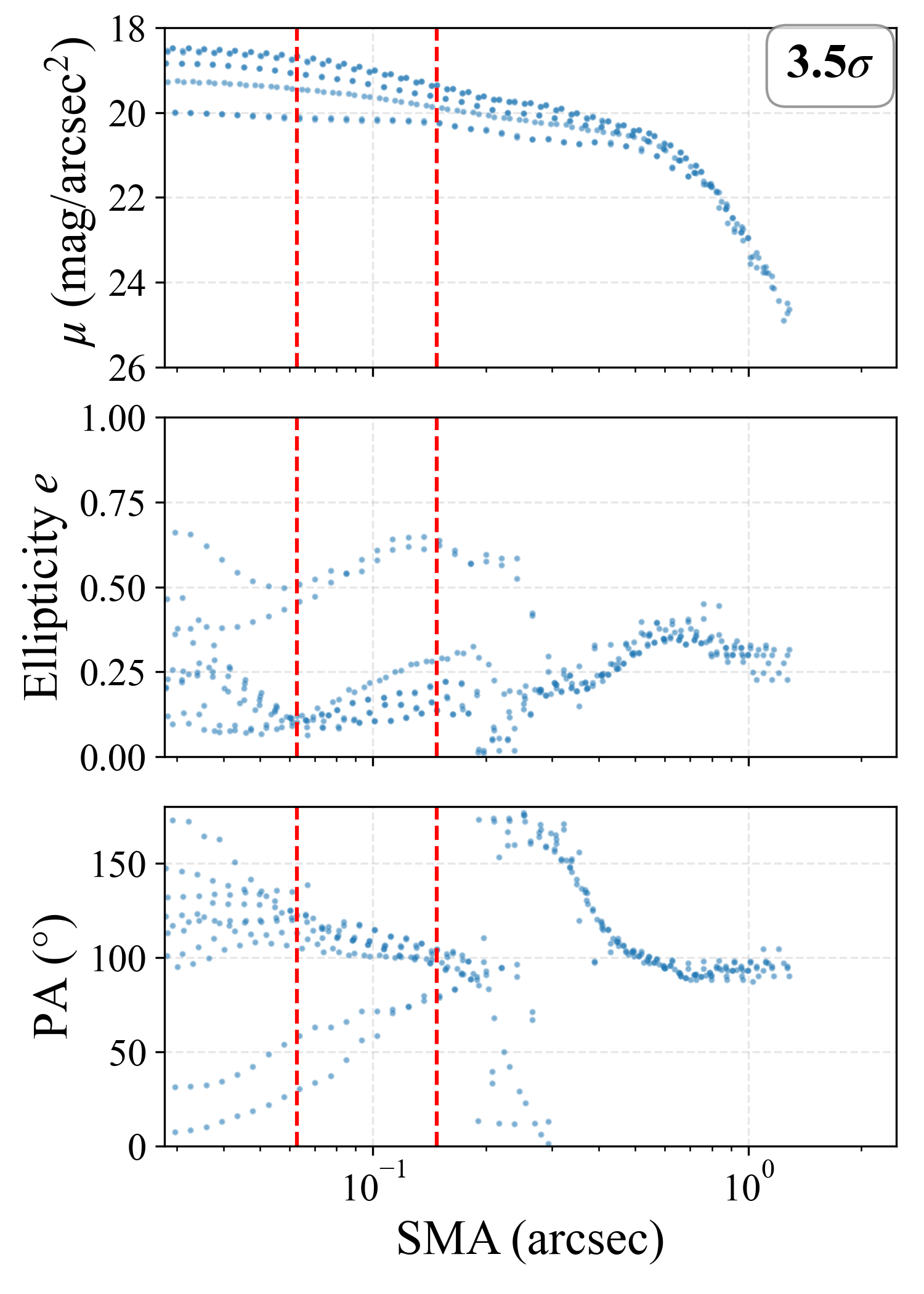}
        }
        \hspace{\fill}
        \subfigure
        {
            \includegraphics[width=0.31\textwidth]{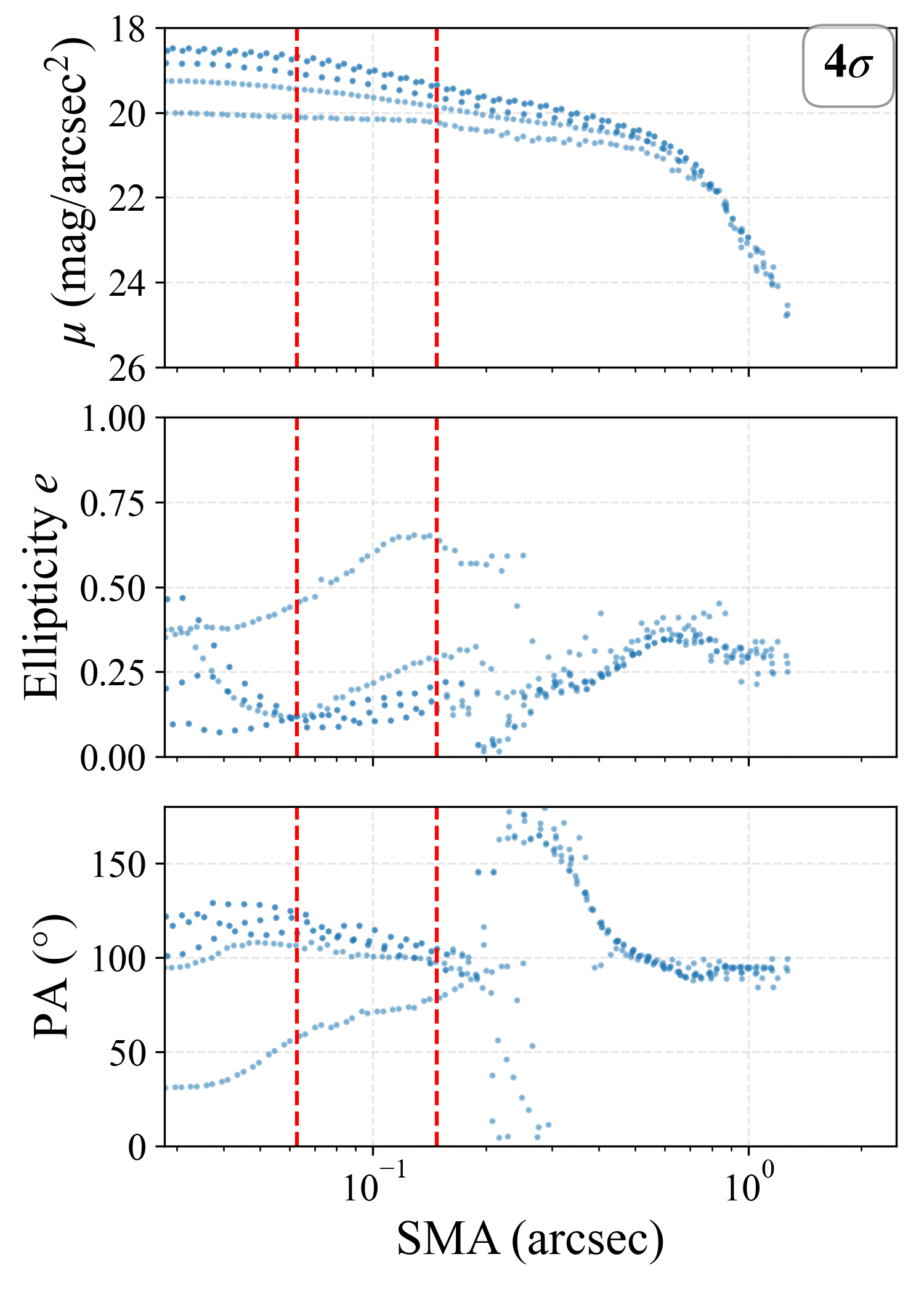}
        }
        \hspace{\fill}
        \subfigure
        {
            \includegraphics[width=0.31\textwidth]{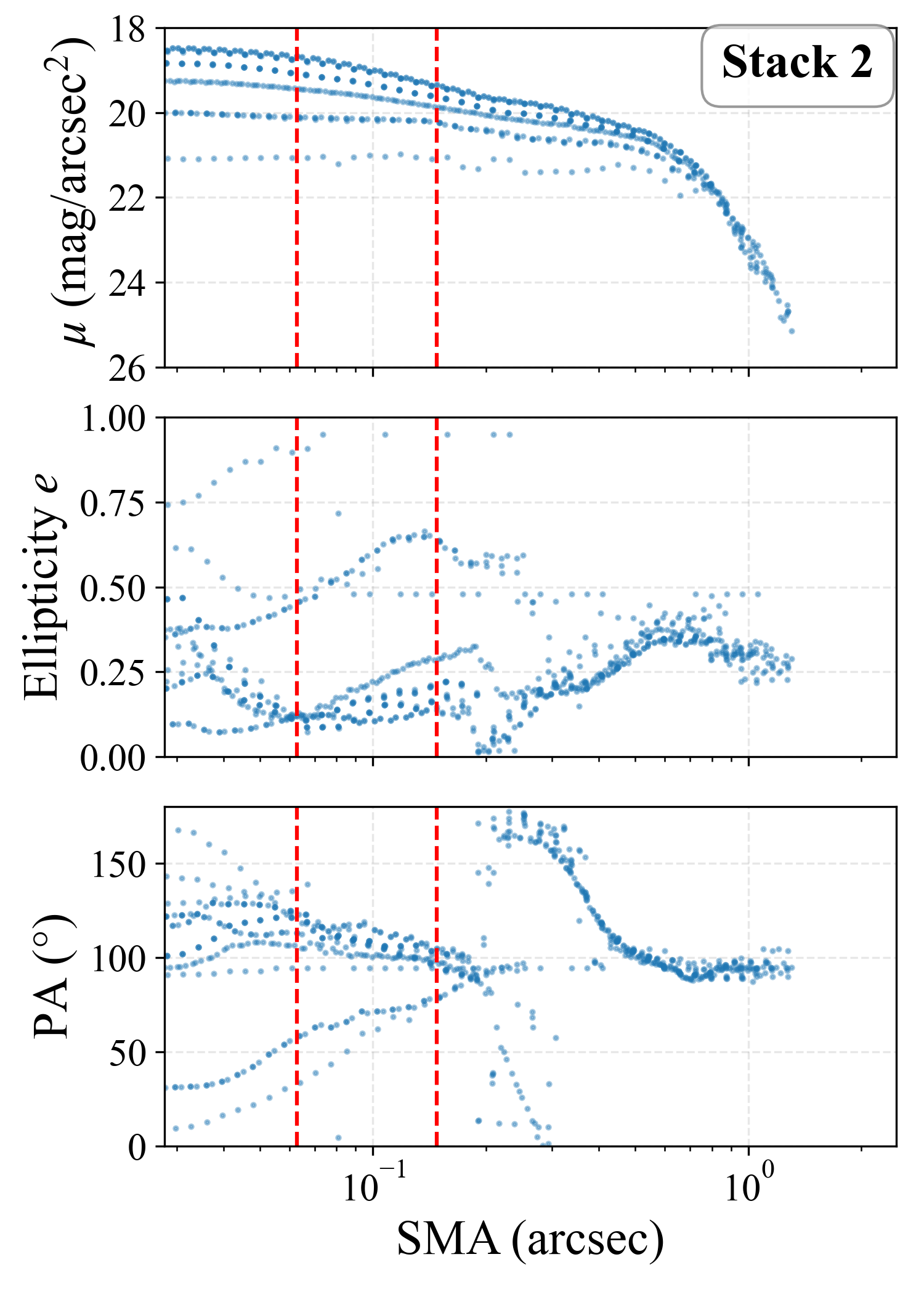}
        }
        \hspace{\fill}
        \subfigure
        {
            \includegraphics[width=0.31\textwidth]{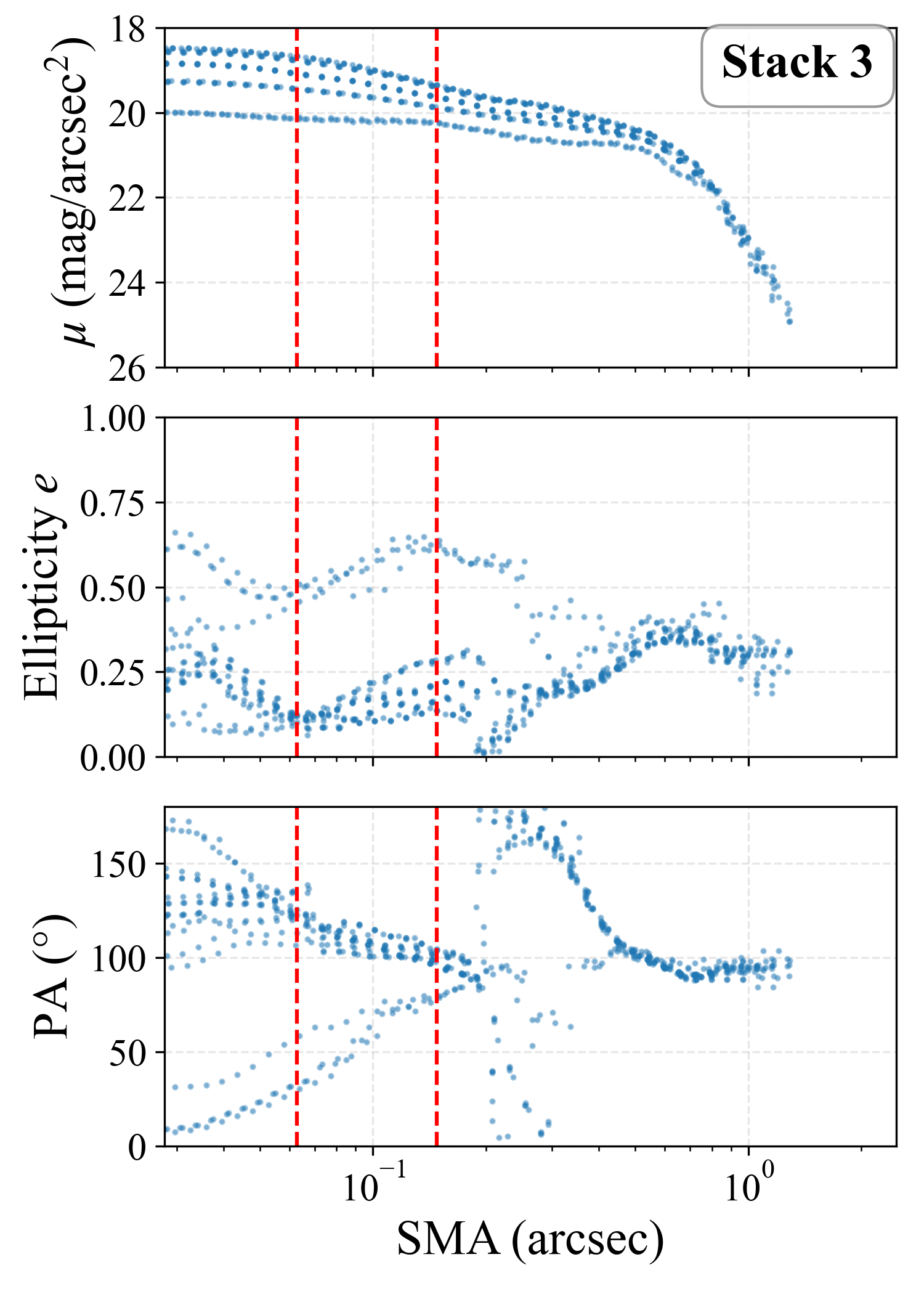}
        }
    
        \caption{As \autoref{fig:isophotes} but now showing the combined data from the two different stacks. There is no significant difference between the stacks or between different noise level cutoffs.}
        \label{fig:isophotes_appendix}
    \end{figure*}

\autoref{fig:isophotes_appendix} shows isophotal plots of all eight masks across the two stacks separated by the noise cutoff and by stack. The different noise cutoffs are largely indistinguishable with the exception of a slightly noisier bar region in the $2.5\sigma$ cutoff and the presence of a successful F150W fit in this mask. The two stacks also show no significant differences among themselves, save for the same 2.5$\sigma$ F150W filter image being fit successfully only in stack 2. The principal difference for the isophotal analysis is that the PA change across the bar region in the F200W filters is somewhat higher, overrunning the $\Delta \theta$ in all but the $4\sigma$ mask, and the F277W filter's drop in ellipticity after the bar region is slightly lower such that the ellipticity drop criterion being fulfilled depends sensitively on the end of the bar region and necessitates shifting its end about an isophote further from the center in accordance with the method for selecting the bar region described in \S\ref{subsec:isophotal}. Note that the selected center used for all isophotes in these plots is the same as that adopted for the Fourier analysis and longer wavelength isophotes in the main text.

\begin{figure*}
        \centering
        \includegraphics[width=0.75\linewidth]{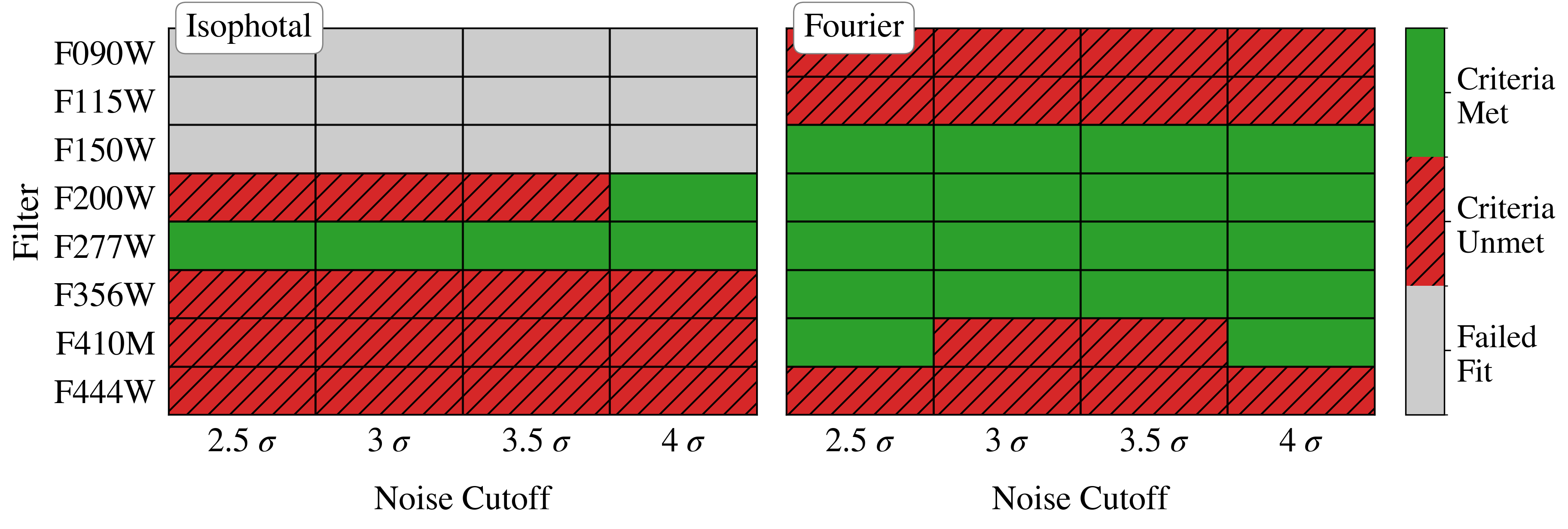}
        \caption{As \autoref{fig:analysis_panels} but for stack 3.}
        \label{fig:analysis_appendix}
    \end{figure*}

The two panels in \autoref{fig:analysis_appendix} show the final results of both the isophotal and Fourier analysis using stack 3. The Fourier analysis is largely unchanged save for a lack of detections in F444W and two non-detections in F410M. The isophotal analysis only has one F200W detection due to the steeper $\Delta \theta$ over the bar region.

We ultimately selected stack 2 for use in our analysis as it was simpler, using only ACS imaging, while producing results that were not significantly different from stack 3 while at the same time including more overall detections across the Fourier and isophotal analysis and a successful isophote fit in the F150W filter. The choice of stack did not affect our conclusions.

\pagebreak

    \clearpage
    
    \clearpage


\section{Supplementary Figures}
\label{app:supp}

\begin{figure*}[h]
        \centering
    
        \subfigure
        {
            \includegraphics[width=0.31\textwidth]{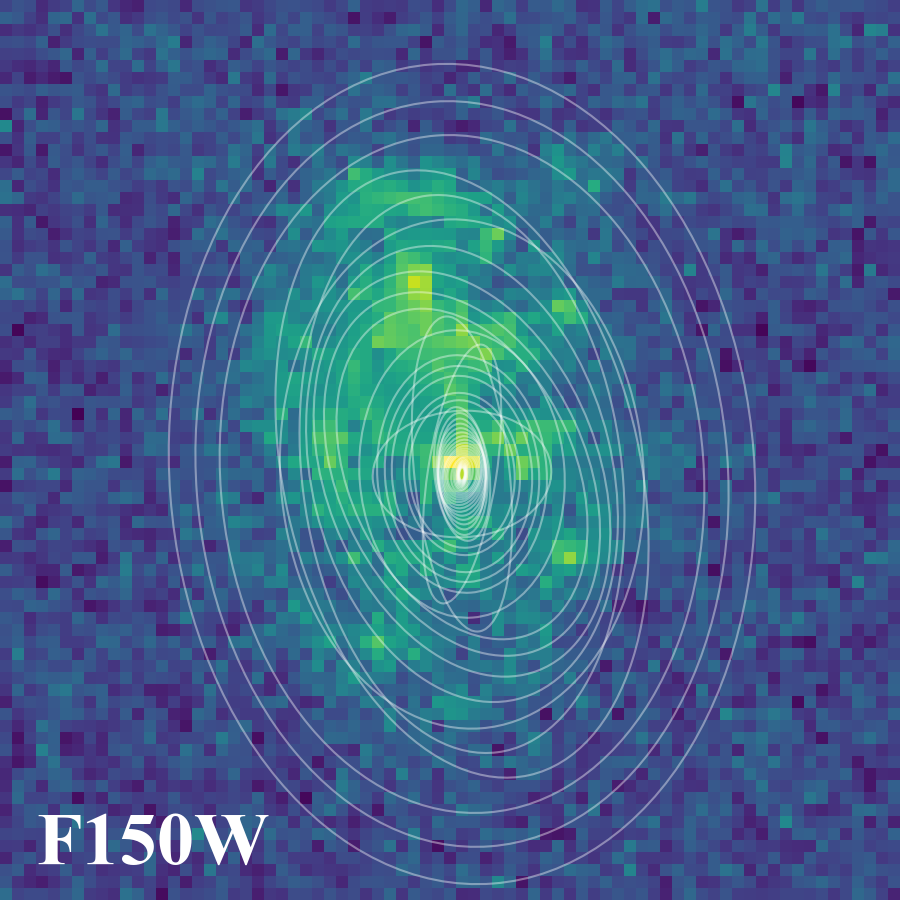}
        }
        \hspace{\fill}
        \subfigure
        {
            \includegraphics[width=0.31\textwidth]{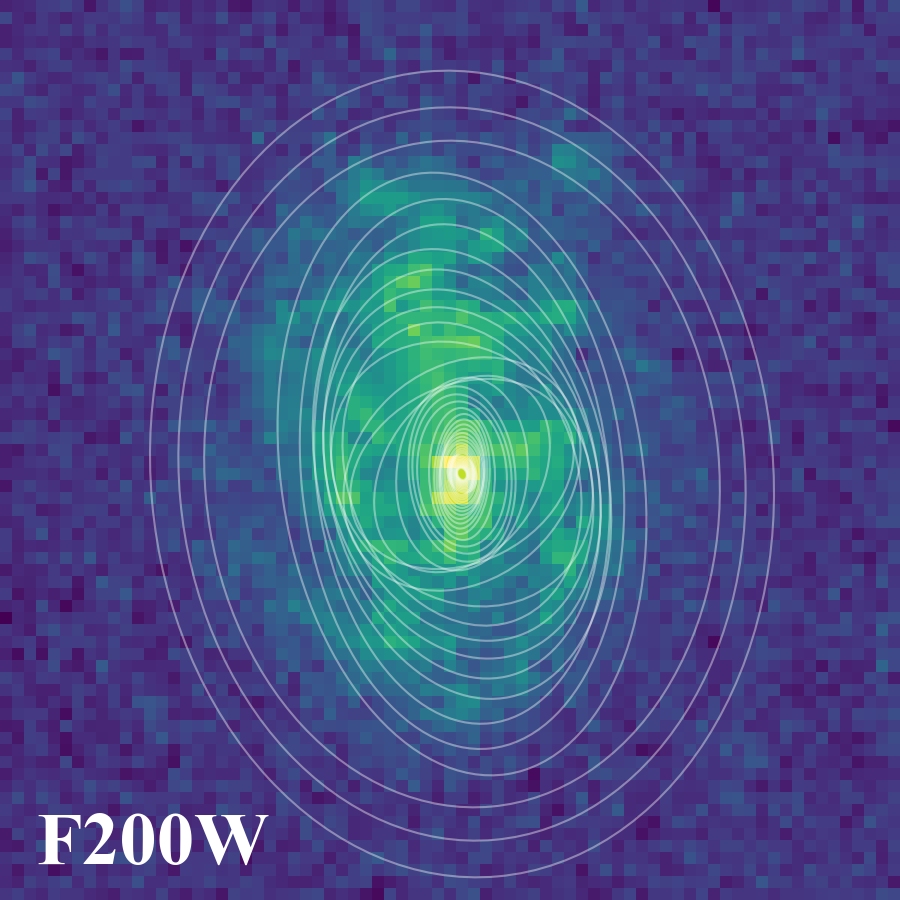}
        }
        \hspace{\fill}
        \subfigure
        {
            \includegraphics[width=0.31\textwidth]{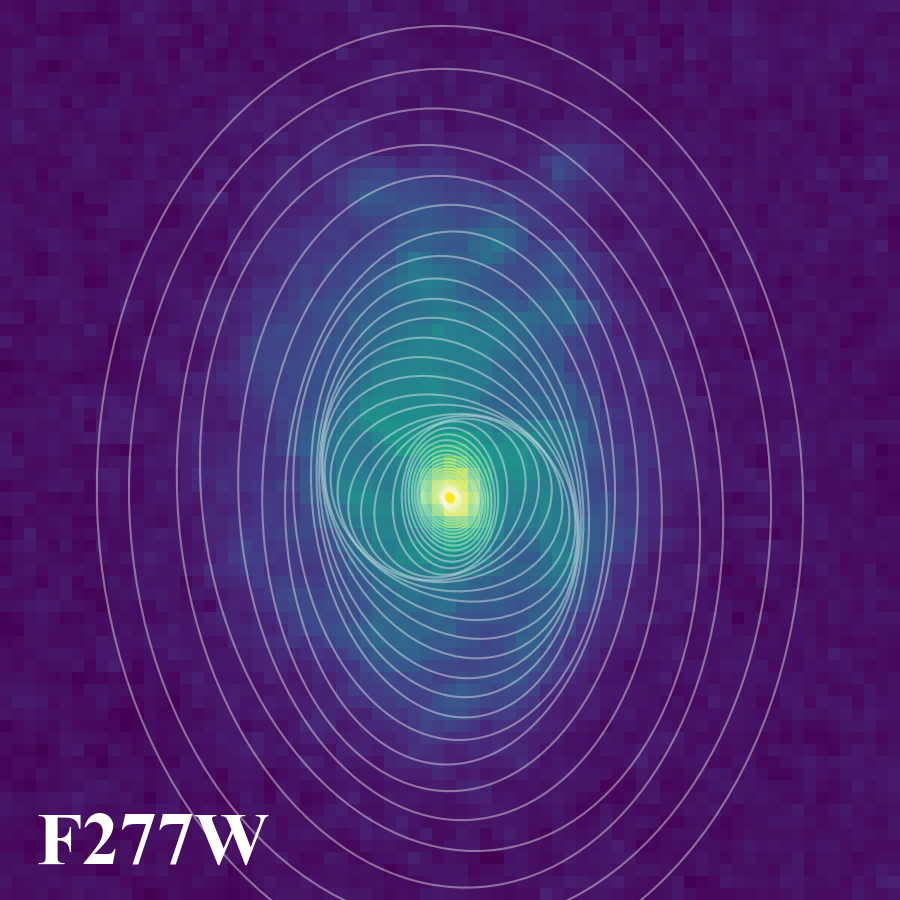}
        }
        \hspace{\fill}
        \subfigure
        {
            \includegraphics[width=0.31\textwidth]{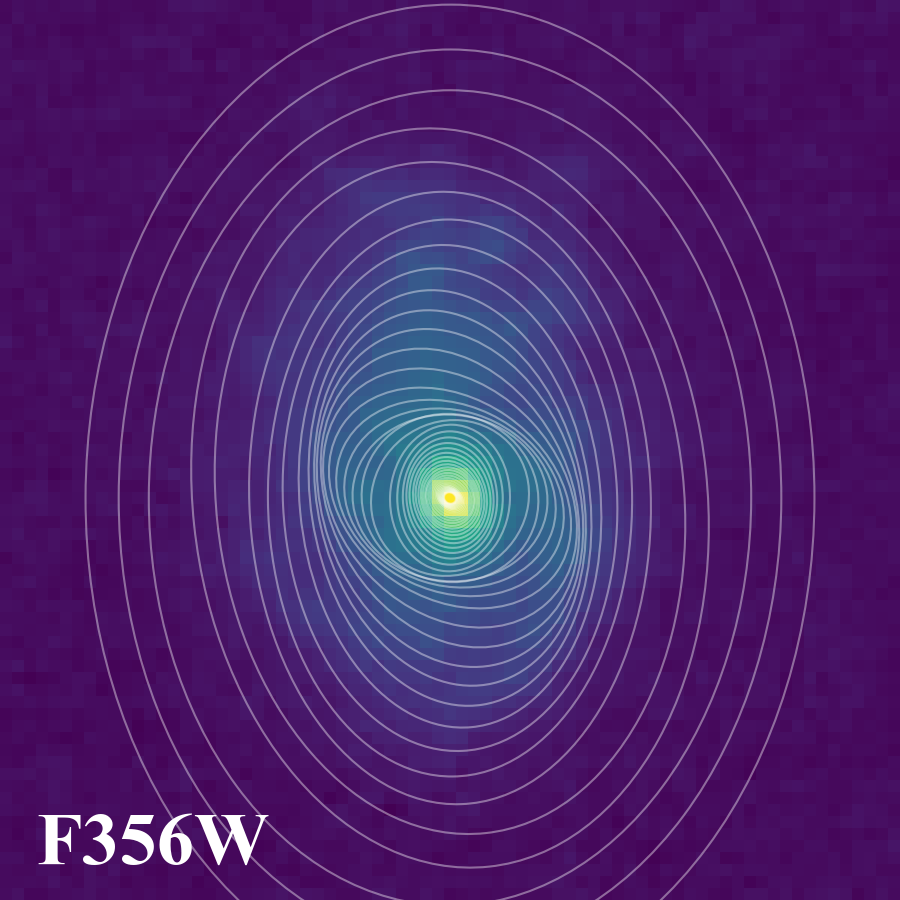}
        }
        \hspace{\fill}
        \subfigure
        {
            \includegraphics[width=0.31\textwidth]{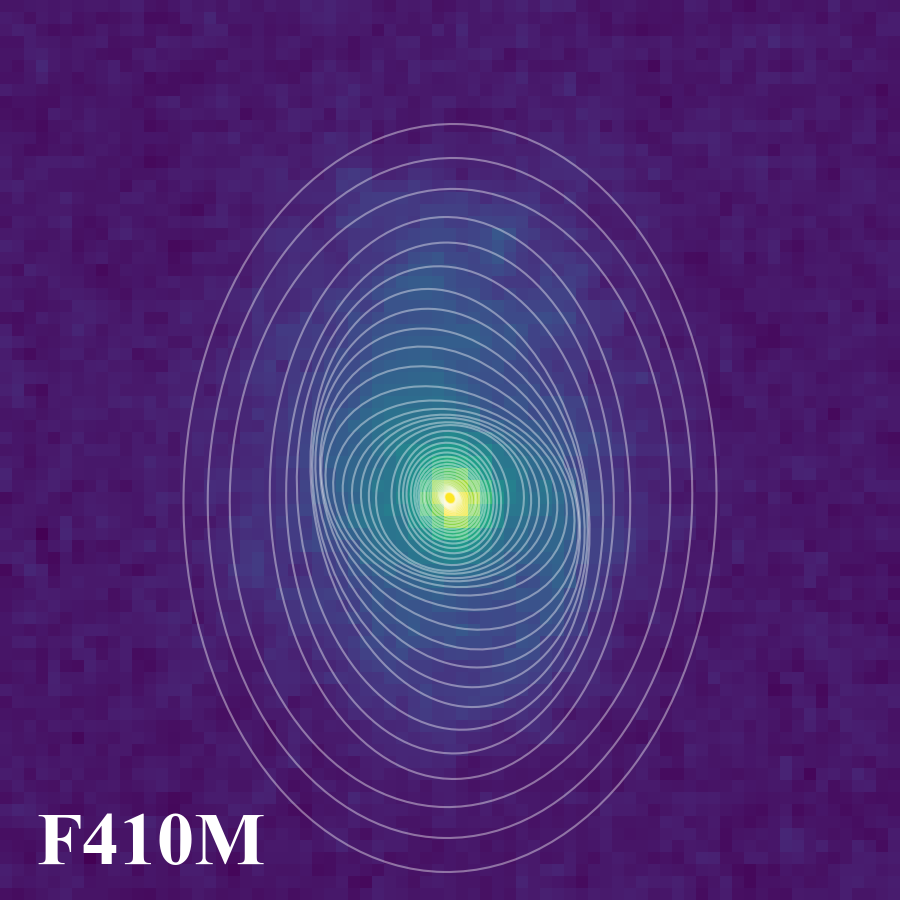}
        }
        \hspace{\fill}
        \subfigure
        {
            \includegraphics[width=0.31\textwidth]{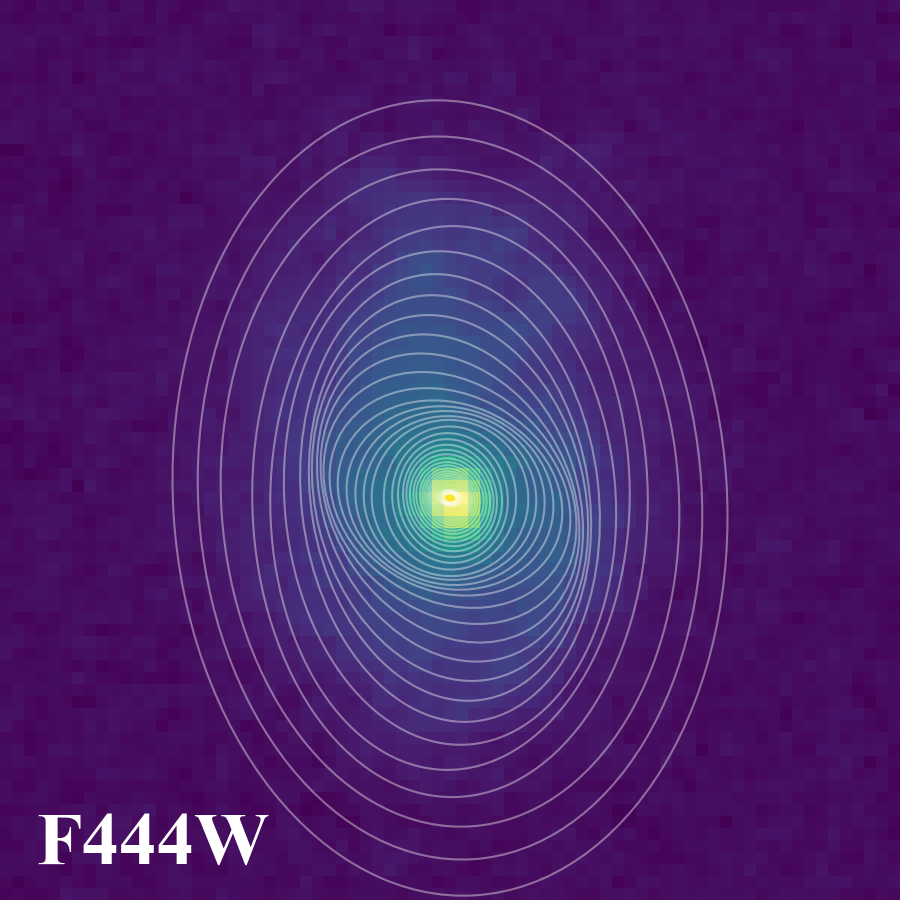}
        }
        \caption{Elliptical isophotes for stack 2 and $2.5\sigma$ masked images in each filter. The transition from the bar region can best be seen at the radius where the position angle of the isophotes sharply changes in the F277W filter. Only this filter satisfies the bar criteria for this mask.}
        \label{fig:isophotes_appendix_imgs}
    \end{figure*}

\begin{figure*}[h]
    \centering

    \subfigure
    {
        \includegraphics[width=0.3\textwidth]{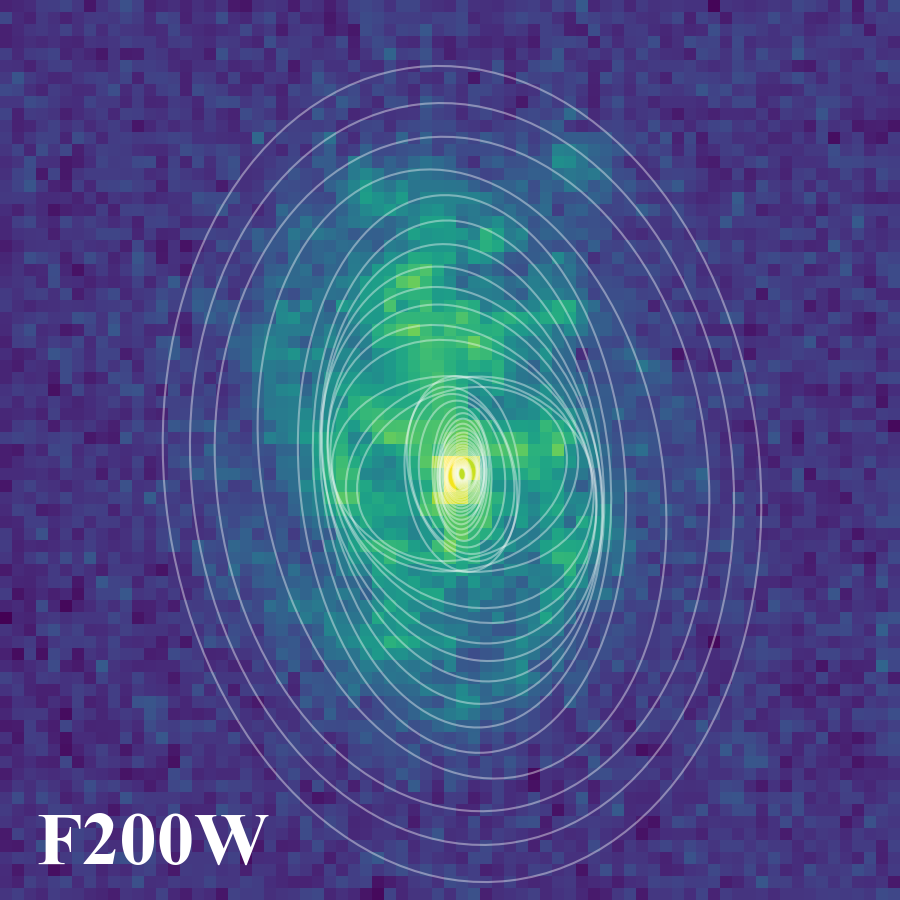}
    }
    \hspace{\fill}
    \subfigure
    {
        \includegraphics[width=0.3\textwidth]{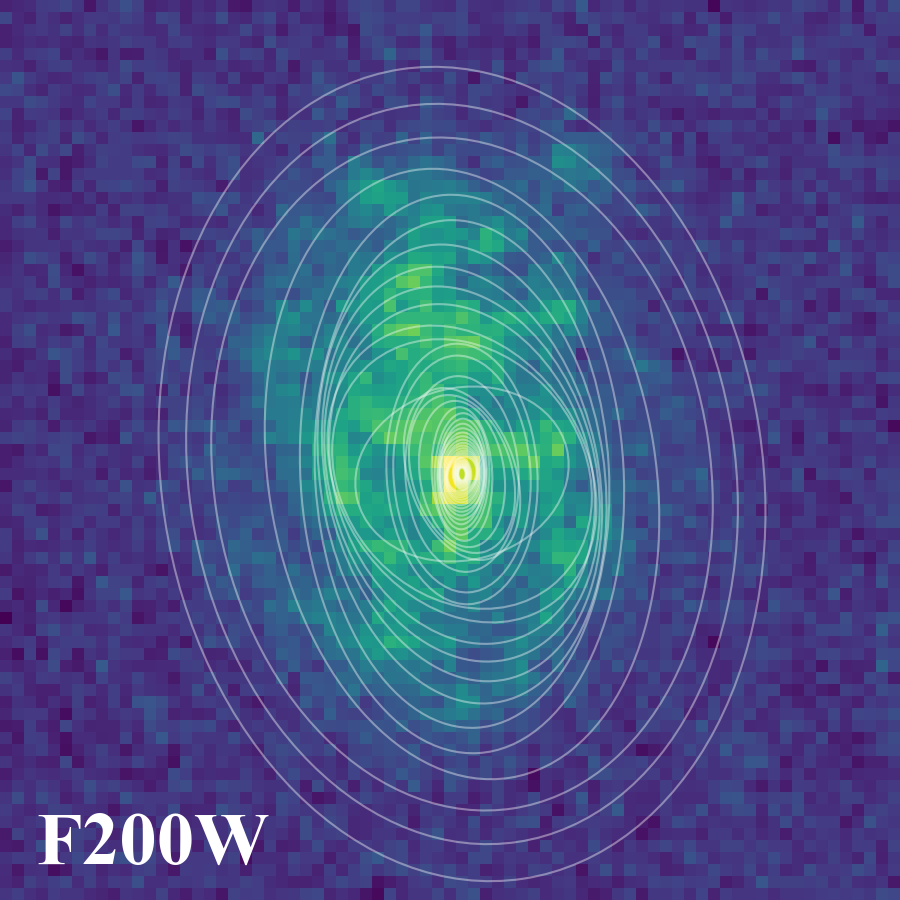}
    }
    \hspace{\fill}
    \subfigure
    {
        \includegraphics[width=0.3\textwidth]{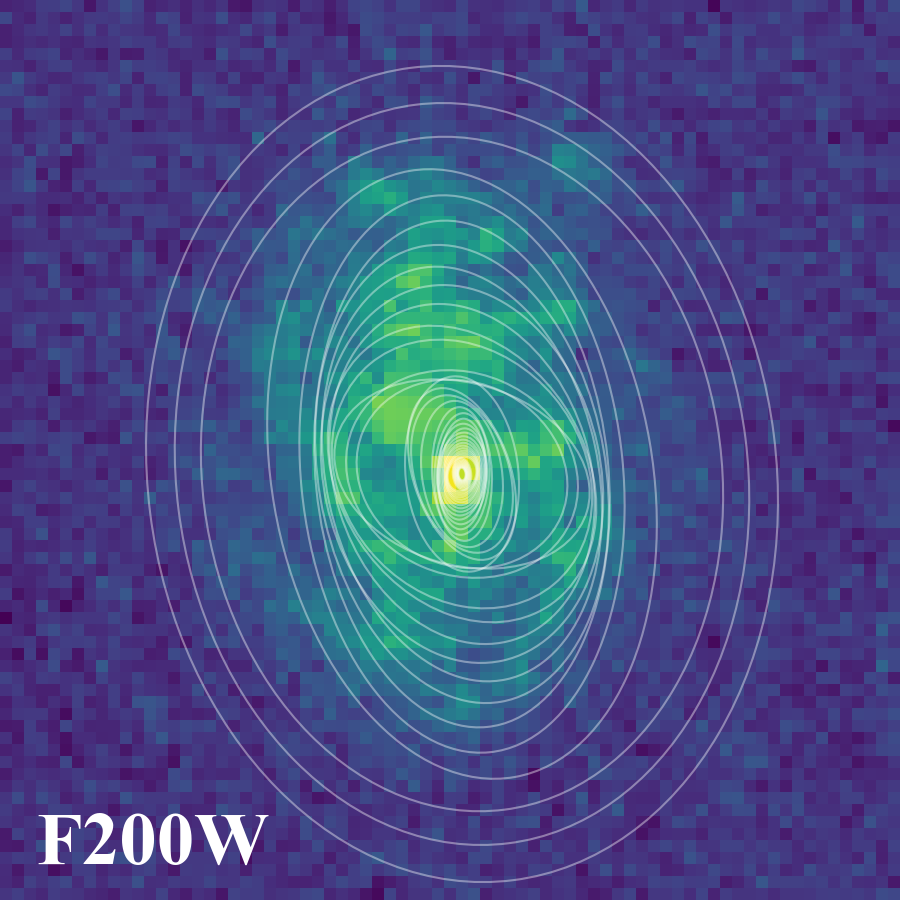}
    }

    \subfigure
    {
        \includegraphics[width=0.3\textwidth]{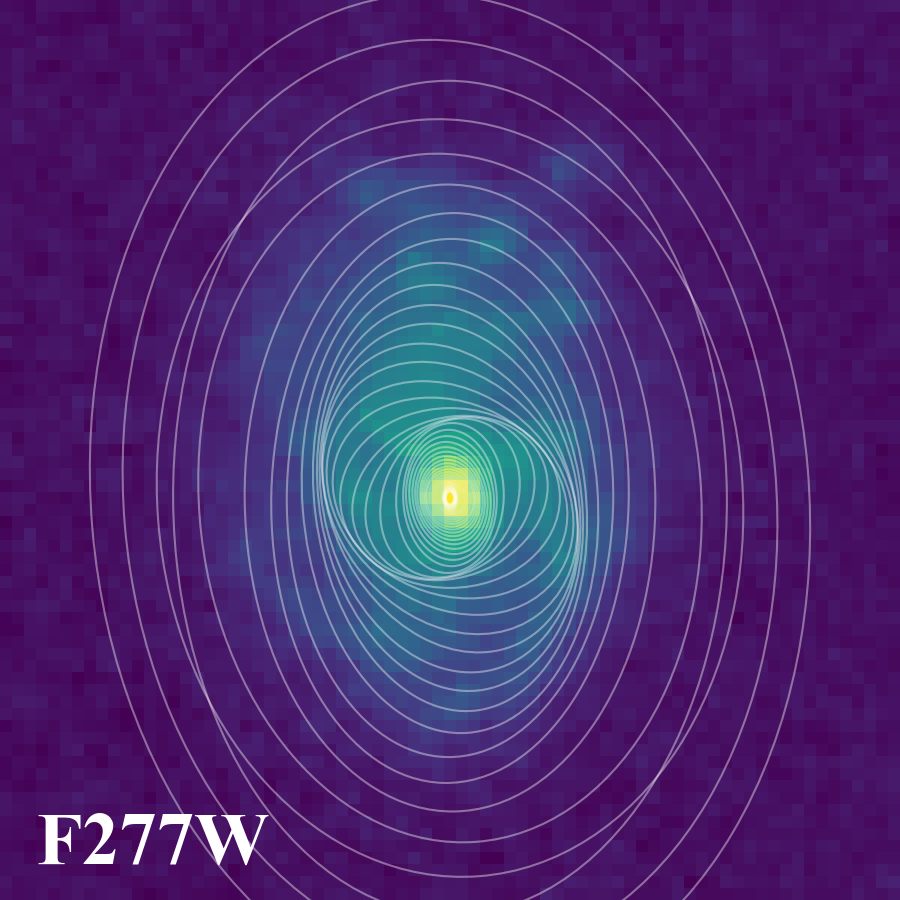}
    }
    \hspace{\fill}
    \subfigure
    {
        \includegraphics[width=0.3\textwidth]{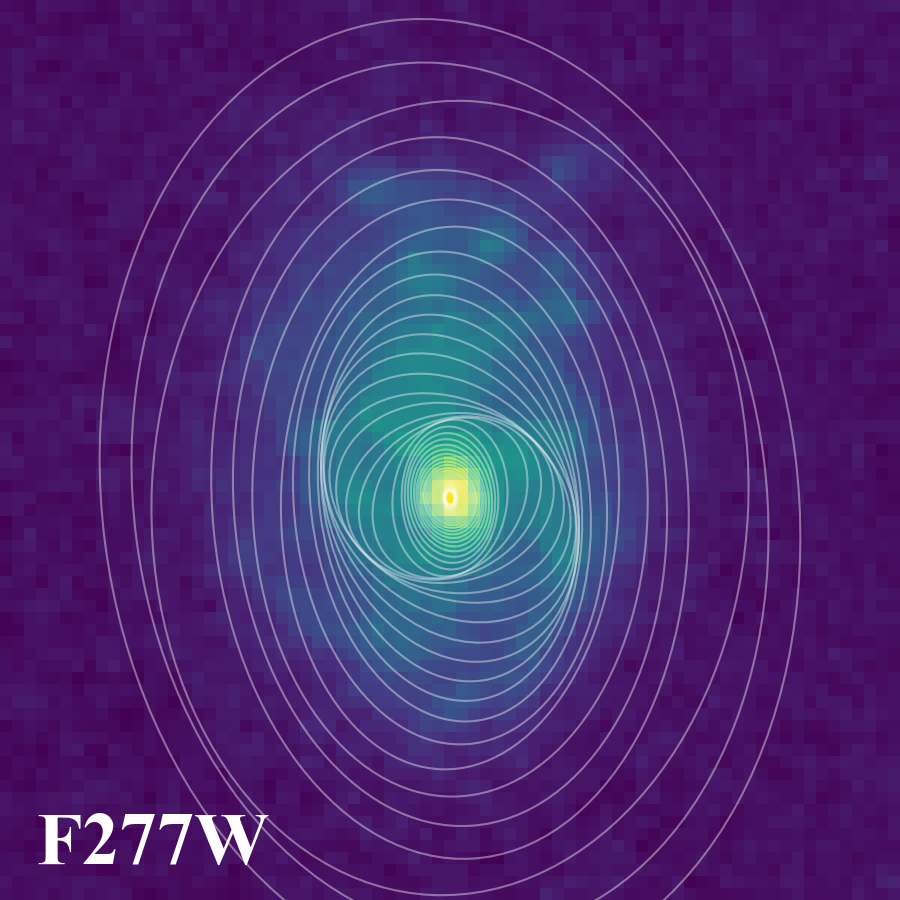}
    }
    \hspace{\fill}
    \subfigure
    {
        \includegraphics[width=0.3\textwidth]{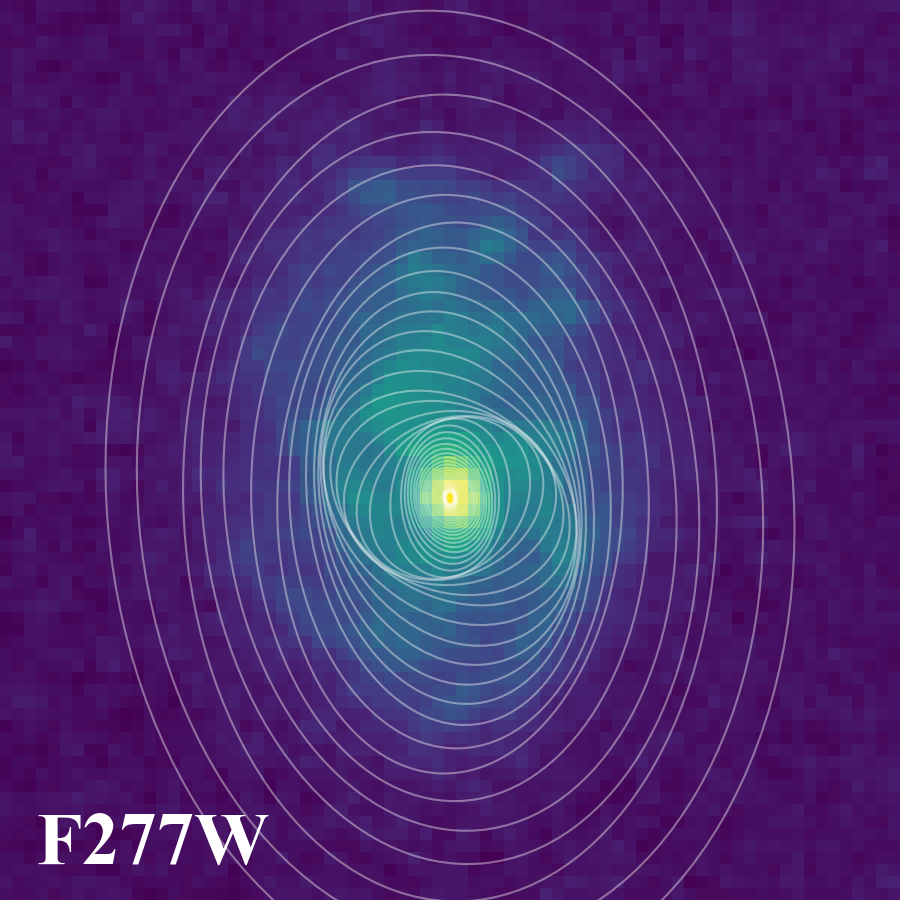}
    }

    \subfigure
    {
        \includegraphics[width=0.3\textwidth]{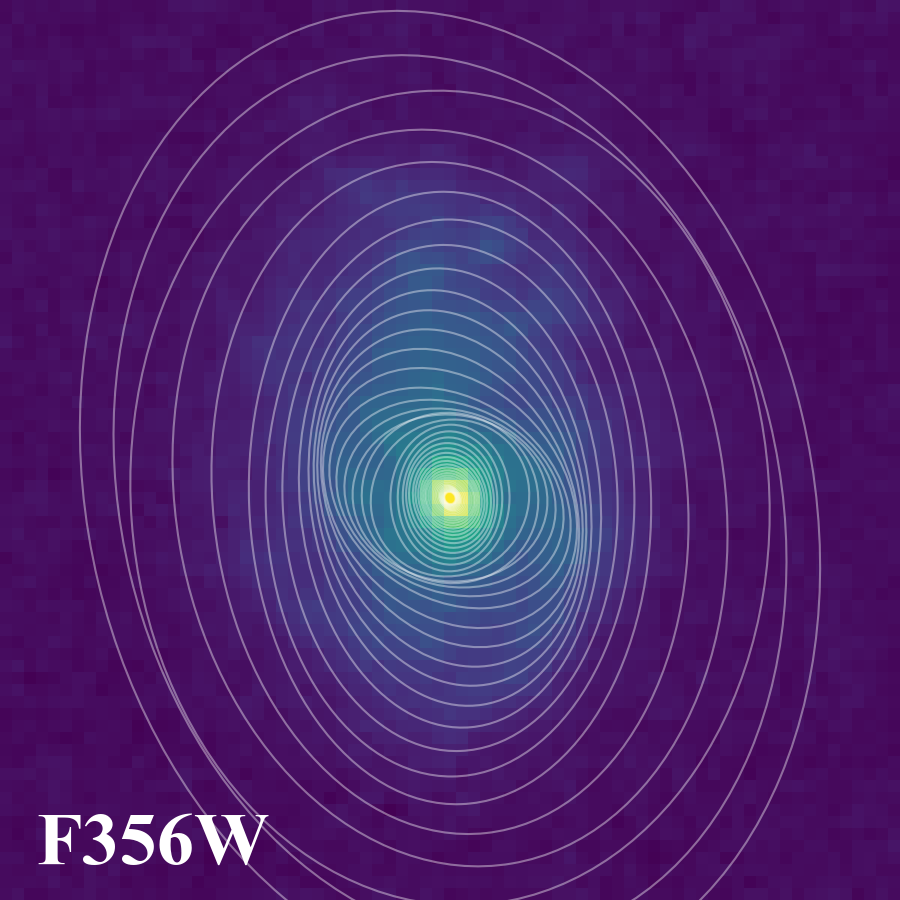}
    }
    \hspace{\fill}
    \subfigure
    {
        \includegraphics[width=0.3\textwidth]{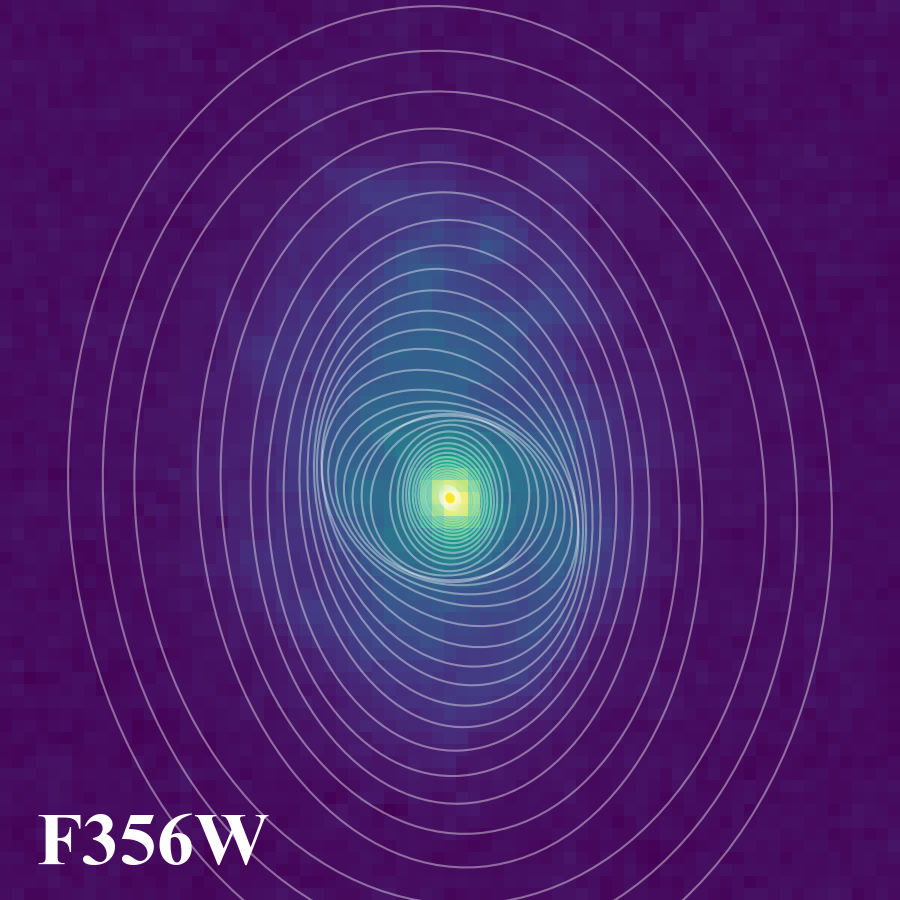}
    }
    \hspace{\fill}
    \subfigure
    {
        \includegraphics[width=0.3\textwidth]{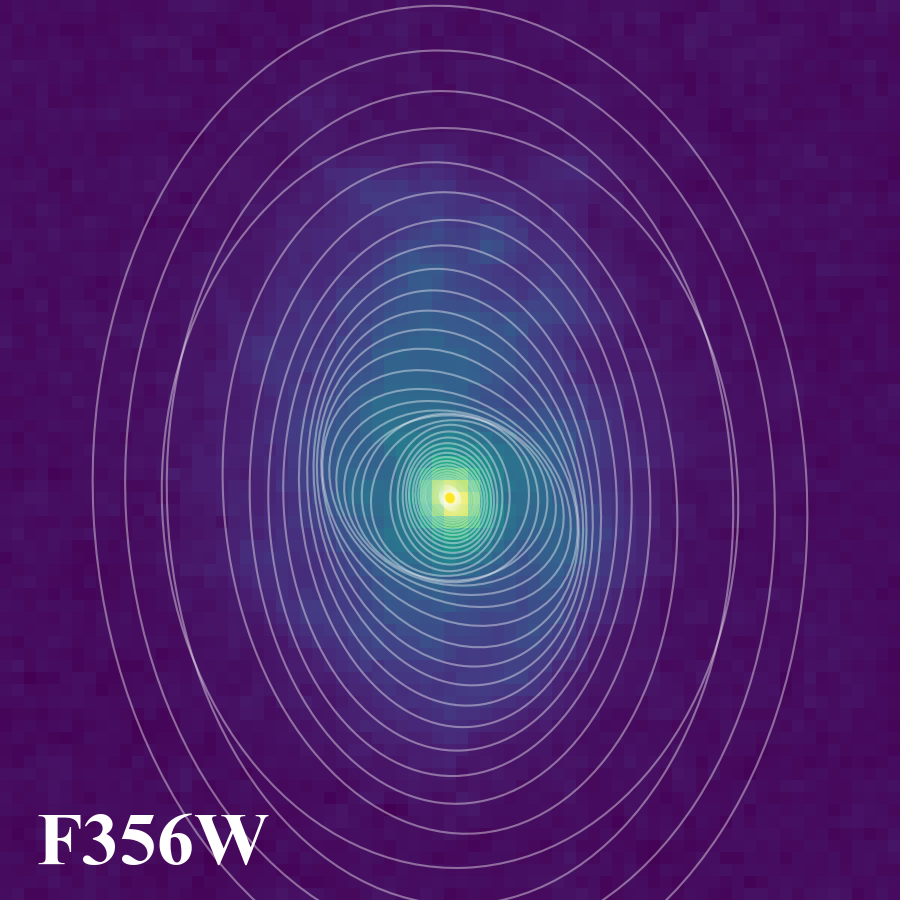}
    }

    \subfigure
    {
        \includegraphics[width=0.3\textwidth]{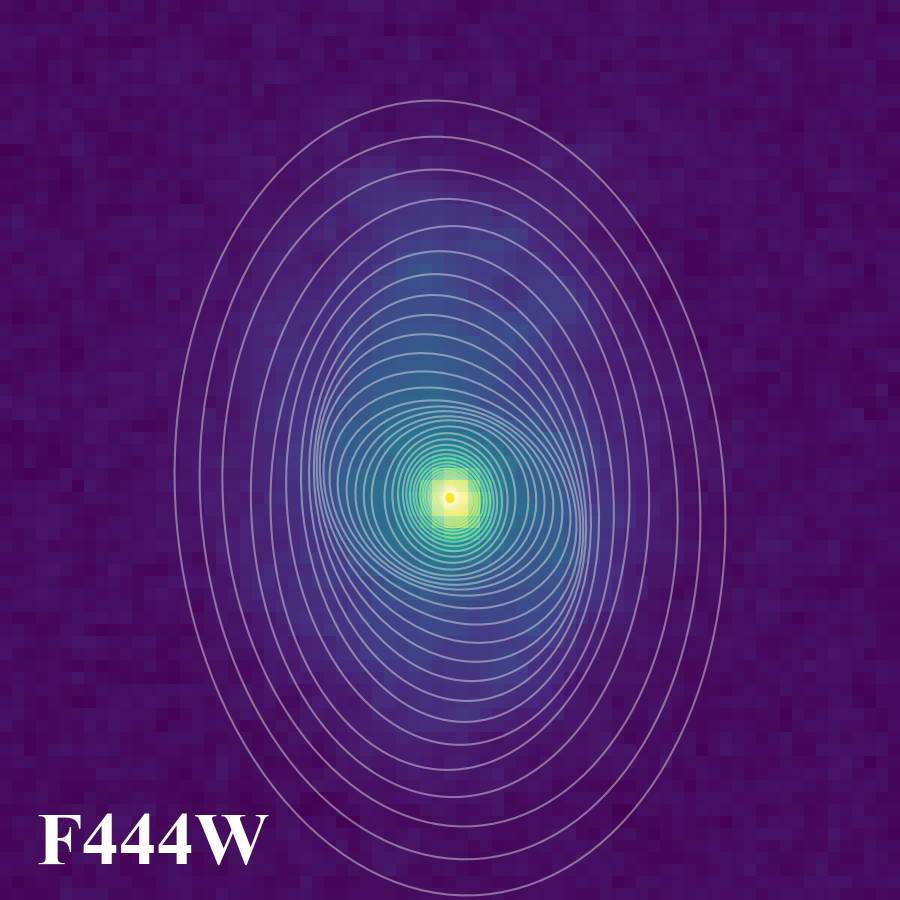}
    }
    \hspace{\fill}
    \subfigure
    {
        \includegraphics[width=0.3\textwidth]{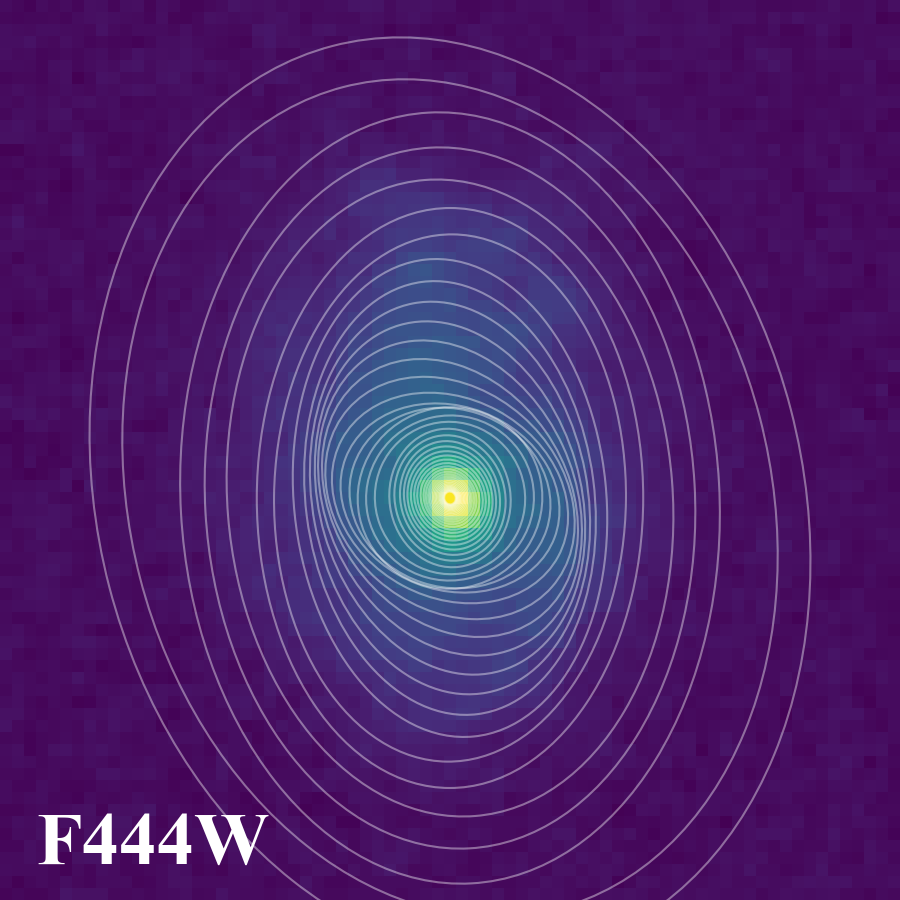}
    }
    \hspace{\fill}
    \subfigure
    {
        \includegraphics[width=0.3\textwidth]{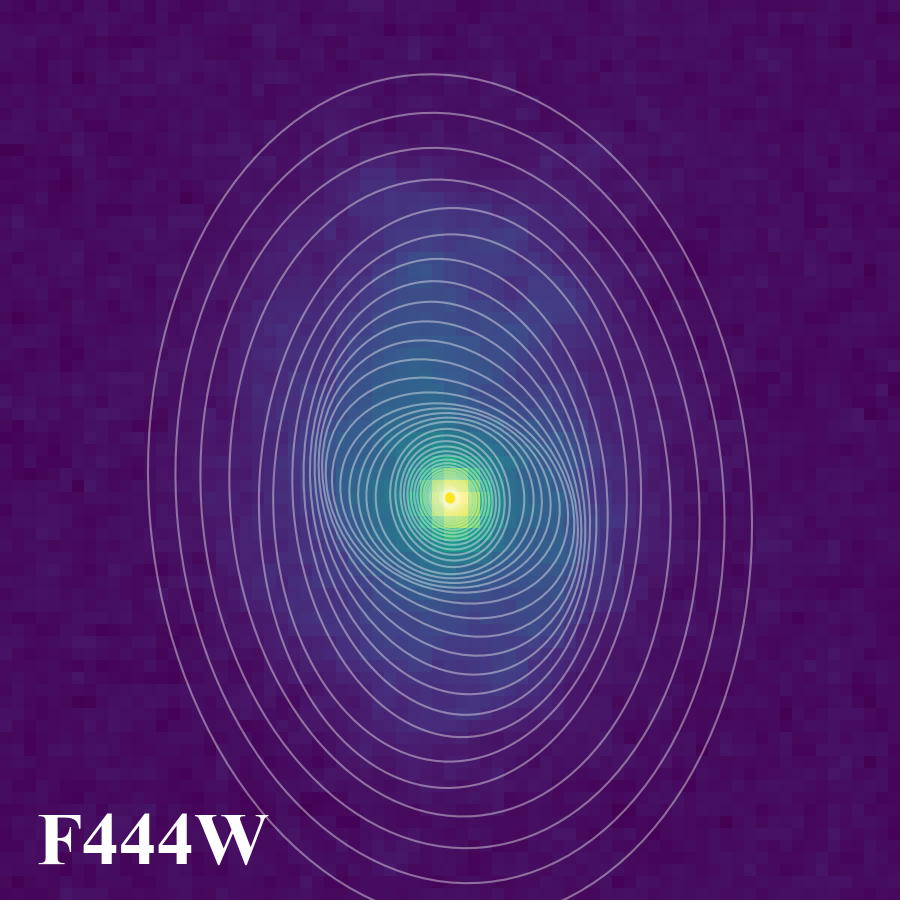}
    }
    \caption{Elliptical isophotes for stack 2 masked images. Rows from top to bottom: F200W, F277W, F356W, F444W. Columns from left to right: $3\sigma$, $3.5\sigma$, $4\sigma$ masks.}
    \label{fig:isophotes_appendix_large}
\end{figure*}

\begin{figure*}[t]
    \centering
    \subfigure
    {
        \includegraphics[width=1.0\textwidth]{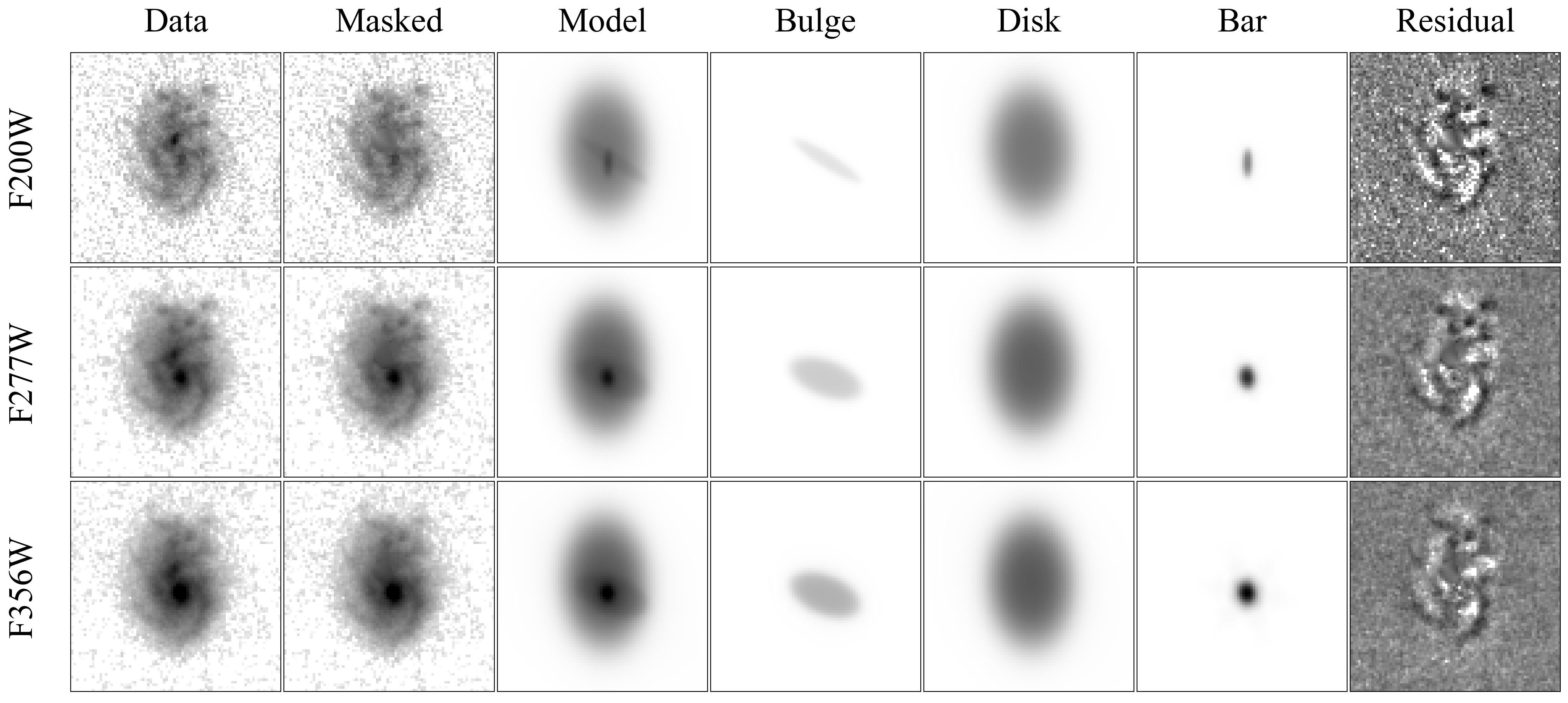}
    }

    \caption{As \autoref{fig:decomp} but for the three-component model. The component intended to correspond to the bulge does not adopt a physically realistic profile. Constraining it to be circular results in boundary-hitting which prevents the MCMC code from adequately exploring the parameter space.}
    \label{fig:morph_3comp}
\end{figure*}

\begin{figure*}[t]
    \centering
    \subfigure
    {
        \includegraphics[width=1.0\textwidth]{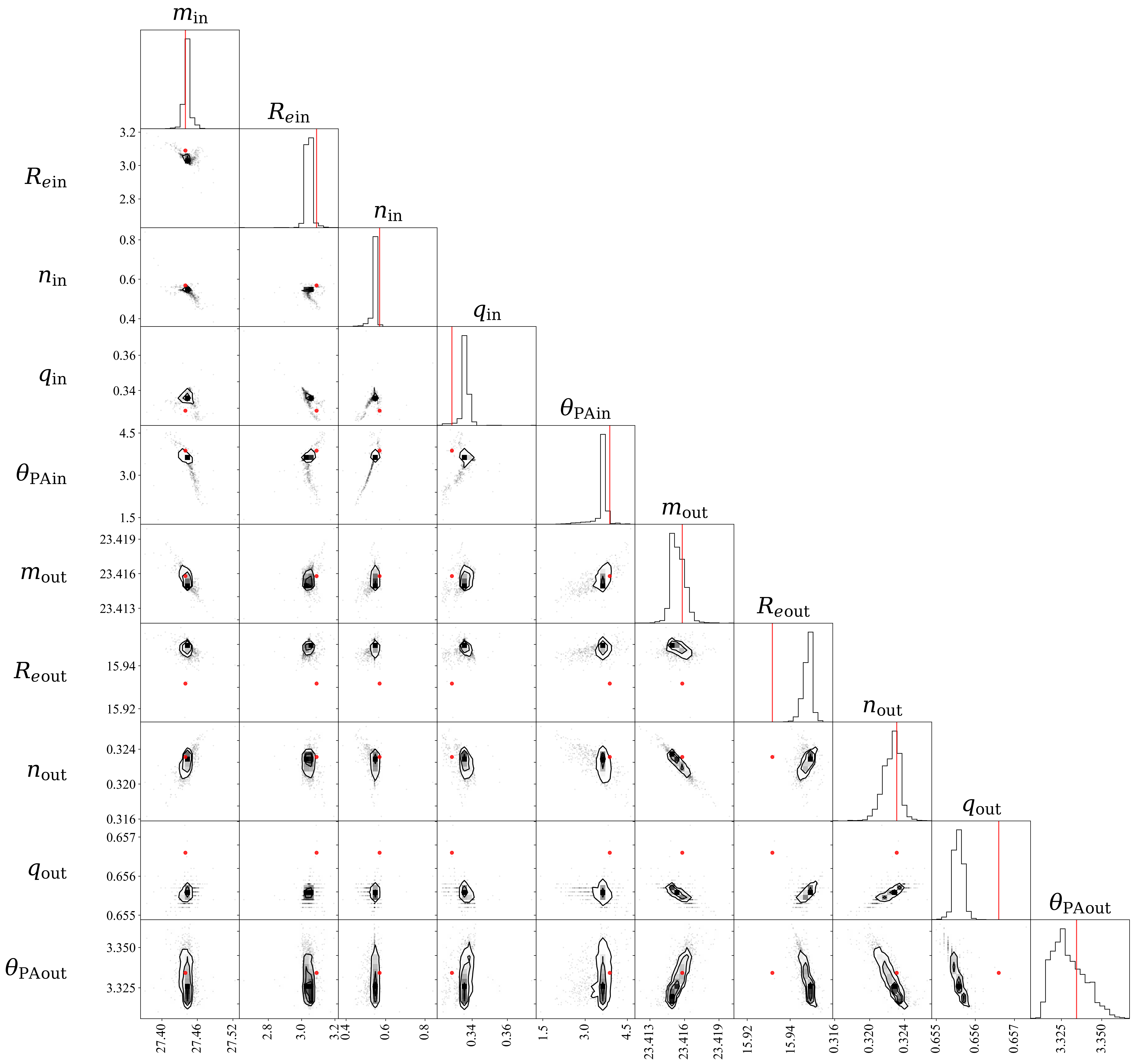}
    }

    \caption{Corner plot of the posterior of the MCMC exploration of the F200W filter parameter space. Red dots and lines indicate the best $\chi^2$.}
    \label{fig:corner_200}
\end{figure*}

\begin{figure*}[t]
    \centering
    \subfigure
    {
        \includegraphics[width=1.0\textwidth]{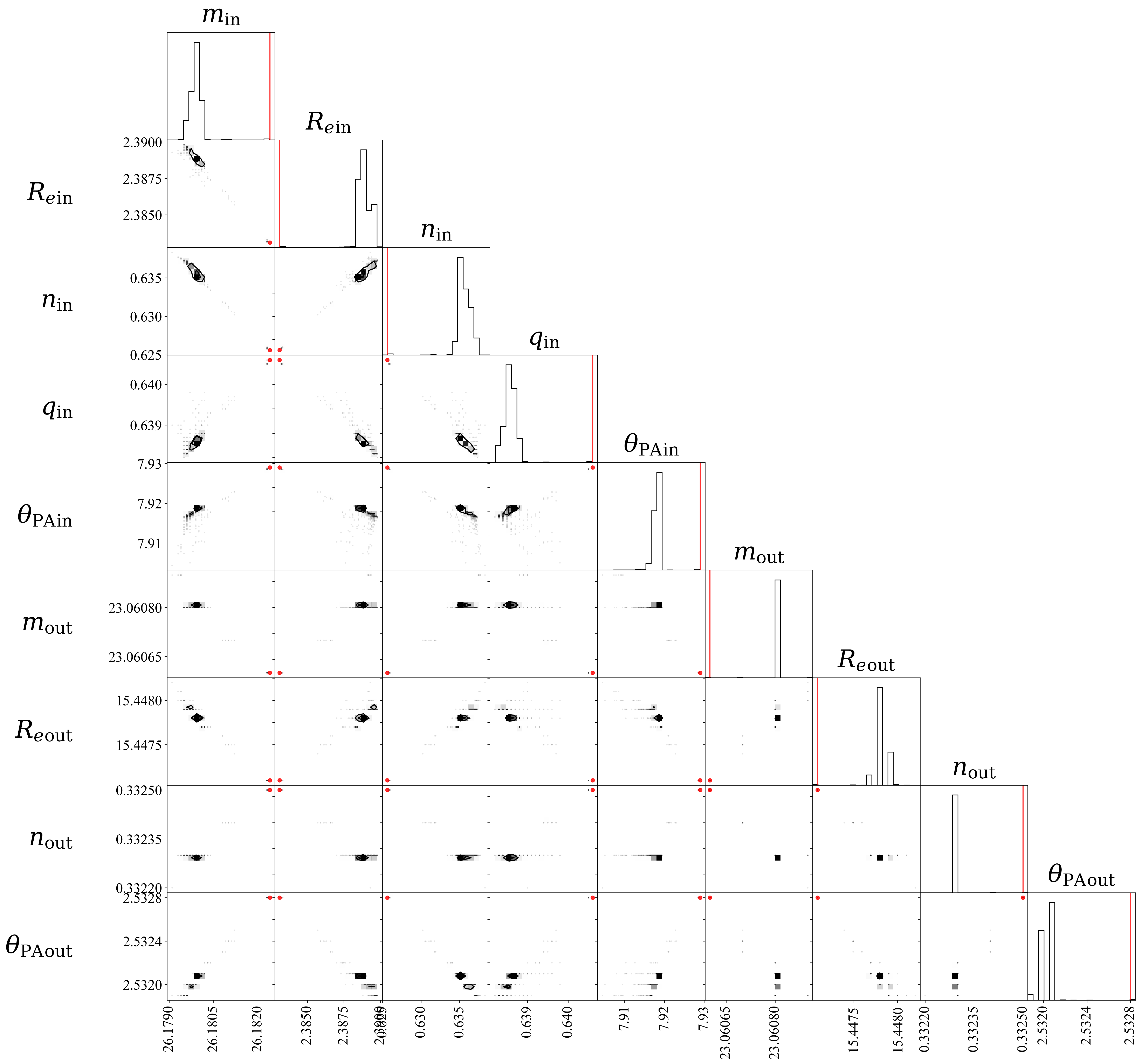}
    }

    \caption{As \autoref{fig:corner_200} but for F277W.}
    \label{fig:corner_277}
\end{figure*}

\begin{figure*}[t]
    \centering
    \subfigure
    {
        \includegraphics[width=1.0\textwidth]{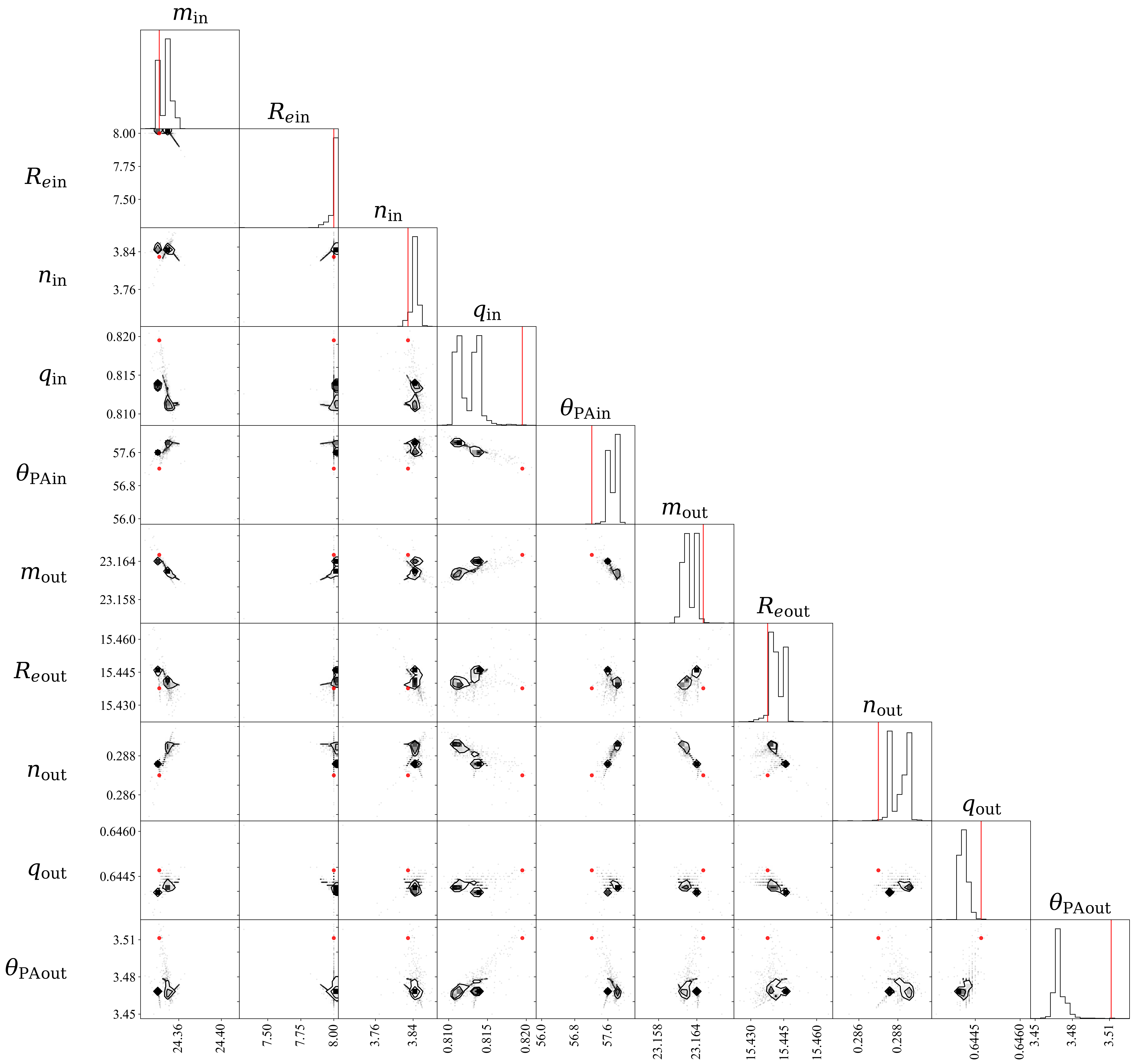}
    }

    \caption{As \autoref{fig:corner_200} but for F356W.}
    \label{fig:corner_356}
\end{figure*}

We produced a number of diagnostic plots for the purposes of verifying our findings. \autoref{fig:isophotes_appendix_imgs} shows COSMOS-74706 in the $2.5\sigma$ mask across six different filters with the elliptical isophotes used in our analysis overlaid. \autoref{fig:isophotes_appendix_large} shows the same but for the $3\sigma$, $3.5\sigma$, and $4\sigma$ masks omitting the F150W filter, which did not have a converged set of isophotes for all three masks. \autoref{fig:morph_3comp} shows the three-component GALFIT decomposition of the target object demonstrating the influence of the inner portion of the spiral arms on the profiles. \autoref{fig:corner_200}, \autoref{fig:corner_277}, and \autoref{fig:corner_356} are corner plots depicting the posterior of the Monte-Carlo meta-exploration of the GALFIT output space across all three filters in which the analysis was performed. \autoref{tab:deproj} lists the bar modulus which results upon the deprojection of COSMOS-74706 at various angles. Finally, \autoref{tab:galfit} lists the constraints placed on the GALFIT fitting algorithm during the course of the parameter exploration.

    \begin{table}[h]
    \centering
  \setlength{\tabcolsep}{18pt}
  \begin{tabular}{ c c }
    \hline
    \textbf{IA} & \textbf{$\boldsymbol{B_{M}}$}  \\
    \hline
    10\degree  & 13.02 \\
    20\degree  & 8.88 \\
    30\degree   & 9.44 \\
    40\degree   & 6.61\\
    50\degree   & 0.73 \\
    60\degree   & 11.35 \\
    70\degree   & 18.45 \\
    \hline
  \end{tabular}
  \caption{Bar modulus corresponding to different deprojections of the 2.5$\sigma$ mask F200W filter image of COSMOS-74706. Only the deprojection assuming a 50\degree inclination angle leads to a bar modulus which falls below the cutoff threshold. Position angle was held constant so that changes in the inclination angle would decrease the ellipticity of the disk.}
  \label{tab:deproj}
\end{table}

\newpage

\clearpage

\begin{table}[h!]
  \centering
  \caption{GALFIT parameter constraints used for the two-component models given as [min, max]. $\Delta x, \Delta y$ represent the range of pixels in both the $x$ and $y$ directions over which the center of the component was allowed to vary from an input position. All parameters are as in \autoref{tab:params_all}. The position angle of both components was left free to vary.}
  \label{tab:galfit}
  \begin{tabular}{l c}
    \hline
    \textbf{Parameter} & \textbf{Constraint} \\
    \hline
    \multicolumn{2}{l}{\textit{Inner Light}} \\
    \quad $\Delta$x, $\Delta$y (px) & [$-$1.0, 1.0] \\
    \quad mag                       & [15.0, 30.0]\\
    \quad $R_e$ (px)               & [1.0, 8.0] \\
    \quad $n$ & [0.1, 6.0] \\
    \quad $q$ (b/a)                 & [0.25, 1.0]    \smallskip \\
    \hline
    \multicolumn{2}{l}{\textit{Outer Disk}} \\
    \quad $\Delta$x, $\Delta$y (px) & [$-$2.5, 2.5] \\
    \quad mag                       &  [15.0, 30.0] \\
    \quad $R_e$ (px)               & [10.0, 30.0] \\
    \quad $n$ & [0.1, 3.0] \\
    \quad $q$ (b/a)                 & [0.15, 1.0]    \smallskip \\
    \hline
  \end{tabular}
\end{table}

    \clearpage


\end{document}